\documentclass[11pt]{article}
\usepackage{amsfonts}
\usepackage{amsmath}
\usepackage{amssymb}
\usepackage{mathtools}
\usepackage{graphicx}
\usepackage{geometry}
\usepackage{listings}
\usepackage{mathrsfs}
\usepackage{natbib}
\usepackage{bm}
\usepackage{subcaption}
\usepackage{multirow}
\usepackage{hyperref}
\usepackage{booktabs, caption, makecell}

\usepackage{threeparttable}
\usepackage{enumitem}
\usepackage{color, colortbl}
\usepackage{float}
\usepackage{hhline}

\usepackage[title]{appendix}
\definecolor{Gray}{gray}{0.5}

\usepackage{algorithm,algcompatible}
\algnewcommand\INPUT{\item[\textbf{Input:}]}%
\algnewcommand\OUTPUT{\item[\textbf{Output:}]}%

\usepackage{xr}
\makeatletter
\newcommand*{\addFileDependency}[1]{
  \typeout{(#1)}
  \@addtofilelist{#1}
  \IfFileExists{#1}{}{\typeout{No file #1.}}
}
\makeatother

\newcommand*{\myexternaldocument}[1]{%
    \externaldocument{#1}%
    \addFileDependency{#1.tex}%
    \addFileDependency{#1.aux}%
}
\myexternaldocument{supplementary}

\providecommand{\U}[1]{\protect\rule{.1in}{.1in}}

\newtheorem{theorem}{Theorem}

\newtheorem{proposition}{Proposition}

\newtheorem{remark}{Remark}

\numberwithin{equation}{section}
\begin{document}

\title{Maximum weighted likelihood estimator for robust heavy-tail modelling of finite mixture models}

\author{Tsz Chai Fung\thanks{Address: Department of Risk Management and Insurance, Georgia State University, Atlanta, GA 30303. Email: \href{mailto:tfung@gsu.edu}{tfung@gsu.edu}.}}
\maketitle

\abstract{In this article, we present the maximum weighted likelihood estimator (MWLE) for robust estimations of heavy-tail finite mixture models (FMM). This is motivated by the complex distributional phenomena of insurance claim severity data, where flexible density estimation tools such as FMM are needed but MLE often produces unstable tail estimates under FMM. Under some regularity conditions, MWLE is proved to be consistent and asymptotically normal. We further prove that the tail index obtained by MWLE is consistent even if the model is misspecified, justifying the robustness of MWLE in estimating the tail part of FMM. With a probabilistic interpretation for MWLE, Generalized Expectation-Maximization (GEM) algorithm is still applicable for efficient parameter estimations. We therefore present and compare two distinctive constructions of complete data to implement the GEM algorithm. By exemplifying our approach on two simulation studies and a real motor insurance data set, we show that comparing to MLE, MWLE produces more appropriate estimations on the tail part of FMM, without much sacrificing the flexibility of FMM in capturing the body part.}

\medskip


\section{Introduction}
Modelling insurance claim sizes is not only essential in various actuarial applications including pricing and risk management, but is also very challenging due to several peculiar characteristics of claim severity distributions. Claim size distributions often exhibit multimodality for small and moderate claims, when there exists unobserved heterogeneities possibly reflected by different claim types and accident causes, or when the observed samples come from a contaminated distribution. Also, the distribution is usually heavy-tail in nature, where very large claims occur with a small but non-negligible probability.
Due to the highly complex distributional characteristics, we have to admit the impossibility to perfectly capture all the distributional features using a parametric model without excessively large number of parameters (which results in over-fitting). When model misspecification is unavoidable, correct specification of the tail part is more important than finely capturing the distributional nodes of smaller claims which are rather immaterial to the insurance portfolio, because the large claims are the losses which can severely damage the portfolio. As a result, we need to specify an appropriate distributional model with a justifiable statistical inference approach which not only preserves sufficient flexibility to appropriately capture the whole severity distribution, but also puts a particular emphasis on robust estimation of the tail.


Existing actuarial loss modelling literature focus a lot on the model specifications, by introducing various distributional classes to capture the peculiar characteristics of claim severity distributions. Notable directions include extreme value distributions (EVD, see e.g. \cite{embrechts1999extreme}) to capture the heavy-tailedness, composite loss modelling (see e.g. \cite{cooray2005modeling}, \cite{scollnik2007composite}, \cite{bakar2015modeling} and \cite{grun2019extending}) to cater for mismatch between body and tail behavior of claim severity distributions, and finite mixture model (FMM, see e.g. \cite{LEE2010modeling} and \cite{MILJKOVIC2016387}) to capture distributional multimodality.
In particular, FMM is becoming an increasingly useful smooth density estimation tool in insurance claim severity modelling perspective, due to its high versatility theoretically justified by denseness theorems (\cite{LEE2010modeling}).
The mismatch between its body and tail behavior can also be easily modelled by FMM by selecting varying distributional classes among mixture component functions (see e.g. \cite{BLOSTEIN201935} and \cite{fung2021mixture}), including both light-tailed and heavy-tailed distributions. In both actuarial research and practice, statistical inferences of FMM are predominantly based on maximum likelihood estimation (MLE) with the use of Expectation-Maximization (EM) algorithm.

Nonetheless, MLE would often cause tail-robustness issues where the tail part of the fitted model is very sensitive to model misspecifications -- when the observations are generated from a perturbed and/or contaminated distribution.
As evidenced by several empirical studies including \cite{fung2021mixture} and \cite{wuthrich2021statistical}, the estimated tail part of the FMM obtained by MLE can be unreliable and highly unstable in most practical cases. This is mainly due to the overlapping density regions between mixture components modelling small to moderate claims (body) and those modelling large claims (tail). Hence, the estimated tail distribution will be heavily influenced by some smaller claims if FMM is not able to fully explain those small claims, which is always the case in practice due to the distributional complexity of real dataset impossible to be perfectly captured even by flexible density approximation tools including FMM. Under MLE approach, FMM may fail to extrapolate well the large claims, and this would lead to serious implications to insurance pricing and risk management perspectives. It is therefore natural to question whether or not MLE is still a plausible approach in modelling actuarial claim severity data, and whether or not there exists an alternative statistical inference tool which better addresses our modelling challenges and outperforms the MLE.

Robust statistical inference methods for heavy tail distributions are relatively scarce in actuarial science literature. 
Notable contributions include \cite{brazauskas2000robust}, \cite{serfling2002efficient}, \cite{brazauskas2003favorable} and \cite{dornheim2007robust} who adopt various kinds of statistical inference tools, such as quantile, trimmed mean, trimmed-M and generalized median estimators, to robustly estimate Gamma, Pareto and Log-normal distributions.
Recent actuarial works study several variations of the method of moments (MoM) for robust estimations of Pareto and Log-normal loss models. Notable contributions in this direction include \cite{brazauskas2009robust}, \cite{poudyal2021robust} (trimmed moments), \cite{zhao2018robust} (winsorized moments) and \cite{poudyal2021truncated} (truncated moments).
Note that these research works address robustness issue against the upper outliers by reducing the influence of few extreme observations to the estimated model parameters. This outlier-robustness issue is however very different from the tail-robustness issue mentioned above as the key motivation of this paper, where the contaminations on the body part affects the tail extrapolations. Very few research works look into this ``non-standard" tail-robustness issue. Notable contributions are \cite{beran2012robust} and \cite{gong2018robust} who propose a huberization of the MLE, which protects against perturbations and misspecifications in the body part of distribution, to robustly estimate the tail index of Pareto and Weibull distributions. 
All of the above existing approaches focus solely on one or two-parameter distributions. A general robust tail estimation strategy under multi-parameter flexible models like FMM is lacking.

Motivated by the aforementioned tail-robustness issue in insurance context, we propose a new maximum weighted likelihood estimation (MWLE) approach for robust heavy-tail modelling of FMM. 
Under the MWLE, an observation-dependent weight function is introduced to the log-likelihood, de-emphasizing the contributions of smaller claims and hence reducing their influence to the estimated tail part of FMM. Down-weighting small claims is also natural in insurance loss modelling perspective, as mentioned in the beginning of this section, accurate modelling of the large claims is more important than the smaller claims. 
To offset the bias caused by the weighting scheme, we also include an adjustment term in the weighted log-likelihood, which can be interpreted as the effects of randomly truncating the observations. With the bias adjustment term, we prove that estimated parameters under the proposed MWLE is consistent and asymptotically normal with any pre-specified choices of weight functions.
Also, under some specific choices of weight functions, we will show that the MWLE tail index, which determines the tail heaviness of a distribution, is consistent, even under model misspecifications where the true model is not FMM. Therefore, MWLE can be regarded as a generalized alternative framework of Hill estimator (\cite{hill1975simple}). Furthermore, with a probabilistic interpretation of the proposed MWLE approach, it is still possible to derive a Generalized EM (GEM) algorithm to efficiently estimate parameters which maximize the weighted log-likelihood function.

Note that the proposed MWLE is different from the existing statistics papers which adopt weighting schemes for likelihood-based inference. The existing literature are mainly motivated by one of the following two aspects very different from the focus of this paper: (i) Robustness against upper and lower outliers, where related research works include e.g. \cite{fieldsmith1994}, \cite{markatou1997weighted}, \cite{markatou2000mixture}, \cite{dupuis2002robust}, \cite{ahmed2005robust}, \cite{wong2014robust} and \cite{aeberhard2021robust}; (ii) Bringing in more relevant observations for statistical inference to increase precision while trading off some biases, studied by e.g. \cite{wang2001maximum}, \cite{hu2002weighted}, \cite{wang2004asymptotic} and \cite{wang2005selecting}. Note also that our proposed MWLE stems differently from the existing statistics literature in terms of mathematical technicality, since none of the above papers incorporate the truncation-based bias adjustment as included in the proposed WMLE.

The rest of this paper is structured as follows. In Section \ref{sec:fmm}, we briefly revisit the class of FMM for heavy tail modelling. Section \ref{sec:MWLE} introduces the proposed MWLE for robust heavy-tail modelling of FMM and explains its motivations in terms of insurance claims modelling. Section \ref{sec:theory} explores several theoretical properties to understand and justify the proposed MWLE. After that, we present in Section \ref{sec:em} two types of GEM algorithms for efficient parameter estimations under the MWLE approach on FMM. In Section \ref{sec:ex}, we analyze the performance of the proposed MWLE through three empirical examples: a toy example, a simulation study and a real insurance claim severity dataset. After showing the superior performance of MWLE compared to MLE, we finally summarize our findings in Section \ref{sec:discussion} with a brief discussion how the proposed MWLE can be extended to a regression framework.

\section{Finite mixture model} \label{sec:fmm}
This section provides a very brief review on finite mixture model (FMM) which serves as a flexible density estimation tool. Suppose that there are $n$ i.i.d. claim severities $\bm{Y}=(Y_1,\ldots, Y_n)$ with realizations $\bm{y}=(y_1,\ldots, y_n)$. $Y_i$ is generated by a probability distribution of $G(\cdot)$ with density function $g(\cdot)$ which is unknown. In insurance context, claim severity distribution often exhibits multimodality, which results from the unobserved heterogeneity stemming from the amalgamation of different types of claims unobserved in advance. Also, claim sizes are often heavy-tail in nature, which can be attributed to a few large losses from a portfolio of policies which usually represent the greatest part of the indemnities paid by the insurance company. The mismatch between body and tail behavior often poses difficulties to fit the data well using only a standard parametric distribution.

Motivated by the challenges of modelling insurance claim severities, we aim to model the dataset using finite mixture model (FMM). Define a class of finite mixture distributions $\mathcal{H}=\{H(\cdot;\bm{\Phi}):\bm{\Phi}\in\Omega\}$, where $\bm{\Phi}=(\psi_1,\ldots,\psi_P)^T$ is a column vector with length $P$ representing the model parameters and $\Omega$ is the parameter space. Its density function $h(y;\bm{\Phi})$ is given by the following form:

\begin{equation} \label{eq:model}
h(y_i;\bm{\Phi})=\sum_{j=1}^{J}\pi_j f_b(y_i;\bm{\varphi}_j)+\pi_{J+1}f_t(y_i;\bm{\eta}),\qquad y_i>0,
\end{equation}
where the parameters $\bm{\Phi}$ can alternatively be written as $\bm{\Phi}=(\bm{\pi},\bm{\varphi},\bm{\eta})$. Here, $\bm{\pi}=(\pi_1,\ldots,\pi_{J+1})$ are the mixture probabilities for each of the $J+1$ components with $\sum_{j=1}^{J+1}\pi_j=1$. $\bm{\varphi}=(\bm{\varphi}_1,\ldots,\bm{\varphi}_J)$ and $\bm{\eta}$ are the parameters for the mixture densities $f_b$ and $f_t$ respectively.

The $J$ mixture components with densities $f_b$ mainly serve as modelling the multimodality for the body part of the distribution. $f_b$ is naturally chosen as a light-tailed distribution like Gamma, Weibull and Inverse-Gaussian. 
The remaining mixture component $f_t$ is designed to capture the large observations and hence the tail distribution can be properly extrapolated. The possible choices of heavy-tail distribution for $f_t$ include Log-normal, Pareto and Inverse-Gamma.

\section{Maximum weighted log-likelihood estimator} \label{sec:MWLE}
With the maximum likelihood estimation (MLE) approach, parameter estimations require maximizing the log-likelihood function

\begin{equation} \label{eq:loglik}
\mathcal{L}_n(\bm{\Phi};\bm{y})=\sum_{i=1}^{n}\log h(y_i;\bm{\Phi})
\end{equation}
with respect to the parameters $\bm{\Phi}$. Under this approach, each claim has the same relative influence to the estimated parameters, but in insurance loss modelling and ratemaking perspective, correct specification and projection of larger claims are more important than those of smaller claims. More importantly, as explained by \cite{fung2021mixture} and \cite{wuthrich2021statistical}, MLE of FMM in Equation (\ref{eq:model}) would fail in most practical cases due to incorrectly estimations of tail heaviness under model misspecification. Precisely, because of the overlapping region between the body parts $f_b$ and tail part $f_t$ of the distribution, the small claims may distort the estimated tail distribution $f_t$ if they are not fully captured by the $J$ mixture densities in the body. However, due to the highly complex multimodality characteristics of the body distribution which often appears in real insurance claim severity data, it is impossible to capture all the body distributional patterns without prohibitively large $J$ which causes over-fitting and loss of model interpretability. Therefore, it is often the case in practice that MLE of FMM would result in unstable estimates of tail distribution, causing unreliable tail extrapolation.

One way to mitigate the aforementioned model misspecification effect is to impose observation-dependent weights to the log-likelihood function, where a larger claim $y$ is assigned to a larger weight. This will reduce the influence of smaller observed values to the estimated tail parameter. For parameter estimations, we propose maximizing the weighted log-likelihood as follows instead

\begin{equation} \label{eq:loglik_weight}
\mathcal{L}^*_n(\bm{\Phi};\bm{y})=\sum_{i=1}^{n}W(y_i)\log \frac{h(y_i;\bm{\Phi})W(y_i)}{\int_{0}^{\infty}h(u;\bm{\Phi})W(u)du},
\end{equation}
where $0\leq W(\cdot)\leq 1$ is the weight of the log-likelihood function. We call the resulting parameters maximum weighted likelihood estimators (MWLE). To allow for greater relative influence of larger claims, we construct $W(u)$ as a monotonically non-decreasing function of $u$. In this case, we may interpret the weighted log-likelihood function as follows: First, we pretend that each claims $y_1,\ldots,y_n$ are only observed respectively by $W(y_1),\ldots,W(y_n)$ times. However, this alone will introduce bias to a heavier estimated tail because this implies more large claims are effectively included due to the weighting effect. To remove such a bias, we pretend to model $y_i$ by a random truncation model $\tilde{h}(y_i;\bm{\Phi}):=h(y_i;\bm{\Phi})W(y_i)/\int_{0}^{\infty}h(u;\bm{\Phi})W(u)du$ instead of the original modelling distribution $h(y_i;\bm{\Phi})$.

\begin{remark}
The proposed MWLE can be viewed as a form of M-estimator, where the optimal parameters are determined through maximizing a function. We here discuss two special cases of MWLE. (i) MLE: When $W(\cdot)=1$, MWLE is reduced to a standard MLE; (ii) Truncated MLE: When $W(y)=1\{y\geq\tau\}$ for some threshold $\tau>0$, then MWLE is reduced to truncated MLE introduced by \cite{marazzi2004adaptively}, where a hard rejection is applied to all samples smaller than $\tau$.
\end{remark}


\section{Theoretical Properties} \label{sec:theory}
This section presents several theoretical properties of the proposed MWLE to theoretically justify the use of the proposed MWLE. Unless specified otherwise, throughout this section the estimated model parameters $\hat{\bm{\Phi}}$ are obtained by maximizing the proposed weighted log-likelihood function given by Equation (\ref{eq:loglik_weight}).
\subsection{Asymptotic behavior with fixed weight function}
\subsubsection{Consistency and asymptotic normality}
We first want to show that the proposed weighted log-likelihood approach leads to correct convergence to true model parameters as $n\rightarrow\infty$. The proof is presented in Section 2 of the supplementary materials.
\begin{theorem} \label{thm:asym_tru}
Suppose that $G(\cdot)=H(\cdot;\bm{\Phi}_0)\in\mathcal{H}$. Assume that the density function $h(y;\bm{\Phi})$ satisfies a set of regularity conditions outlined in Section 1 of supplementary materials\footnote{Note that the set of regularity conditions are equivalent to those required for consistent and asymptotic normal estimations of MLE.}. Then, there exists a local maximizer $\hat{\bm{\Phi}}_n$ of the weighted log-likelihood $\mathcal{L}^{*}_n(\bm{\Phi};\bm{y})$ such that
\begin{equation}
\sqrt{n}(\hat{\bm{\Phi}}_n-\bm{\Phi}_0)\overset{d}{\rightarrow}\mathcal{N}(\bm{0},\bm{\Sigma}),
\end{equation}
where $\bm{\Sigma}=\bm{\Gamma}^{-1}\bm{\Lambda}\bm{\Gamma}^{-1}$, with $\bm{\Lambda}$ and $\bm{\Gamma}$ being $P\times P$ matrices given by
\begin{align} \label{eq:asym:lambda}
\bm{\Lambda} 
&=E_{\bm{\Phi}_0}\left[W(Y)^2\left[\frac{\partial}{\partial\bm{\Phi}}\log h(Y;\bm{\Phi})\right]\left[\frac{\partial}{\partial\bm{\Phi}}\log h(Y;\bm{\Phi})\right]^T\Bigg|_{\bm{\Phi}=\bm{\Phi}_0}\right]\nonumber\\
&\quad-\frac{1}{E_{\bm{\Phi}_0}\left[W(Y)\right]}\Bigg\{E_{\bm{\Phi}_0}\left[W(Y)^2\frac{\partial}{\partial\bm{\Phi}}\log h(Y;\bm{\Phi})\Bigg|_{\bm{\Phi}=\bm{\Phi}_0}\right]E_{\bm{\Phi}_0}\left[W(Y)\frac{\partial}{\partial\bm{\Phi}}\log h(Y;\bm{\Phi})\Bigg|_{\bm{\Phi}=\bm{\Phi}_0}\right]^T\nonumber\\
&\hspace{8em}+E_{\bm{\Phi}_0}\left[W(Y)\frac{\partial}{\partial\bm{\Phi}}\log h(Y;\bm{\Phi})\Bigg|_{\bm{\Phi}=\bm{\Phi}_0}\right]E_{\bm{\Phi}_0}\left[W(Y)^2\frac{\partial}{\partial\bm{\Phi}}\log h(Y;\bm{\Phi})\Bigg|_{\bm{\Phi}=\bm{\Phi}_0}\right]^T\Bigg\}\nonumber\\
&\quad+\frac{E_{\bm{\Phi}_0}\left[W(Y)^2\right]}{E_{\bm{\Phi}_0}\left[W(Y)\right]^2}E_{\bm{\Phi}_0}\left[W(Y)\frac{\partial}{\partial\bm{\Phi}}\log h(Y;\bm{\Phi})\Bigg|_{\bm{\Phi}=\bm{\Phi}_0}\right]E_{\bm{\Phi}_0}\left[W(Y)\frac{\partial}{\partial\bm{\Phi}}\log h(Y;\bm{\Phi})\Bigg|_{\bm{\Phi}=\bm{\Phi}_0}\right]^T
\end{align}
and
\begin{align} \label{eq:asym:gamma}
\bm{\Gamma}
&=-E_{\bm{\Phi}_0}\left[W(Y)\left[\frac{\partial}{\partial\bm{\Phi}}\log h(Y;\bm{\Phi})\right]\left[\frac{\partial}{\partial\bm{\Phi}}\log h(Y;\bm{\Phi})\right]^T\Bigg|_{\bm{\Phi}=\bm{\Phi}_0}\right]\nonumber\\
&\quad+\frac{1}{E_{\bm{\Phi}_0}\left[W(Y)\right]}E_{\bm{\Phi}_0}\left[W(Y)\frac{\partial}{\partial\bm{\Phi}}\log h(Y;\bm{\Phi})\Bigg|_{\bm{\Phi}=\bm{\Phi}_0}\right]E_{\bm{\Phi}_0}\left[W(Y)\frac{\partial}{\partial\bm{\Phi}}\log h(Y;\bm{\Phi})\Bigg|_{\bm{\Phi}=\bm{\Phi}_0}\right]^T,
\end{align}
where $E_{\bm{\Phi}_0}[Q(Y)]=\int_{0}^{\infty}Q(u)h(u;\bm{\Phi}_0)du$ represents the expectation under density $h(\cdot;\bm{\Phi}_0)$ for any functions $Q$, and the derivative ${\partial}/{\partial\bm{\Phi}}\log h(Y;\bm{\Phi})$ is assumed to be a column vector with length $P$.
\end{theorem}

\begin{remark}
When $W(\cdot)=1$, all except the first term in the right hand side of Equations (\ref{eq:asym:lambda}) and (\ref{eq:asym:gamma}) vanish. As a result, the asymptotic variance $\bm{\Sigma}=\bm{\Gamma}^{-1}\bm{\Lambda}\bm{\Gamma}^{-1}$ is reduced to the inverse of Fisher information matrix under standard MLE approach.
\end{remark}

\begin{remark}
Theorem \ref{eq:asym:gamma} only asserts the existence of local maximizer instead of global maximizer, because in FMM it is common that the likelihood function has multiple critical points and/or is unbounded (\cite{Mclachlan2004Finite}).
\end{remark}

The above theorem suggest that for large sample size, the estimated parameters are approximately unbiased and we may approximate the variance of estimated parameters as
\begin{equation} \label{eq:asym_var}
\widehat{\text{Var}}(\hat{\bm{\Phi}}_{n})\approx \frac{1}{n}\hat{\bm{\Gamma}}_n^{-1}\hat{\bm{\Lambda}}_n\hat{\bm{\Gamma}}_n^{-1}
\end{equation}
where $\hat{\bm{\Lambda}}_n$ and $\hat{\bm{\Gamma}}_n$ are given by $\bm{\Lambda}$ and $\bm{\Gamma}$ in Equations (\ref{eq:asym:lambda}) and (\ref{eq:asym:gamma}) except that the expectations are changed to empirical means and $\bm{\Phi}_0$ is changed to $\hat{\bm{\Phi}}_n$. Then, it is easy to construct a two-sided Wald-type confidence interval (CI) for $\psi_p$ ($p=1,\ldots,P$) as
\begin{equation} \label{eq:asym_CI}
\left[\hat{\psi}_{n,p}-\frac{z_{1-\kappa/2}}{\sqrt{n}}\sqrt{\left[\hat{\bm{\Gamma}}_n^{-1}\hat{\bm{\Lambda}}_n\hat{\bm{\Gamma}}_n^{-1}\right]_{p,p}},\hat{\psi}_{n,p}+\frac{z_{\kappa/2}}{\sqrt{n}}\sqrt{\left[\hat{\bm{\Gamma}}_n^{-1}\hat{\bm{\Lambda}}_n\hat{\bm{\Gamma}}_n^{-1}\right]_{p,p}}\right],
\end{equation}
where $\hat{\psi}_{n,p}$ is the estimated $\psi_p$, $z_{\kappa}$ is the $\kappa$-quantile of the standard normal distribution and $\left[\bm{M}\right]_{p,p}$ is the $(p,p)$-th element of $\bm{M}$ for some matrices $\bm{M}$. For other quantities of interest (e.g.~mean, VaR and CTE of claim amounts), one may apply a delta method or simulate parameters from ${\cal N}(\hat{\bm{\Phi}}_{n},\widehat{\text{Var}}(\hat{\bm{\Phi}}_{n}))$ to analytically or empirically approximate their CIs.

\medskip

Next, we examine the asymptotic property of MWLE dropping the assumption that $G(\cdot)\in\mathcal{H}$ (i.e. we may misspecify the model class).

\begin{theorem} \label{thm:asym_mis}
Assume that the density function $h(y;\bm{\Phi})$ satisfies the same set of regularity conditions as in the previous theorem. Further assume that there is a local maximizer $\bm{\Phi}_0^{*}$ of
\begin{equation}
\tilde{E}\left[\mathcal{L}^{*}(\bm{\Phi};Y)\right]:=
\tilde{E}\left[W(Y)\log \frac{h(Y;\bm{\Phi})W(Y)}{\int_{0}^{\infty}h(u;\bm{\Phi})W(u)du}\right],
\end{equation}
where $\tilde{E}\left[Q(Y)\right]=\int_0^{\infty}Q(u)dG(u)$ represents the expectation under distribution $G(y)$ for any functions $Q$. Then, there exists a local maximizer $\hat{\bm{\Phi}}_n$ of the weighted log-likelihood $\mathcal{L}^{*}_n(\bm{\Phi};\bm{y})$ such that
\begin{equation}
\sqrt{n}(\hat{\bm{\Phi}}_n-\bm{\Phi}_0^{*})\overset{d}{\rightarrow}\mathcal{N}(\bm{0},\tilde{\bm{\Sigma}}),
\end{equation}
where $\tilde{\bm{\Sigma}}=\tilde{\bm{\Gamma}}^{-1}\tilde{\bm{\Lambda}}\tilde{\bm{\Gamma}}^{-1}$, with $\tilde{\bm{\Lambda}}$ and $\tilde{\bm{\Gamma}}$ given by
\begin{align} \label{eq:asym:lambda_mis}
\tilde{\bm{\Lambda}} 
&=\tilde{E}\left[W(Y)^2\left[\frac{\partial}{\partial\bm{\Phi}}\log h(Y;\bm{\Phi})\right]\left[\frac{\partial}{\partial\bm{\Phi}}\log h(Y;\bm{\Phi})\right]^T\Bigg|_{\bm{\Phi}=\bm{\Phi}_0}\right]\nonumber\\
&\quad-\frac{1}{E_{\bm{\Phi}_0}\left[W(Y)\right]}\Bigg\{\tilde{E}\left[W(Y)^2\frac{\partial}{\partial\bm{\Phi}}\log h(Y;\bm{\Phi})\Bigg|_{\bm{\Phi}=\bm{\Phi}_0}\right]E_{\bm{\Phi}_0}\left[W(Y)\frac{\partial}{\partial\bm{\Phi}}\log h(Y;\bm{\Phi})\Bigg|_{\bm{\Phi}=\bm{\Phi}_0}\right]^T\nonumber\\
&\hspace{8em}+E_{\bm{\Phi}_0}\left[W(Y)\frac{\partial}{\partial\bm{\Phi}}\log h(Y;\bm{\Phi})\Bigg|_{\bm{\Phi}=\bm{\Phi}_0}\right]\tilde{E}\left[W(Y)^2\frac{\partial}{\partial\bm{\Phi}}\log h(Y;\bm{\Phi})\Bigg|_{\bm{\Phi}=\bm{\Phi}_0}\right]^T\Bigg\}\nonumber\\
&\quad+\frac{\tilde{E}\left[W(Y)^2\right]}{E_{\bm{\Phi}_0}\left[W(Y)\right]^2}E_{\bm{\Phi}_0}\left[W(Y)\frac{\partial}{\partial\bm{\Phi}}\log h(Y;\bm{\Phi})\Bigg|_{\bm{\Phi}=\bm{\Phi}_0}\right]E_{\bm{\Phi}_0}\left[W(Y)\frac{\partial}{\partial\bm{\Phi}}\log h(Y;\bm{\Phi})\Bigg|_{\bm{\Phi}=\bm{\Phi}_0}\right]^T
\end{align}
and
\begin{align} \label{eq:asym:gamma_mis}
\tilde{\bm{\Gamma}}
&=\tilde{E}\left[W(Y)\frac{\partial^2}{\partial\bm{\Phi}\partial\bm{\Phi}^T}\log h(Y;\bm{\Phi})\right]
-\frac{\tilde{E}\left[W(Y)\right]}{E_{\bm{\Phi}_0}\left[W(Y)\right]}E_{\bm{\Phi}_0}\left[W(Y)\frac{\partial^2}{\partial\bm{\Phi}\partial\bm{\Phi}^T}\log h(Y;\bm{\Phi})\right]\nonumber\\
&\quad-\frac{\tilde{E}\left[W(Y)\right]}{E_{\bm{\Phi}_0}\left[W(Y)\right]}E_{\bm{\Phi}_0}\left[W(Y)\left[\frac{\partial}{\partial\bm{\Phi}}\log h(Y;\bm{\Phi})\right]\left[\frac{\partial}{\partial\bm{\Phi}}\log h(Y;\bm{\Phi})\right]^T\Bigg|_{\bm{\Phi}=\bm{\Phi}_0}\right]\nonumber\\
&\quad+\frac{\tilde{E}\left[W(Y)\right]}{E_{\bm{\Phi}_0}\left[W(Y)\right]^2}E_{\bm{\Phi}_0}\left[W(Y)\frac{\partial}{\partial\bm{\Phi}}\log h(Y;\bm{\Phi})\Bigg|_{\bm{\Phi}=\bm{\Phi}_0}\right]E_{\bm{\Phi}_0}\left[W(Y)\frac{\partial}{\partial\bm{\Phi}}\log h(Y;\bm{\Phi})\Bigg|_{\bm{\Phi}=\bm{\Phi}_0}\right]^T.
\end{align}
\end{theorem}

As shown by the above theorem, the MWLE is still asymptotically convergent and normally distributed. As a result, it is still justifiable to evaluate the parameter uncertainties and CI of parameters in the forms of Equations (\ref{eq:asym_var}) and (\ref{eq:asym_CI}). However, in the context of modelling heavy-tail distributions as an example, there could be an asymptotic bias on the estimated tail index. As a result, it is important to theoretically examine how the choice of weight functions influence the impacts of model misspecifications. These will be leveraged to the next subsections on the robustness studies and asymptotics under varying weight functions.

\subsubsection{Robustness}
It is well known that MLE is the most efficient estimator under all asymptotically unbiased estimators. Therefore, with an attempt to reduce the bias of estimated tail distribution under misspecified models through MWLE approach with $W(\cdot)\neq 1$, there will be a trade-off between bias regularizations and loss in efficiencies. This subsection will analyze such a trade-off, which may provide guidance to choose an appropriate weight function $W(\cdot)$. In light of Theorem \ref{thm:asym_tru}, it is easy to show the following proposition by applying delta method.

\begin{proposition}
Suppose that $G(\cdot)=H(\cdot;\bm{\Phi}_0)\in\mathcal{H}$ with the same set of regularity conditions as previous theorems. Define $\hat{\bm{\Phi}}_n$ and $\hat{\bm{\Phi}}_n^{(0)}$ as the MWLE and MLE respectively. Then, for some differentiable functions $U(\cdot)$, the relative asymptotic efficiency (AEFF) of $U(\hat{\bm{\Phi}}_n)$ is given by
\begin{equation}
\text{AEFF}(W;\bm{\Phi}_0):=\lim_{n\rightarrow\infty}\frac{\text{Var}(\hat{\bm{\Phi}}_n^{(0)})}{\text{Var}(\hat{\bm{\Phi}}_n)}=\frac{U'(\bm{\Phi}_0)^T\bm{\Sigma}^{(0)}U'(\bm{\Phi}_0)}{U'(\bm{\Phi}_0)^T\bm{\Sigma}U'(\bm{\Phi}_0)},
\end{equation}
where $U'(\bm{\Phi})$ is the gradient of $U(\bm{\Phi})$ w.r.t. $\bm{\Phi}$, and $\bm{\Sigma}^{(0)}$ is the inverse of Fisher information matrix under standard MLE approach.
\end{proposition}

Next, we need to quantify robustness by some statistical measures. In a theoretical setting, we follow e.g. \cite{huber1981robust}, \cite{beran2012robust} and \cite{gong2018robust} to consider the case that $Y_i$ is generated by a contamination model, given by

\begin{equation} \label{eq:asym_contam}
G(y):=G(y;\epsilon,M,\bm{\Phi}_0)=(1-\epsilon)H(y;\bm{\Phi}_0)+\epsilon M(y), \quad y>0,
\end{equation}
for some contamination distribution function $M$. Then, the asymptotic bias can be analyzed through evaluating the influence function (IF), a column vector with length $P$ given by

\begin{equation}
\text{IF}(\bm{\Phi}_0; H, M)=\lim_{\epsilon\rightarrow 0}\frac{\tilde{\bm{\Phi}}^{\epsilon,M}-\bm{\Phi}_0}{\epsilon},
\end{equation}
where $\tilde{\bm{\Phi}}^{\epsilon,M}$ is the asymptotic estimated parameters if $H^{\epsilon,M}$ given by Equation (\ref{eq:asym_contam}) is distribution generating $Y_i$, contrasting to $\bm{\Phi}_0$ which are the true model parameters. IF can be interpreted as the infinitesimal asymptotic bias of estimated parameters by perturbing the model generating $Y_i$. Smaller $|\text{IF}_p(\bm{\Phi}_0; H, M)|$ (with $\text{IF}_p$ being the $p^{\text{th}}$ element of $\text{IF}$) means a more robust estimation of $\phi_p$ under model misspecification. Our goal is to demonstrate the potential of the proposed MWLE to reduce such a bias and hence improve the robustness. We have the following proposition which derives the IF under the MWLE approach:

\begin{proposition}
The IF is given by
\begin{align}
\text{IF}(\bm{\Phi}_0; H, M)
&=-\bm{\Gamma}^{-1}\Bigg\{E_M\left[W(Y)\frac{\partial}{\partial\bm{\Phi}}\log h(Y;\bm{\Phi})\Bigg|_{\bm{\Phi}=\bm{\Phi}_0}\right]\nonumber\\
&\hspace{5em}-\frac{E_M\left[W(Y)\right]}{E_{\bm{\Phi}_0}\left[W(Y)\right]}E_{\bm{\Phi}_0}\left[W(Y)\frac{\partial}{\partial\bm{\Phi}}\log h(Y;\bm{\Phi})\Bigg|_{\bm{\Phi}=\bm{\Phi}_0}\right]\Bigg\},
\end{align}
where $E_M[Q(Y)]=\int_0^{\infty}Q(u)dM(u)$ for some functions $Q$, and $\bm{\Gamma}$ is given by Equation (\ref{eq:asym:gamma}).
\end{proposition}

We will show empirically how the choice of weight functions $W(\cdot)$ affects the AEFF and IF in Section \ref{sec:ex:toy}, which will help us understand the bias-variance tradeoff of our proposed MWLE approach.

\subsection{Asymptotic behavior of tail index with varying weight functions} \label{sec:theory:tail_idx}

Tail index measures the tail-heaviness of a probability distribution. Correctly specifying the tail index is a critical task of modelling insurance data with heavy-tail nature, as insurance companies often care more about large claims which are more material than small ones. In this section, we show that under some sequences of weight functions $W_n(\cdot)$ which depend on the number of observations $n$, the estimated tail index will be consistent under the proposed MWLE even if the model class is misspecified. 
This result theoretically justifies how the proposed MWLE addresses the tail-robustness issue caused by model misspecification and distributional contamination, by showing that reduced influence of smaller claims through downweighting can be useful for producing a plausible tail index estimate. 
Also, the result may provide some theoretical guidance on selecting an appropriate weight function.

Denote $\mathcal{R}_{-\gamma}$ be a class of regularly varying distributions with tail index $\gamma>0$, such that $\bar{H}\in\mathcal{R}_{-\gamma}$ if and only if $\bar{H}(y)\sim y^{-\gamma}L_0(y)$ as $y\rightarrow\infty$ for some slowly varying functions $L_0(y)\in\mathcal{R}_0$ satisfying $L_0(ty)/L_0(y)\rightarrow 1$ as $y\rightarrow\infty$ for any $t>0$. Smaller $\gamma$ implies heavier tail. Note that regularly varying distributions include many distributions that capture heavy-tail behaviors of loss random variables and we refer the readers to \citet{cooke2014fat} for more explanation on these distributions. Also define the following transformed density functions
\begin{equation}
\tilde{g}_{n}(y)=\frac{g(y)W_n(y)}{\int_{0}^{\infty}g(u)W_n(u)du},\qquad \tilde{h}_{n}(y;\bm{\Phi})=\frac{h(y;\bm{\Phi})W_n(y)}{\int_{0}^{\infty}h(u;\bm{\Phi})W_n(u)du},
\end{equation}
and $\tilde{G}_{n}(\cdot)$ and $\tilde{H}_{n}(\cdot)$ are the corresponding distribution functions. We further put a bar to any function $Q$ to denote its survival function (i.e. $\bar{Q}:=1-Q$). We then make the following assumptions:

\begin{enumerate}[font={\bfseries},label={A\arabic*.}]
\item $\bar{G}\in\mathcal{R}_{-\gamma_0}$ with tail index $\gamma_0>0$.
\item $\bar{H}(y;\bm{\Phi})=y^{-\gamma}L(y;\bm{\Phi})$ for some slowly varying functions $L$, so that $\bar{H}\in\mathcal{R}_{-\gamma}$. Here, $\gamma$ is the only model parameter within $\bm{\Phi}$ that governs the tail index. Also, both $L(yt;\bm{\Phi})/L(y;\bm{\Phi})$ and its derivative w.r.t. $\bm{\Phi}$ converges uniformly as $y\rightarrow\infty$ for any fixed $t>1$.
\item There exists some sequences of thresholds $\{\tau_n\}_{n=1,2,\ldots}$ with $\tau_n\rightarrow\infty$ as $n\rightarrow\infty$ such that $\tau_nW_n(\tau_n)\rightarrow 0$ as $n\rightarrow\infty$.
\item $\tilde{E}_n[(\log \tilde{h}_{n}(Y;\bm{\Phi}))^2]/(n\tilde{E}[W_n(Y)])\rightarrow 0$ as $n\rightarrow\infty$, where $\tilde{E}_n[Q(Y)]=\int_{0}^{\infty}Q(u)d\tilde{G}_{n}$ and $\tilde{E}[Q(Y)]=\int_{0}^{\infty}Q(u)dG$ for some functions $Q$.
\item The density functions $h(y;\bm{\Phi})$ and $g(y)$ are ultimately monotone (i.e. monotone on $y\in (z,\infty)$ for some $z>0$), uniformly on $\bm{\Phi}$.
\end{enumerate}

Assumptions \textbf{A1} and \textbf{A2} ensure that both the model generating the observations and the fitted model class are heavy tail in nature, with tail heaviness quantified by tail indices $\gamma_0$ and $\gamma$ respectively. In finite mixture context, see Section \ref{sec:fmm}, \textbf{A2} can be easily satisfied choosing $h$ in Equation (\ref{eq:model}) as any standard regularly varying distributions such as Pareto and Inverse-Gamma with compact parameter space. Assumption \textbf{A3} asserts that all observations other than the extreme ones are greatly down-weighted. This assumption provides a theoretical guidance of choosing the weight function such that small to moderate claims should only be allocated by small weights, while substantial weights should be assigned only to large claims. \textbf{A4} requires that the effective number of MWLE observations $n\tilde{E}[W_n(Y)]\rightarrow\infty$ such that large sample theories hold. The numerator $\tilde{E}_n[(\log \tilde{h}_{n}(Y;\bm{\Phi}))^2]$ grows much slower than the denominator as a logarithm is involved. Assumption \textbf{A5} is of no practical concern. Now, we have the following theorem which asserts the consistency of estimated tail index. The proof is leveraged to Section 3 of the supplementary material.

\begin{theorem} \label{thm:asym:tail_idx}
Assume \textbf{A1} to \textbf{A5} hold for the settings under the MWLE, and the regularity conditions outlined in Section 1 of Supplementary materials are satisfied. Then, there exists a local maximizer $\hat{\bm{\Phi}}_n$ of the weighted log-likelihood function $\mathcal{L}_n^{*}(\bm{\Phi};\bm{y})$ with the estimated tail index $\hat{\gamma}_n$ such that $\hat{\gamma}_n\rightarrow\gamma_0$ as $n\rightarrow\infty$. Further, the local maximizer $\hat{\gamma}_n$ is unique in probability as $n\rightarrow\infty$.
\end{theorem}


\begin{remark}
Consider a special case where: (i) the weight functions $W_n(y)=1\{y>\tau_n\}$ are step functions for some sequences of $\tau_n\rightarrow\infty$; and (ii) the fitted model class $H(y;\bm{\Phi})$ is chosen as a Generalized Pareto distribution (GPD) or equivalently Lomax distribution which will be described in Section \ref{sec:ex} (i.e. $H(y;\bm{\Phi})$ is an FMM in Equation (\ref{eq:model}) with $J=0$ and $f_t$ is a GPD). Theorem \ref{thm:asym:tail_idx} is then asserting the consistency of tail index obtained by excess over threshold method on GPD (\cite{smith1987estimating}), which has a very close connection with the consistency property of the Hill estimator (\cite{hill1975simple}) (see Section 4 of \cite{smith1987estimating}). Therefore, we can regard the proposed MWLE approach as a generalized framework of the Hill-type estimator by \cite{hill1975simple}.
\end{remark}

\section{Parameter estimation} \label{sec:em}
\subsection{GEM algorithm} \label{sec:em:gem}
Since there is a probabilistic interpretation of the weighted log-likelihood given by Equation (\ref{eq:loglik_weight}), it is feasible to construct a generalized Expectation-Maximization (GEM) algorithm for efficient parameter estimations. In this paper, we will present two distinctive approaches of complete data constructions which result to two different kinds of GEM algorithms.

\subsubsection{Method 1: Hypothetical data approach}
\paragraph{Construction of complete data}
To address the challenges of optimizing directly the ``observed data" weighted log-likelihood in Equation (\ref{eq:loglik_weight}), we extend the ``hypothetical complete data" method proposed by \cite{FUNG2020MoECensTrun}, by defining the complete data
\begin{equation}
\mathcal{D}^{\text{com}}=\{(y_i,\bm{z}_i,k_i,\{\bm{z}'_{is},y'_{is}\}_{s=1,\ldots,k_i})\}_{i=1,\ldots,n},
\end{equation}
where $k_i$ is the number of missing sample points ``generated" by observation $i$, due to the probabilistic interpretation that each sample $i$ is removed with a probability of $1-W(y_i)$. As an auxiliary tool for efficient computations we assume that $k_i$ follows geometric distribution with probability mass function
\begin{equation} \label{eq:em:k}
p(k_i;\bm{\Phi})=\left[1-\int_0^{\infty}h(u;\bm{\Phi})W(u)du\right]^{k_i}\int_0^{\infty}h(u;\bm{\Phi})W(u)du, \qquad k_i=0,1,\ldots,
\end{equation}
and $\{y'_{is}\}_{s=1,\ldots,k_i}$ are i.i.d. variables representing the missing samples. We assume that $Y'_{is}$ (with realization $y'_{is}$) is independent of $y_i$ and $k_i$, and follows a distribution with the following density function
\begin{equation} \label{eq:em:y}
\tilde{h}^{*}(y'_{is};\bm{\Phi})=\frac{h(y'_{is};\bm{\Phi})(1-W(y'_{is}))}{\int_0^{\infty}h(u;\bm{\Phi})(1-W(u))du},\qquad y'_{is}>0.
\end{equation}

Further, $\bm{z}_i=(z_{i1},\ldots,z_{i(J+1)})$ are the latent mixture components assignment labels, where $z_{ij}=1$ if observation $i$ belongs to the $j^{\text{th}}$ latent class and $z_{ij}=0$ otherwise. Similarly, $\bm{z}'_i=(z'_{is1},\ldots,z'_{is(J+1)})$ are the labels for missing data, where $z'_{isj}=1$ if the $s^{\text{th}}$ missing sample generated by observation $i$ belongs to the $j^{\text{th}}$ latent class, and $z'_{isj}=0$ otherwise. 

The complete data weighted log-likelihood function is then given by

\begingroup
\allowdisplaybreaks
\begin{align}
\tilde{\mathcal{L}}^{*}_n(\bm{\Phi};\mathcal{D}^{\text{com}})
&=\sum_{i=1}^{n}W(y_i)\log\Bigg\{\frac{\left\{\prod_{j=1}^{J}[\pi_jf_b(y_i;\bm{\varphi}_j)]^{z_{ij}}\right\}\left(\pi_{J+1}f_t(y_i;\bm{\eta})\right)^{z_{i(J+1)}}W(y_i)}{\int_{0}^{\infty}h(u;\bm{\Phi})W(u)du} \nonumber\\
&\hspace{8em} \times \left[1-\int_0^{\infty}h(u;\bm{\Phi})W(u)du\right]^{k_i}\int_0^{\infty}h(u;\bm{\Phi})W(u)du \nonumber\\
&\hspace{8em} \times \prod_{s=1}^{k_i}\frac{\left\{\prod_{j=1}^{J}[\pi_jf_b(y'_{is};\bm{\varphi}_j)]^{z'_{isj}}\right\}\left(\pi_{J+1}f_t(y'_{is};\bm{\eta})\right)^{z'_{is(J+1)}}W(y_i)}{\int_0^{\infty}h(u;\bm{\Phi})(1-W(u))du}\Bigg\} \nonumber\\
&=\sum_{i=1}^nW(y_i)\left\{\left[\sum_{j=1}^{J}z_{ij}\log \pi_jf_b(y_i;\bm{\varphi}_j)\right]+z_{i(J+1)}\log\pi_{J+1} f_t(y_i;\bm{\eta})\right\}\nonumber\\
&\quad +\sum_{i=1}^{n}\sum_{s=1}^{k_i}W(y_i)\left\{\left[\sum_{j=1}^J z'_{ijs}\log\pi_j f_b(y'_{is};\bm{\varphi}_j)\right]+z'_{i(J+1)s}\log\pi_{J+1} f_t(y'_{is};\bm{\eta})\right\}+\text{const.},
\end{align}
\endgroup
which is more computationally tractable. In the following we will omit the constant term which is irrelevant for calculations.

\paragraph{Iterative procedures}
In the $l^{\text{th}}$ iteration of the E-step, we compute the expectation of the complete data weighted log-likelihood as follows:
\begin{align} \label{eq:em:q}
Q^{*}(\bm{\Phi}|\bm{\Phi}^{(l-1)})
&=\sum_{i=1}^nW(y_i)\left\{\left[\sum_{j=1}^{J}z_{ij}^{(l)}\log \pi_jf_b(y_i;\bm{\varphi}_j)\right]+z_{i(J+1)}^{(l)}\log\pi_{J+1} f_t(y_i;\bm{\eta})\right\}\nonumber\\
&\quad +\sum_{i=1}^{n}k_i^{(l)}W(y_i)\Big\{\left[\sum_{j=1}^J {z'}^{(l)}_{ij}\left(\log\pi_j +E\left[\log f_b(Y';\bm{\varphi}_j)|\mathcal{D}^{\text{com}},\bm{\Phi}^{(l-1)}\right]\right)\right] \nonumber\\
& \hspace{8em} +{z'}^{(l)}_{i(J+1)}\left(\log\pi_{J+1}+ E\left[\log f_t(Y';\bm{\eta})|\mathcal{D}^{\text{com}},\bm{\Phi}^{(l-1)}\right]\right)\Big\},
\end{align}
where $z_{ij}^{(l)}=E[z_{ij}|\mathcal{D}^{\text{com}},\bm{\Phi}^{(l-1)}]$, ${z'}^{(l)}_{ij}=E[z'_{ijs}|\mathcal{D}^{\text{com}},\bm{\Phi}^{(l-1)}]$ and $k_i^{(l)}=E[K_i|\mathcal{D}^{\text{com}},\bm{\Phi}^{(l-1)}]$. Also, $K$ follows $p(\cdot;\bm{\Phi}^{(l-1)})$ in Equation (\ref{eq:em:k}) and $Y'$ follows $\tilde{h}^{*}(\cdot;\bm{\Phi}^{(l-1)})$ in Equation (\ref{eq:em:y}). The precise expressions of the above expectations are presented in Section 4.2 of supplementary materials, under a particular specification of Gamma distribution for $f_b$ and Lomax distribution for $f_t$. This specification will also be studied in the illustrating examples (Section \ref{sec:ex}).

In the M-step, we attempt to find the updated parameters $\bm{\Phi}^{(l)}$ such that $Q^{*}(\bm{\Phi}^{(l)}|\bm{\Phi}^{(l-1)})\geq Q^{*}(\bm{\Phi}^{(l-1)}|\bm{\Phi}^{(l-1)})$. Note in Equation (\ref{eq:em:q}) that $Q^{*}(\bm{\Phi}|\bm{\Phi}^{(l-1)})$ is linearly separable w.r.t. parameters $(\bm{\pi}, \bm{\varphi}_1,\ldots,\bm{\varphi}_J,\bm{\eta})$. Therefore, the optimization can be done separately w.r.t. each subset of parameters. Details are leveraged to Section 4.3 of supplementary materials. 

\subsubsection{Method 2: Parameter transformation approach}
\paragraph{Construction of complete data}
Motivated by the mixture probability transformation approach adopted by e.g. \cite{lee2012algorithms} and \cite{VERBELEN2015Censor} for truncated data, we here rewrite the random truncation distribution $\tilde{h}(y_i;\bm{\Phi})$ in Equation (\ref{eq:loglik_weight}) as
\begin{equation}
\tilde{h}(y_i;\bm{\Phi})
=\frac{h(y_i;\bm{\Phi})W(y_i)}{\int_{0}^{\infty}h(u;\bm{\Phi})W(u)du}
=\sum_{j=1}^J\pi_j^{*}\frac{f_b(y_i;\bm{\varphi}_j)W(y_i)}{\int_0^{\infty}f_b(u;\bm{\varphi}_j)W(u)du}+\pi_{J+1}^{*}\frac{f_t(y_i;\bm{\eta})W(y_i)}{\int_0^{\infty}f_t(u;\bm{\eta})W(u)du},
\end{equation}
where $\bm{\pi}^{*}:=(\pi_1^{*},\ldots,\pi_{J+1}^{*})$ are the transformed mixing weight parameters given by
\begin{equation} \label{eq:em:pi_trans}
\pi_j^{*}=\frac{\pi_j\int_0^{\infty}f_b(u;\bm{\varphi}_j)W(u)du}{\int_{0}^{\infty}h(u;\bm{\Phi})W(u)du},~j=1,\ldots,J;\qquad
\pi_{J+1}^{*}=\frac{\pi_{J+1}\int_0^{\infty}f_t(u;\bm{\eta})W(u)du}{\int_{0}^{\infty}h(u;\bm{\Phi})W(u)du}.
\end{equation}

As a result, the problem is reduced to maximizing the weighted log-likelihood of finite mixture of random truncated distributions. In this case, define the complete data
\begin{equation}
\mathcal{D}^{\text{com}}=\{(y_i,\bm{z}_i^{*})\}_{i=1,\ldots,n},
\end{equation}
where $\bm{z}_i^{*}=(z_{i1}^{*},\ldots,z_{i(J+1)}^{*})$ are the labels where $z_{ij}^{*}=1$ if observation $i$ belongs to the $j^{\text{th}}$ (transformed) latent mixture component and $z_{ij}^{*}=0$ otherwise. The complete data weighted log-likelihood function is reduced to
\begin{align}
\tilde{\mathcal{L}}^{*}_n(\bm{\Phi};\mathcal{D}^{\text{com}})
&=\sum_{i=1}^{n}W(y_i)\Bigg\{\left[\sum_{j=1}^{J}z_{ij}^{*}\left(\log\pi_j^{*}+\log\frac{f_b(y_i;\bm{\varphi}_j)W(y_i)}{\int_0^{\infty}f_b(u;\bm{\varphi}_j)W(u)du}\right)\right]\nonumber\\
&\hspace{8em}+z_{i(J+1)}^{*}\left(\log\pi_{J+1}^{*}+\log\frac{f_t(y_i;\bm{\eta})W(y_i)}{\int_0^{\infty}f_t(u;\bm{\eta})W(u)du}\right)\Bigg\}.
\end{align}

\paragraph{Iterative procedures}
In the $l^{\text{th}}$ iteration of the E-step, the expectation of the complete data weighted log-likelihood is:
\begin{align} \label{eq:em:q2}
Q^{*}(\bm{\Phi}|\bm{\Phi}^{(l-1)})
&=\sum_{i=1}^{n}W(y_i)\Bigg\{\left[\sum_{j=1}^{J}z_{ij}^{*(l)}\left(\log\pi_j^{*}+\log\frac{f_b(y_i;\bm{\varphi}_j)W(y_i)}{\int_0^{\infty}f_b(u;\bm{\varphi}_j)W(u)du}\right)\right]\nonumber\\
&\hspace{8em}+z_{i(J+1)}^{*(l)}\left(\log\pi_{J+1}^{*}+\log\frac{f_t(y_i;\bm{\eta})W(y_i)}{\int_0^{\infty}f_t(u;\bm{\eta})W(u)du}\right)\Bigg\},
\end{align}
where $z_{ij}^{*(l)}=E[z_{ij}^{*}|\mathcal{D}^{\text{com}},\bm{\Phi}^{(l-1)}]$ is provided in Section 5.2 of supplementary materials.

In the M-step, similar to Method 1 that $Q^{*}(\bm{\Phi}|\bm{\Phi}^{(l-1)})$ is linearly separable w.r.t. parameters $(\bm{\pi}^{*}, \bm{\varphi}_1,\ldots,\bm{\varphi}_J,\bm{\eta})$, we can maximize $Q^{*}(\bm{\Phi}|\bm{\Phi}^{(l-1)})$ sequentially w.r.t. each subset of parameters. Details are presented in Section 5.3 of supplementary materials. Note that the M-step of this approach is slightly more computationally more intensive than Method 1 as the target function $Q^{*}(\bm{\Phi}|\bm{\Phi}^{(l-1)})$ here involves numerical integrals.

After completing the iterative procedures, we will obtain an estimate of the transformed mixing weights $\bm{\pi}^{*}$ instead of $\bm{\pi}$. One can revert Equation (\ref{eq:em:pi_trans}) to get back the estimated original mixing weights as follows:
\begin{equation}
\pi_j=\frac{\pi_j^{*}[\int_0^{\infty}f_b(u;\bm{\varphi}_j)W(u)du]^{-1}}{\pi_j^{*}\sum_{j'=1}^J[\int_0^{\infty}f_b(u;\bm{\varphi}_{j'})W(u)du]^{-1}+\pi_{J+1}^{*}[\int_{0}^{\infty}f_t(u;\bm{\eta})W(u)du]^{-1}},~j=1,\ldots,J;
\end{equation}
\begin{equation}
\pi_{J+1}=\frac{\pi_{J+1}^{*}[\int_{0}^{\infty}f_t(u;\bm{\eta})W(u)du]^{-1}}{\pi_j^{*}\sum_{j'=1}^J[\int_0^{\infty}f_b(u;\bm{\varphi}_{j'})W(u)du]^{-1}+\pi_{J+1}^{*}[\int_{0}^{\infty}f_t(u;\bm{\eta})W(u)du]^{-1}}.
\end{equation}

\subsection{Ascending property of the GEM algorithm}
It is well known from \cite{DEMPSTER1977EM} that an increase of complete data log-likelihood implies an increase of observed data log-likelihood (Equation (\ref{eq:loglik})). This can be analogously extended to the proposed weighted log-likelihood framework where we have the following proposition. The proof is leveraged to Section 6 of the supplementary material.

\begin{proposition} \label{prop:ascend}
If the expected complete data weighted log-likelihood is increased during the $l^{\text{th}}$ iteration (i.e. $Q^{*}(\bm{\Phi}^{(l)}|\bm{\Phi}^{(l-1)})\geq Q^{*}(\bm{\Phi}^{(l-1)}|\bm{\Phi}^{(l-1)})$), then the observed data weighted log-likelihood is also increased (i.e. $\mathcal{L}^{*}_n(\bm{\Phi}^{(l)};\bm{y})\geq \mathcal{L}^{*}_n(\bm{\Phi}^{(l-1)};\bm{y})$).
\end{proposition}

\subsection{Parameter Initialization, convergence acceleration and stopping criterion} \label{sec:em:init}
Initialization of parameters is an important issue, in the sense that poor initializations may lead to slow convergence, numerical instability and even convergence to spurious local maximum. We suggest to determine the initial parameters $\bm{\Phi}^{(0)}$ using a modified version of clusterized method of moments (CMM) approach by \cite{gui2018fit}. Under this approach, we first determine a threshold $\tau$ which classifies observations $y_i$ into either body ($y_i\leq\tau$) or tail ($y_i>\tau$) part of the distribution. We then apply a $K$-means clustering method to assign ``body" observations $y_i$ with $y_i\leq\tau$ to one of the $J$ mixture components for the body, with moment matching method for each mixture components to determine the initial parameters $(\bm{\pi}^{(0)},\bm{\varphi}^{(0)})$. Moment matching technique is also applied to ``tail" observations $y_i$ with $y_i>\tau$ to initialize $\bm{\eta}^{(0)}$. For details, we direct readers to Section 7 of the supplementary material.

As EM algorithm often converges slowly with small step sizes, we further apply a step lengthening procedure for every two GEM iterations to accelerate the algorithm. This is described by \cite{jamshidian1997acceleration} and its references therein as a ``pure accelerator" for the EM algorithm.

\sloppy The GEM algorithm is iterated until the relative change of iterated parameters $\Delta^{\text{rel}}\bm{\Phi}^{(l)}:=|\log(\bm{\Phi}^{(l)}/\bm{\Phi}^{(l-1)})|/P$ is smaller than a threshold of $10^{-5}$ or the maximum number of iterations of 1000 is reached.

\subsection{Specification of weight function}
Our proposed MWLE is rather flexible by allowing us to pre-specify any weight functions $W(\cdot)$ prior to fitting the GEM algorithm. The appropriate choice of $W(\cdot)$ depends on some decision rules beyond what statistical inference can do. In insurance loss modelling perspective, such decision rule includes the relative importance of insurance company to correctly specify the tail distribution (to evaluate some tail measures such as Value-at-risk (VaR)) compared to that of more accurately modelling the smaller attritional claims. If accurate extrapolation of huge claims are way more important than modelling the smaller claims, then one may consider $W(y)$ to be close to zero unless $y$ is large, aligning with Assumption \textbf{A3} in Section \ref{sec:theory:tail_idx} to ensure near-consistent tail index estimations (Theorem \ref{thm:asym:tail_idx}). Otherwise, one may consider a flatter $W(y)$ across $y$.

Throughout the entire paper, we analyze the following general form of weight function
\begin{equation} \label{eq:em:wgt_func}
W(y):=W(y;\xi,\tilde{\mu},\tilde{\phi})=\xi+(1-\xi)\int_0^y\frac{(\tilde{\phi}\tilde{\mu})^{-1/\tilde{\phi}}}{\Gamma(1/\tilde{\phi})}u^{1/\tilde{\phi}-1}e^{-u/(\tilde{\phi}\tilde{\mu})}du,\quad y>0,
\end{equation}
which is the distribution function of a zero-inflated Gamma distribution. The above weight function has the following characteristics:
\begin{itemize}
\item $W(y)$ is a non-decreasing function of $y$, meaning that smaller observations are down-weighted.
\item $\xi\in[0,1]$ is the minimum weight assigned to each observation.
\item $\tilde{\mu}$ and $\tilde{\phi}$ are the location and dispersion hyperparameters of Gamma distribution respectively. Larger $\tilde{\mu}$ means more (small to moderate) claims are under-weighted by a larger extent, while $\tilde{\phi}$ controls the shape of weight function, or how the observations are under-weighted.
\item If $\xi=1$ or $\tilde{\mu}=0$, then the weight function is reduced to $W(\cdot)=1$, leading to standard MLE approach.
\item If $\xi=0$ and $\tilde{\phi}\rightarrow 0$, then $W(y)=1\{y\geq\tilde{\mu}\}$, meaning that only observations greater than $\tilde{\mu}$ are informative in determining the estimated parameters.
\end{itemize}

Overall, smaller $\xi$, larger $\tilde{\mu}$ and smaller $\tilde{\phi}$ represent greater under-weightings to more small claims, where we will expect more robust tail estimation by sacrificing more efficiencies on body estimations. In this paper, instead of quantifying decision rules to select the hyperparameters, in the subsequent sections we empirically test various (wide range) combinations of $(\xi,\tilde{\mu},\tilde{\phi})$ to study how these hyperparameters affect the trade-off between tail-robustness and estimation efficiency. These provide practical guidance and assessments to determine the suitable hyperparameters.

\begin{remark}
There are many possible ways to quantify the decision rule to select the ``optimal" weight function hyperparameters. We here briefly discuss two possible ways: (1) Consider a goodness-of-fit test statistic for heavy-tailed distributions, such as the modified AD test (\cite{ahmad1988assessment}). Then select weight function hyperparameters which optimizes the test statistic; (2) Define an acceptable range of estimated parameter uncertainty of tail index, e.g. two times as the uncertainty obtained by MLE. Then select the hyperparameters with the greatest distortion metric (e.g. the average downweighting factor $\sum_{i=1}^{n}(1-W(y_i))/n$) where the tail index uncertainty is still within the acceptable range.
\end{remark}

\subsection{Choice of model complexity} \label{sec:em:complex}
The above GEM algorithm assumes a fixed number of mixture component $J$. However, it is important to control the model complexity by choosing an appropriate $J$ which allows enough flexibility to capture the distributional characteristics without over-fitting. 

The first criterion is motivated by maximizing the expected weighted log-likelihood
\begin{equation} \label{eq:em:e_wgt_ll}
n\times\tilde{E}[\mathcal{L}^{*}(\bm{\Phi};\bm{Y})]=n\times\tilde{E}\left[W(Y)\log\frac{h(Y;\bm{\Phi})W(Y)}{\int_0^{\infty}h(u;\bm{\Phi})W(u)du}\right],
\end{equation}
where the expectation is taken on $Y$ under the true model generating the observations. Without knowing the true model (in real data applications), Equation (\ref{eq:em:e_wgt_ll}) is approximated by $\mathcal{L}_n^{*}(\hat{\bm{\Phi}};\bm{y})$ in Equation (\ref{eq:loglik_weight}) with fitted model parameters $\hat{\bm{\Phi}}$. Note that it is positively biased with correction term $\text{tr}(-\bm{\Gamma}^{-1}\bm{\Lambda})$ shown by \cite{konishi1996generalised}. This leads to a robustified AIC
\begin{equation}
\text{RAIC}=-2\times\mathcal{L}_n^{*}(\hat{\bm{\Phi}};\bm{y})+2\times\text{tr}(-\hat{\bm{\Gamma}}^{-1}\hat{\bm{\Lambda}}).
\end{equation}

Analogous and naturally, since AIC-type criteria often choose excessively complex models, we also consider the robustified BIC given by
\begin{equation}
\text{RBIC}=-2\times\mathcal{L}_n^{*}(\hat{\bm{\Phi}};\bm{y})+(\log n)\times\text{tr}(-\hat{\bm{\Gamma}}^{-1}\hat{\bm{\Lambda}}).
\end{equation}

We choose $J$ that minimizes either the RAIC or RBIC, and the $(p,p)$-th element of $-\hat{\bm{\Gamma}}^{-1}\hat{\bm{\Lambda}}$ can be interpreted as the effective number of parameter attributed by the $p^{\text{th}}$ parameter.

Insurance loss dataset is often characterized by very complicated and multimodal distribution on very small claims, yet it is not meaningful to capture all these small nodes by choosing an overly complex mixture distribution with large $J$. However, the above RAIC and RBIC cannot effectively reduce those mixture components as the effective number of parameters for those capturing the smaller claims could be very small if $W(\cdot)$ is chosen very small over the region of small claims. To effectively remove components which excessively capture the small claims, we propose treating all parameters as ``full parameters", which results to the following truncated AIC and BIC:
\begin{equation}
\text{TAIC}=-2\times\mathcal{L}_n^{*}(\hat{\bm{\Phi}};\bm{y})+2\times P,
\end{equation}
\begin{equation}
\text{TBIC}=-2\times\mathcal{L}_n^{*}(\hat{\bm{\Phi}};\bm{y})+\left(\log \sum_{i=1}^{n}W(y_i)\right)\times P.
\end{equation}

\begin{remark} \label{rmk:tic}
The above TAIC and TBIC are motivated by the bias of approximating $n\times\tilde{E}[\mathcal{L}^{*}(\bm{\Phi};\bm{Y})]$ by the empirical truncated log-likelihood $\mathcal{L}^{**}_n(\hat{\bm{\Phi}};\bm{y}):=\sum_{i=1}^{n}V_i(y_i)\log \frac{h(y_i;\bm{\Phi})W(y_i)}{\int_0^{\infty}h(u;\bm{\Phi})W(u)du}$ instead of $\mathcal{L}_n^{*}(\bm{\Phi};\bm{y})$, where $V_i(y)\sim\text{Bernoulli}(W(y))$ is an indicator randomly discarding some observations. It can be easily shown (details in Section 8 of supplementary material) that the asymptotic bias is simply $P$ with effective number of observations $\sum_{i=1}^{n}W(y_i)$. Note also that the weighted log-likelihood $\mathcal{L}_n^{*}(\bm{\Phi};\bm{y})$ is asymptotically equivalent to the truncated log-likelihood $\mathcal{L}^{**}_n(\bm{\Phi};\bm{y})$, except that the former produces more accurate estimated parameters than the latter. This motivates why in TAIC and TBIC we choose to evaluate $\mathcal{L}_n^{*}(\bm{\Phi};\bm{y})$ instead of $\mathcal{L}^{**}_n(\bm{\Phi};\bm{y})$.
\end{remark}

\section{Illustrating examples} \label{sec:ex}
In this section, we analyze the performance of our proposed MWLE approach (Equation (\ref{eq:loglik_weight})) on FMM given by Equation (\ref{eq:model}). In the following examples, we select Gamma density for the body components $f_b$, a light-tailed distribution to capture the distributional multimodality of small to moderate claims, and Lomax density for the tail component $f_t$ to extrapolate well the tail-heaviness of larger observations. Then, Equation (\ref{eq:model}) becomes
\begin{equation} \label{eq:em:density_mixture}
h(y;\bm{\Phi})
=\sum_{j=1}^J\pi_jf_b(y;\mu_j,\phi_j)+\pi_{J+1}f_t(y;\theta,\gamma),
\end{equation}
where the parameter set is re-expressed as $\bm{\Phi}=(\bm{\pi},\bm{\mu},\bm{\phi},\gamma)$ while $\bm{\varphi}_j=(\mu_j,\phi_j)$ and $\bm{\eta}=(\theta,\gamma)$, and the Gamma and Lomax densities $f_b$ and $f_t$ are respectively given by
\begin{equation} \label{eq:em:comp}
f_b(y;\mu,\phi)=\frac{(\phi\mu)^{-1/\phi}}{\Gamma(1/\phi)}y^{1/\phi-1}e^{-y/(\phi\mu)}\quad \text{and} \quad f_t(y;\theta,\gamma)=\frac{\gamma\theta^{\gamma}}{\left(y+\theta\right)^{\gamma+1}},
\end{equation}
where $\mu$ and $\phi$ are the mean and dispersion parameters of Gamma distribution, while $\gamma$ is the tail index parameter for the Lomax distribution. $\theta$ is scale of the Lomax distribution. Note that the above model is a regular varying distribution with the tail behavior predominately explained by the tail index $\gamma$. As a result, tail-robustness is highly determined by how stable and accurate the estimated tail index $\gamma$ is.

The specifications of body and tail component functions are mainly motivated by the key characteristics of insurance claim severity distributions (multimodal distribution of small claim, existence of extremely large claims, mismatch between body and tail behavior etc.) which will be illustrated in the real insurance data application section. While we do not preclude the existence of other specifications, such as Weibull for the body and Inverse-Gamma for the tail, plausible for insurance applications, in this section we simply focus on studying Gamma-Lomax combination to focus on the scope of this paper -- demonstrating the usefulness of the proposed MWLE, instead of performing distributional comparisons under FMM.

\subsection{Toy example} \label{sec:ex:toy}

We demonstrate how the proposed MWLE framework works through a simple toy example of one-parameter Lomax distribution $H(y;\gamma)=1-(y+1)^{-\gamma}$ ($y>0$), which is a special case of Equation (\ref{eq:em:density_mixture}) with $J=0$ and $\theta=1$.

\begin{figure}[!h]
\begin{center}
\begin{subfigure}[h]{0.49\linewidth}
\includegraphics[width=\linewidth]{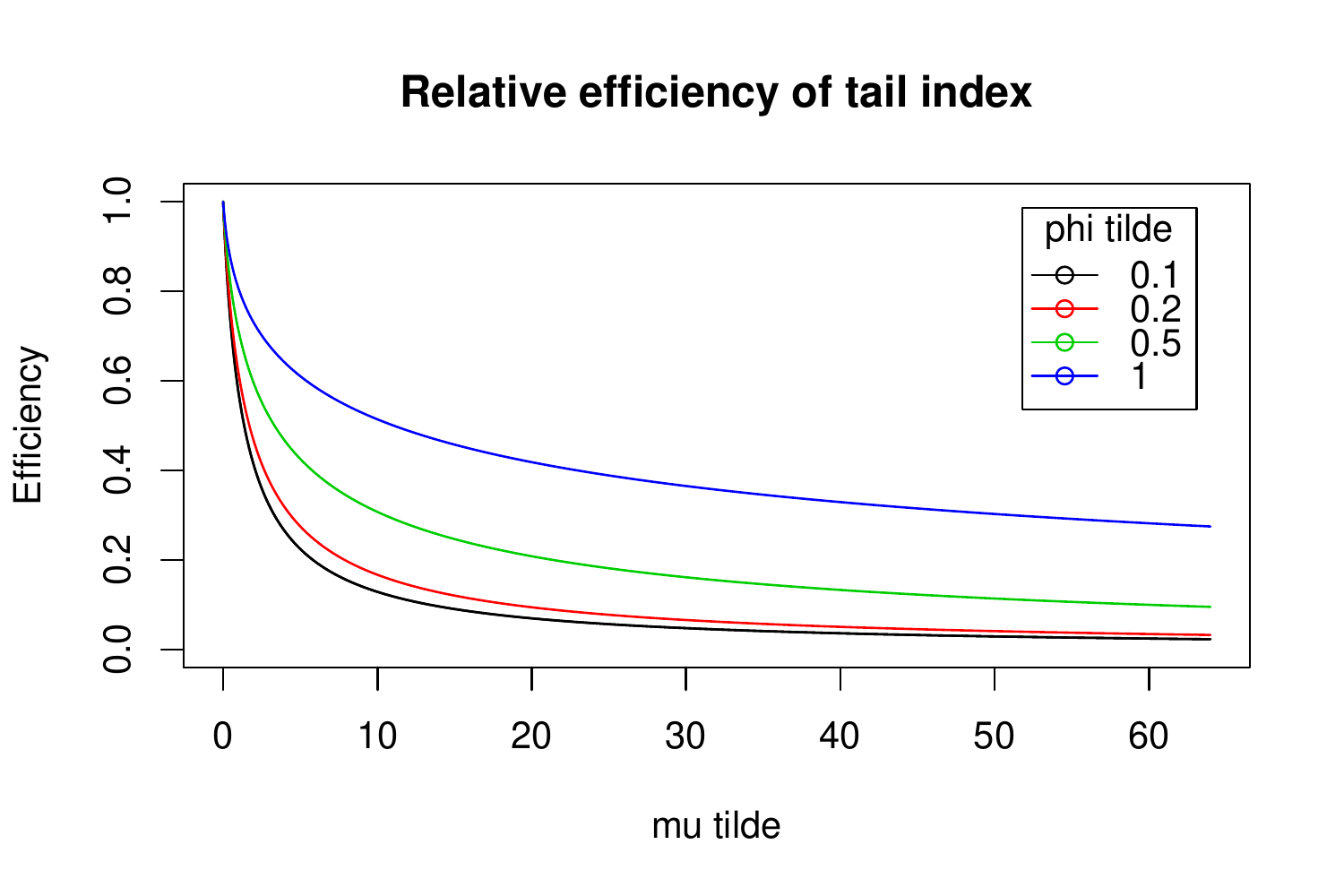}
\end{subfigure}
\hfill
\begin{subfigure}[h]{0.49\linewidth}
\includegraphics[width=\linewidth]{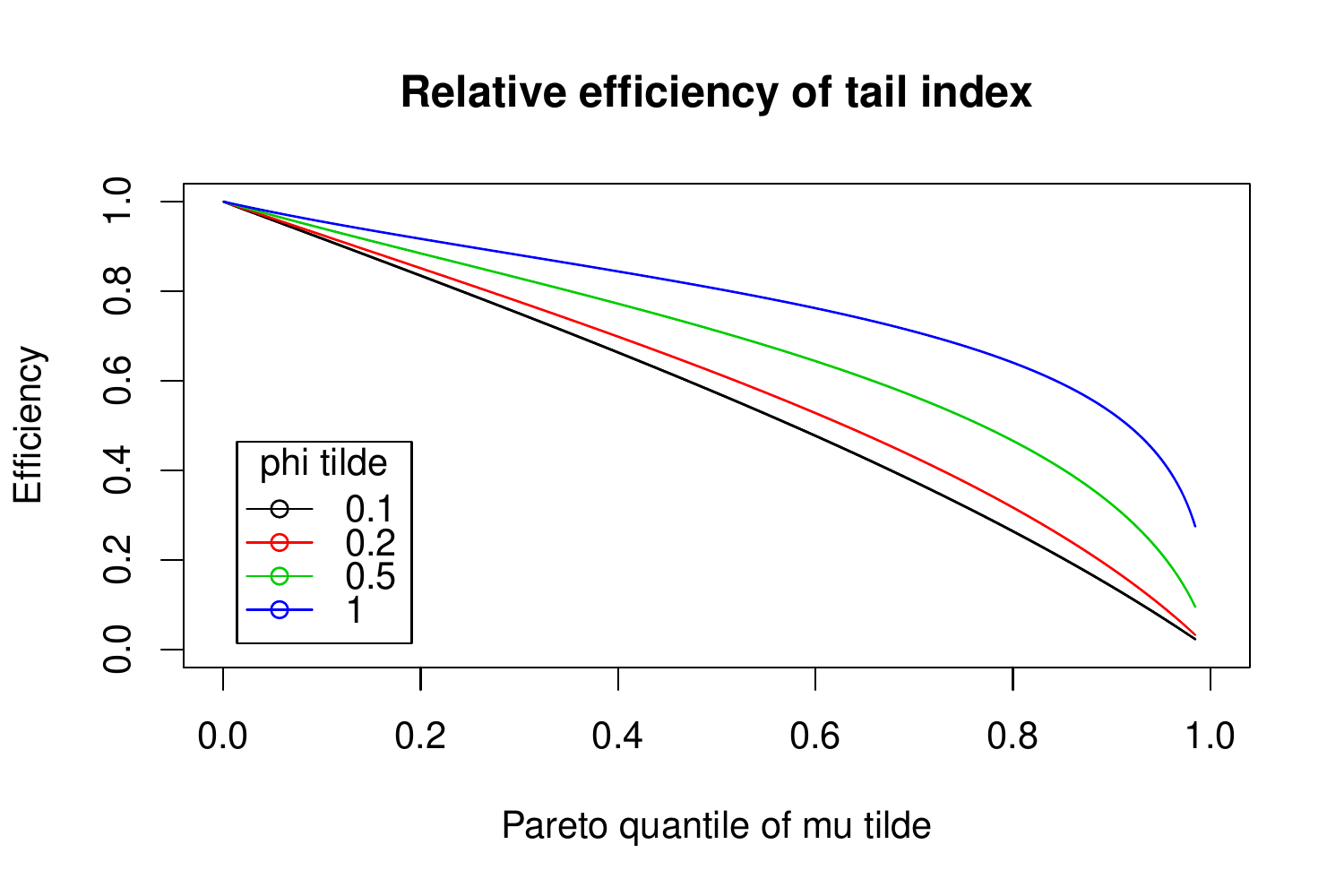}
\end{subfigure}
\end{center}
\caption{AEFF as a function of the weight location hyperparameter $\tilde{\mu}$ (left panel) or Pareto (Lomax) quantile of $\tilde{\mu}$ (right panel) under Lomax true model.}
\label{fig:thm_aeff}
\end{figure}

Consider the first case where the true model $G(\cdot)$ is a Lomax with $\gamma=\gamma_0=1$. For the weight function for the MWLE, we consider the form of Equation (\ref{eq:em:wgt_func}) with $\xi=0$ for simplicity. We will test across a wide range of $\tilde{\mu}$ and across $\tilde{\phi}\in\{0.1, 0.2, 0.5, 1\}$. Figure \ref{fig:thm_aeff} presents how the choices of these hyperparameters affect the AEFF. Starting from $\text{AEFF}=1$ when $\tilde{\mu}=0$ which is equivalent to standard MLE, the AEFF decrease monotonically as $\tilde{\mu}$ increases. This is intuitive because under-weighting smaller observations with MWLE means effectively discarding some observed information, leading to larger parameter uncertainties compared to MLE. Since the MLE estimated tail index is unbiased under the true model, there is obviously no benefit of using the proposed MWLE to fit the true model.

\begin{figure}[!h]
\begin{center}
\begin{subfigure}[h]{0.49\linewidth}
\includegraphics[width=\linewidth]{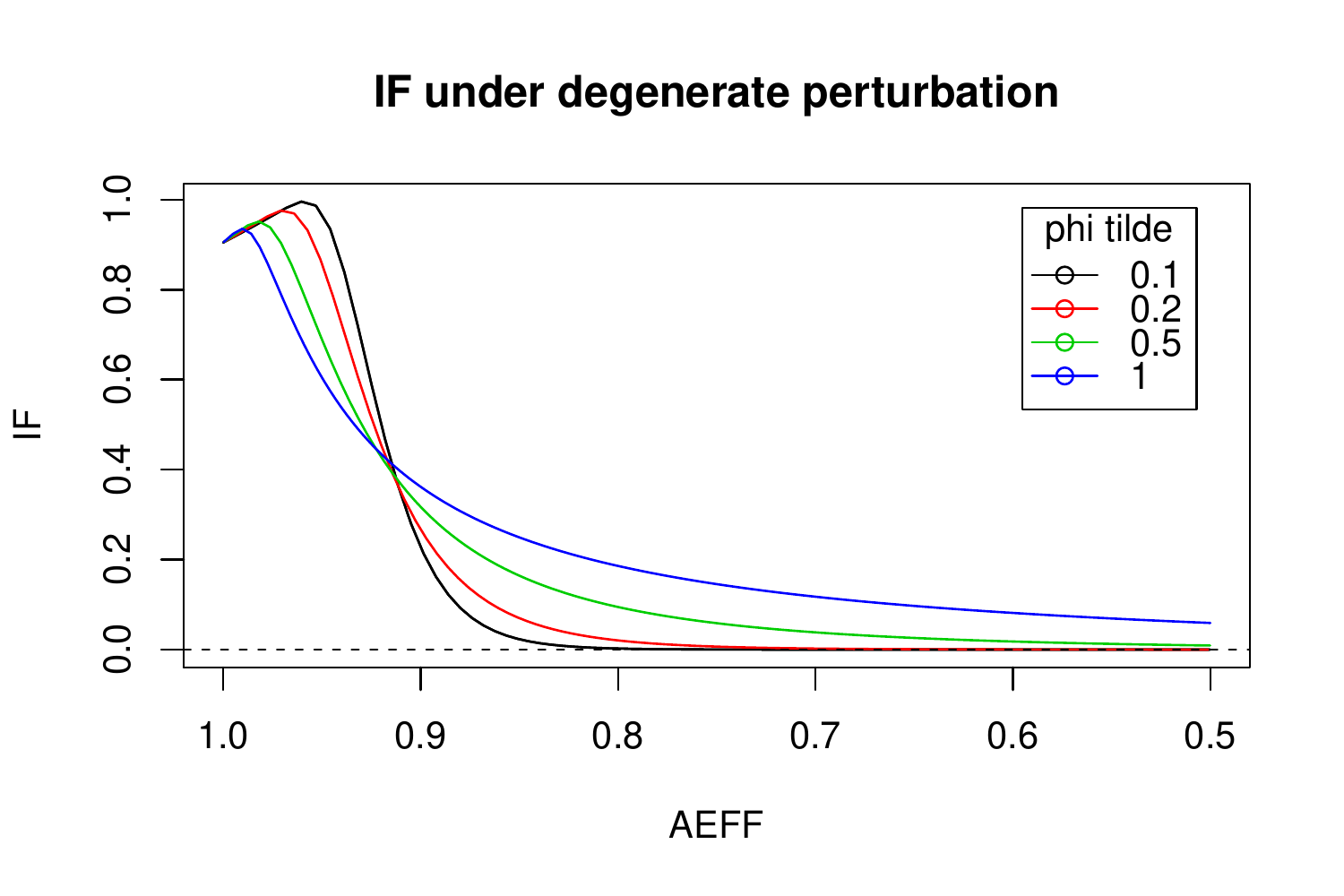}
\end{subfigure}
\hfill
\begin{subfigure}[h]{0.49\linewidth}
\includegraphics[width=\linewidth]{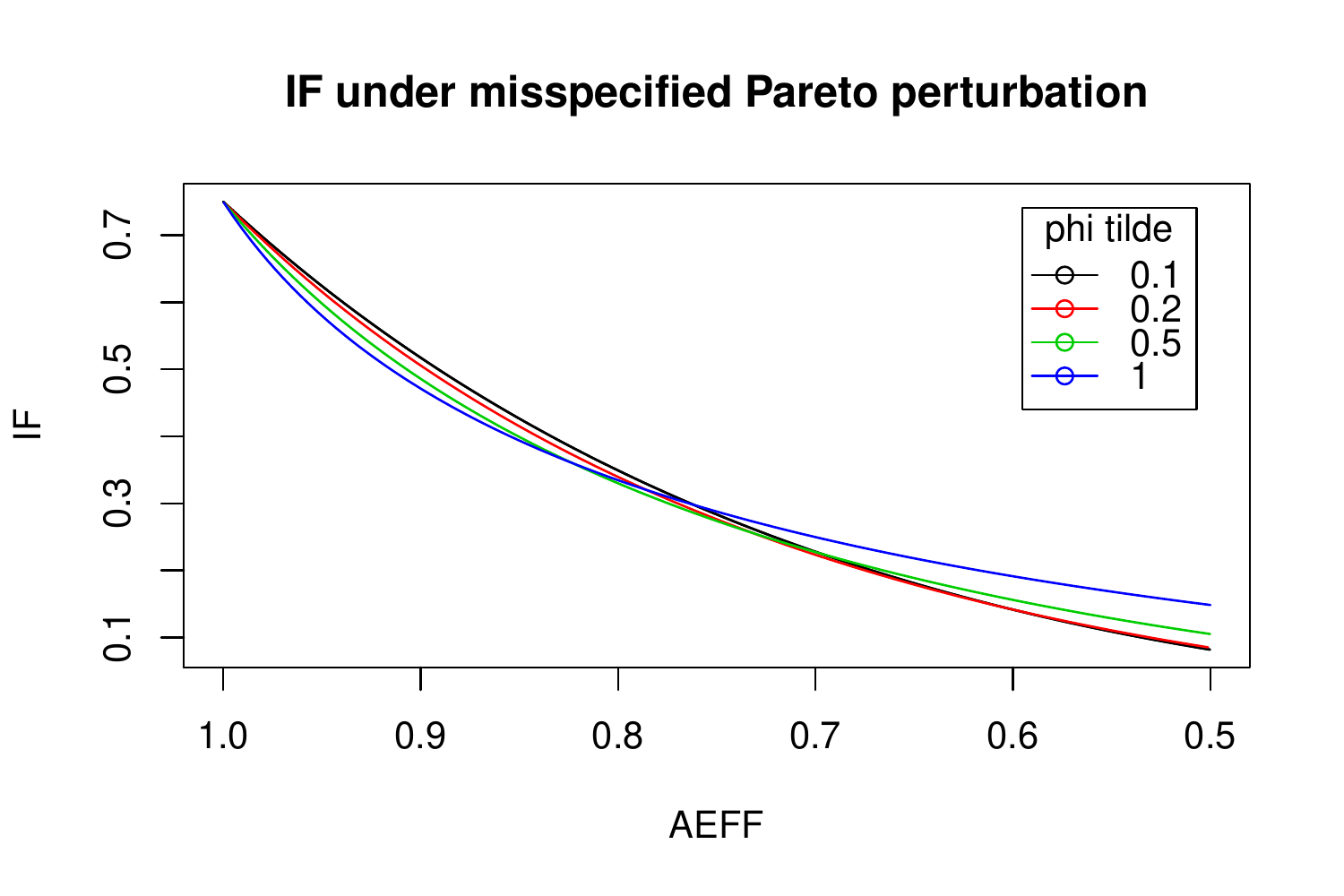}
\end{subfigure}
\end{center}
\caption{IF as a function of AEFF under degenerate (left panel) and Pareto (right panel) contaminations.}
\label{fig:thm_if}
\end{figure}

Now, consider the second case where the true model is perturbed by the contamination function $M$, as presented in Equation (\ref{eq:asym_contam}). In this demonstration example, we consider two following two choices for the contamination function $M$:
\begin{itemize}
\item Degenerate perturbation: One-point distribution on $y=1/4$
\item Pareto perturbation: Lomax distribution with tail index $\gamma=\gamma^{*}=4>\gamma_0$
\end{itemize}
Note that the contamination function is relatively lighter tailed and hence it would not affect the tail behavior of the perturbed distribution. In Figure \ref{fig:thm_if}, we present the IF as a function of the AEFF (determined as a function of chosen $\tilde{\mu}$) under the two choices of $M$. We find that as the AEFF reduces (by choosing a larger $\tilde{\mu}$), the IF would shrink towards zero. This reflects that a more robust estimation of tail index can be achieved using the proposed MWLE approach by trading off some efficiencies.

\subsection{Simulation studies} \label{sec:ex:sim}
\subsubsection{Simulation settings}
We here simulate $n=10,000$ claims (the sample size is motivated by the size of a typical insurance portfolio) from the aforementioned $J$-Gamma Lomax distribution for each of the following two parameter settings with $\theta=1000$:
\begin{itemize}
    \item Model 1: $J=2$, $\bm{\pi}=(0.4,0.4,0.2)$, $\bm{\mu}=(100,300)$, $\bm{\phi}=(0.25,0.25)$ and $\gamma=2$.
    \item Model 2: $J=3$, $\bm{\pi}=(0.4,0.3,0.1,0.2)$, $\bm{\mu}=(50,200,600)$, $\bm{\phi}=(0.2,0.2,0.2)$ and $\gamma=2$.
\end{itemize}

We also consider the zero-inflated Gamma distribution given by Equation (\ref{eq:em:wgt_func}) as the weight function, with $\tilde{\mu}\in\{q_{0},q_{0.9},q_{0.95},q_{0.99},q_{0.995}\}$, $\tilde{\phi}\in\{0.025,0.1,0.25,1\}$ and $\xi\in\{0.001,0.01,0.05,0.25\}$, where $q_{\alpha}$ is the empirical quantile of the data with $0\leq \alpha\leq 1$. Recall that the choice of $\tilde{\mu}=q_0=0$ implies that $W(y;\xi,\tilde{\mu},\tilde{\phi})= 1$ and hence the MWLE is equivalent to standard MLE. For each combinations of models and weight function hyperparameters, the simulations of sample points are repeated by 100 times to enable thorough analysis of the results using the proposed weighted log-likelihood approach under various settings. Each simulated sample is then fitted to the $J$-Gamma Lomax mixture in Equation (\ref{eq:em:density_mixture}) with $J=2$. Note that for simplicity, in the simulation studies we do not examine the choice of $J$ as outlined by Section \ref{sec:em:complex}. As a result, we have the following research goals in the simulation studies:
\begin{itemize}
\item Under Model 1, the data is fitted to the true class of models. Hence, we empirically verify the consistencies of estimating model parameters (Theorem \ref{thm:asym_tru}) using the MWLE. We also study how the selection of weight function hyperparameters affect the estimated parameter uncertainties. Further, we compare the computational efficiency of the two kinds of proposed GEM algorithms.
\item Under Model 2, the data is fitted to a misspecified class of models. Hence, we demonstrate how this would distort the estimation of the tail under the MLE, and study how the proposed MWLE produces a more robust tail estimation.
\end{itemize}

\subsubsection{Results of fitting Model 1 (true model)}

\begin{figure}[!h]
\begin{center}
\begin{subfigure}[h]{0.49\linewidth}
\includegraphics[width=\linewidth]{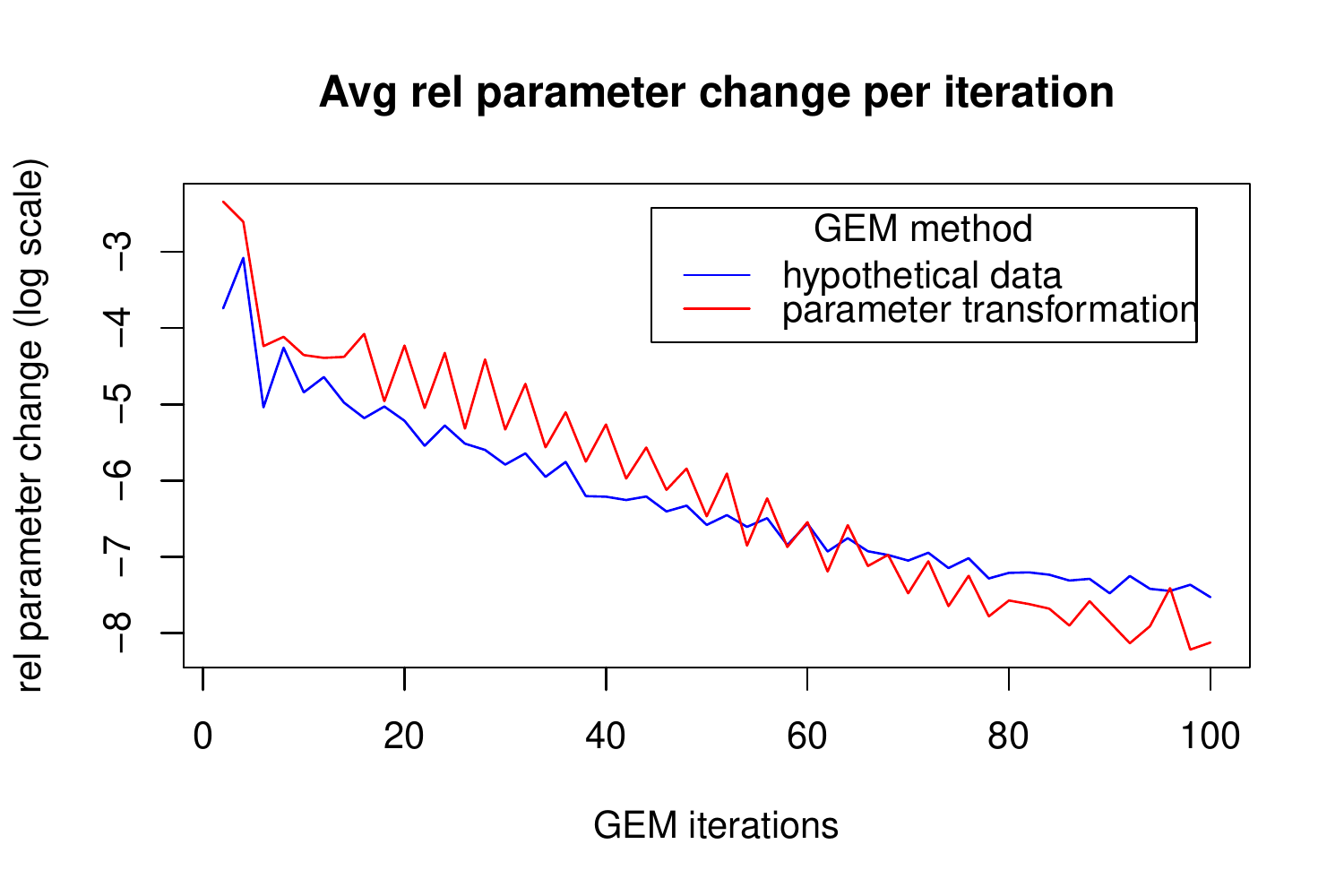}
\end{subfigure}
\hfill
\begin{subfigure}[h]{0.49\linewidth}
\includegraphics[width=\linewidth]{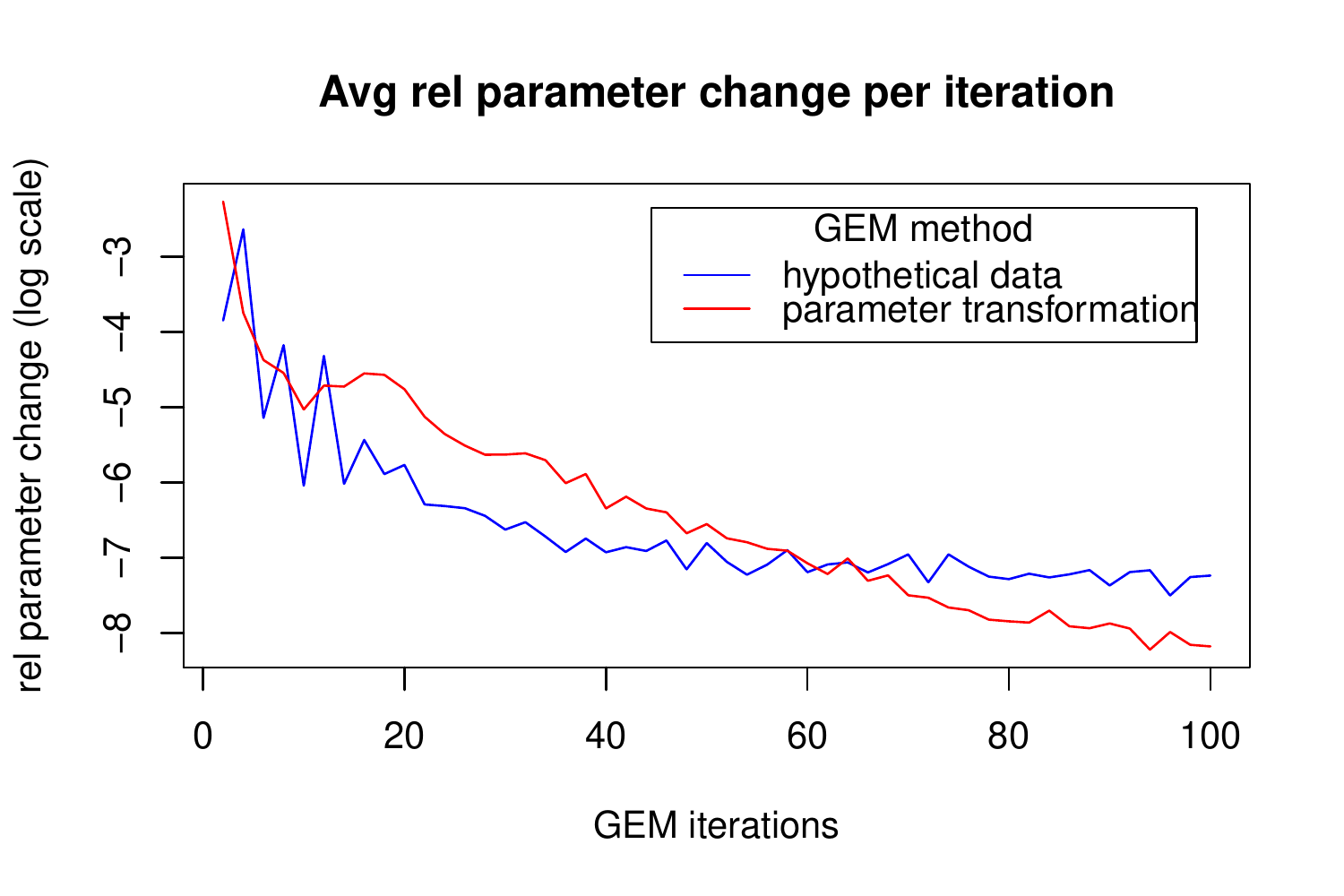}
\end{subfigure}
\end{center}
\caption{The relative change of iterated parameters in the first 100 GEM iterations under two example choices of weight function hyperparameters: Left panel -- $(\xi,\tilde{\mu},\tilde{\phi})=(0.01,q_{0.95},0.1)$; Right panel -- $(\xi,\tilde{\mu},\tilde{\phi})=(0.05,q_{0.99},0.25)$.}
\label{fig:sim_tru_rel}
\end{figure}

Considering the case where we fit the true class of model to the data generated by Model 1, we first compare the computational efficiencies between the two construction methods of the GEM algorithm as presented by Section \ref{sec:em:gem}. In general, around 100 iterations are needed under parameter transformation approach (Method 2), as compared to at least 300 iterations under hypothetical data approach (Method 1), revealing relatively faster convergences under Method 2. Figure \ref{fig:sim_tru_rel} plots the relative change of iterated parameters $\Delta^{\text{rel}}\bm{\Phi}^{(l)}:=|\log(\bm{\Phi}^{(l)}/\bm{\Phi}^{(l-1)})|/P$ versus the GEM iteration $l$ under two example choices of weight function hyperparameters, where the division operator is applied element-wise to the vector of parameters. It is apparent that the curve drops much faster under Method 2 than Method 1, confirming faster convergence under Method 2. The main reason is that the construction of hypothetical missing observations under Method 1 will generally effectively reduce the learning rates of the optimization algorithms. As both methods produce very similar estimated parameters while Method 2 is more computationally efficient, from now on we only present the results produced by the GEM algorithm under Method 2.

\begin{figure}[!h]
\begin{center}
\begin{subfigure}[h]{0.49\linewidth}
\includegraphics[width=\linewidth]{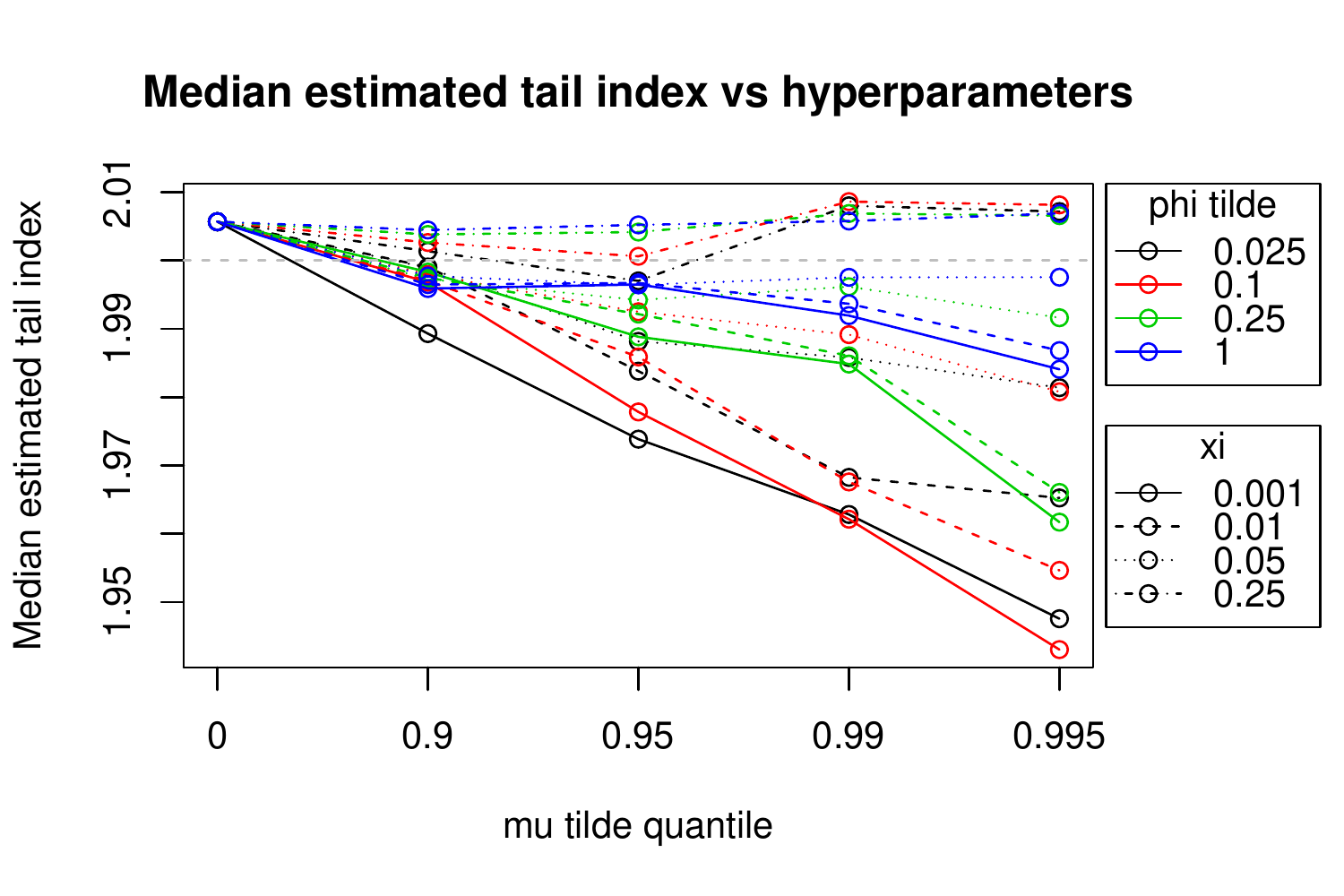}
\end{subfigure}
\hfill
\begin{subfigure}[h]{0.49\linewidth}
\includegraphics[width=\linewidth]{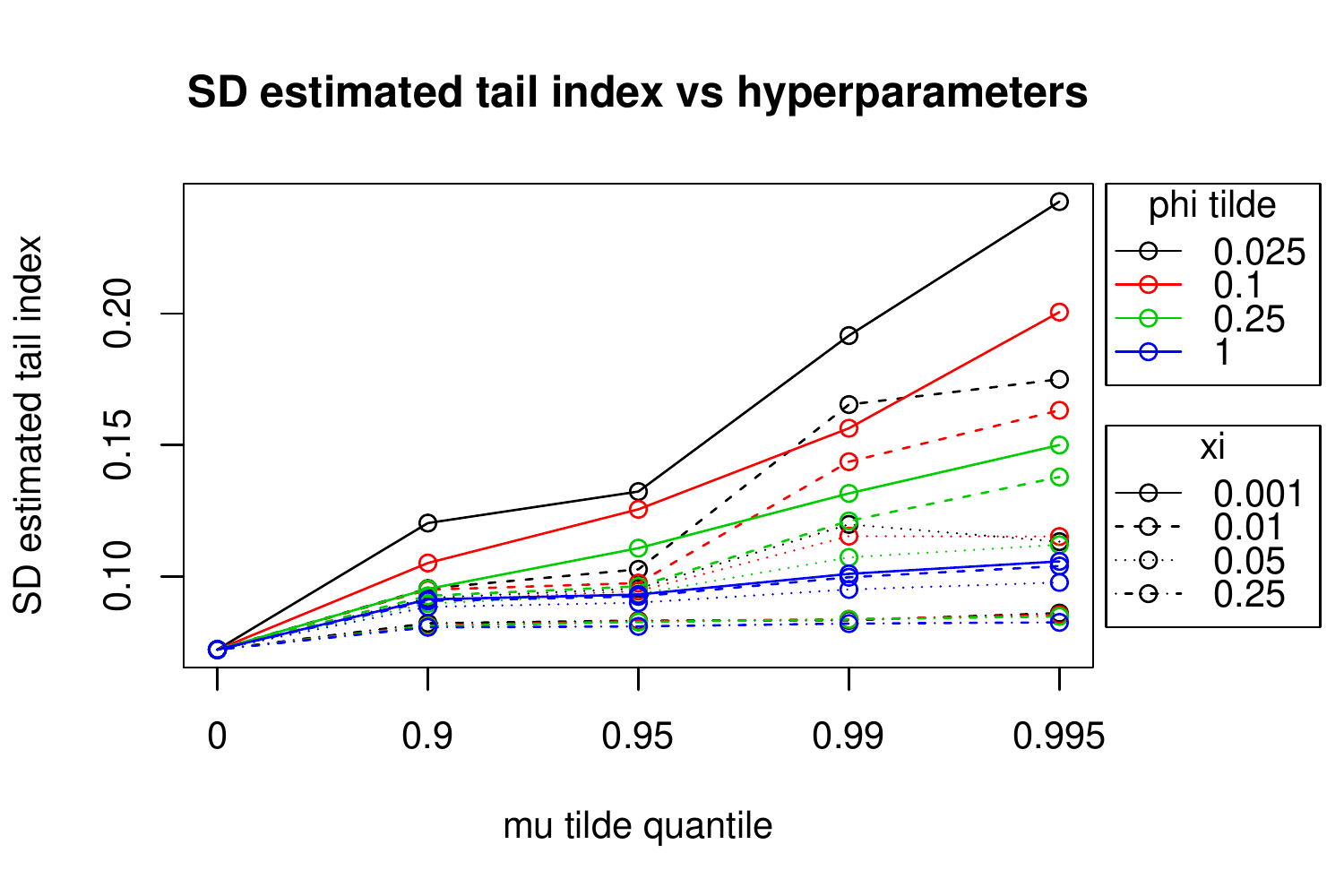}
\end{subfigure}
\end{center}
\caption{Median and standard deviation of the estimated tail index $\hat{\gamma}$ versus various weight function hyperparameters under the true model.}
\label{fig:sim_tru_gamma}
\end{figure}

Figure \ref{fig:sim_tru_gamma} demonstrates the how the biasedness and uncertainty of the estimated tail index $\hat{\gamma}$ differ among various choices of weight functions and their corresponding hyperparameters $(\xi,\tilde{\mu},\tilde{\phi})$. From the left panel, the median estimated parameters are very close to the true model parameters (differ by less than 1-2\%) under most settings of the weight functions, except for few extreme cases where both $\xi$ and $\tilde{\phi}$ are chosen to be very small. This empirically justifies the asymptotic unbiasedness of the MWLE. As expected from the right panel, the uncertainties of MLE parameters are the smallest, verifying that MLE is the asymptotically most efficient estimator among all unbiased estimators if we are fitting the correct model class. The parameter uncertainties generally slightly increase as we choose larger $\tilde{\mu}$ to de-emphasize the impacts of smaller observations. In some extreme cases where $\xi$ and $\tilde{\phi}$ very small and $\tilde{\mu}$ is very large, the standard error can grow dramatically, reflecting that a lot of information are effectively discarded. 

\begin{figure}[!h]
\begin{center}
\begin{subfigure}[h]{0.49\linewidth}
\includegraphics[width=\linewidth]{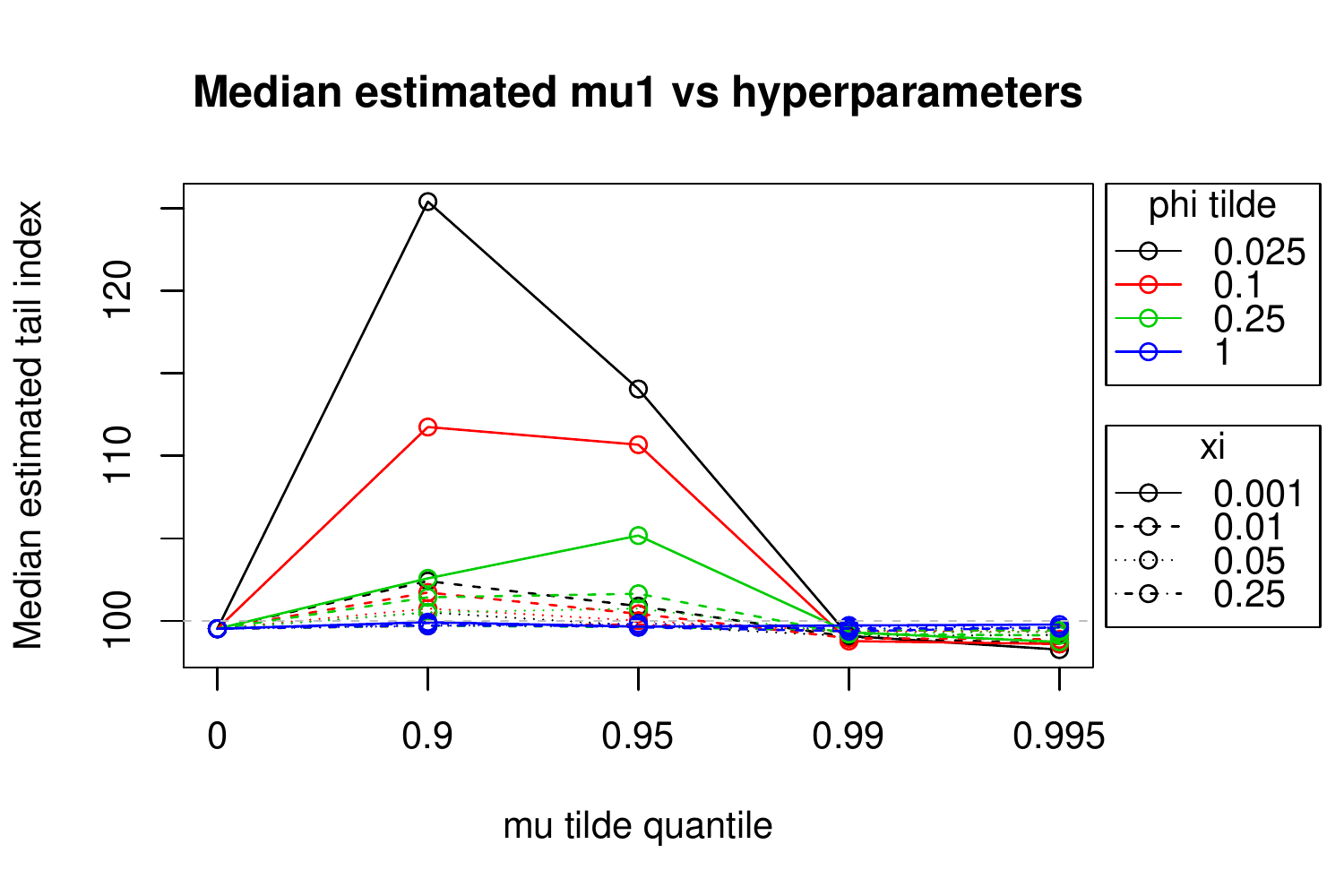}
\end{subfigure}
\hfill
\begin{subfigure}[h]{0.49\linewidth}
\includegraphics[width=\linewidth]{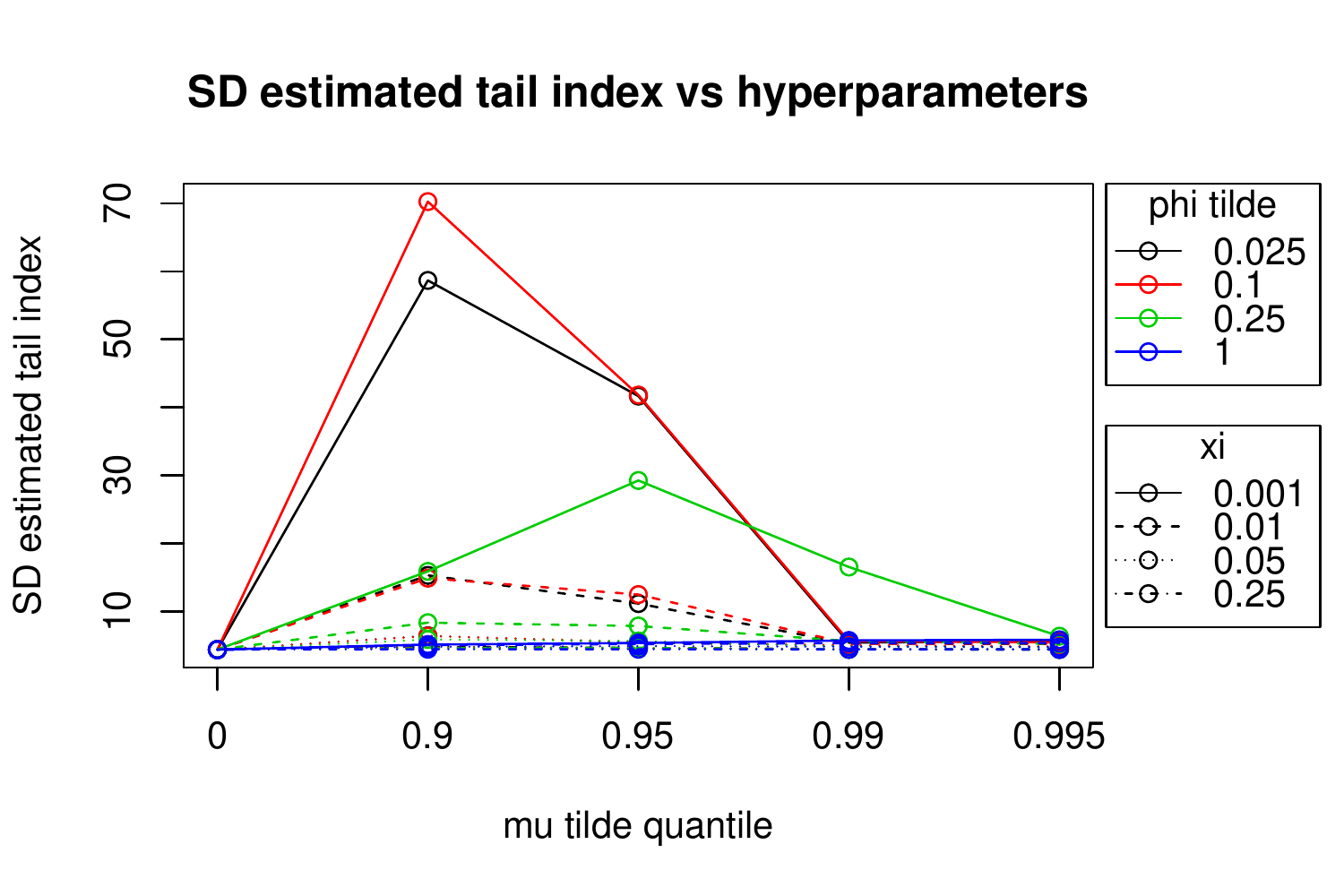}
\end{subfigure}
\end{center}
\caption{Median and standard deviation of the estimated mean parameter of the first mixture component $\hat{\mu_1}$ versus various weight function hyperparameters under the true model.}
\label{fig:sim_tru_mu1}
\end{figure}

Similarly in Figure \ref{fig:sim_tru_mu1} where the biasedness and uncertainty of an estimated mean parameter $\hat{\mu_1}$ from the body distribution are displayed, we observe that the proposed MWLE approach behaves properly for fitting the body distribution unless when $\xi$ and $\tilde{\phi}$ are both chosen to be extremely small (in those cases, the estimated body parameters would become unstable with inflated uncertainties). Hence, these extreme choices of hyperparameters are deemed to be inappropriate.

\subsubsection{Results of fitting Model 2 (misspecified model)}

\begin{figure}[!h]
\begin{center}
\begin{subfigure}[h]{0.49\linewidth}
\includegraphics[width=\linewidth]{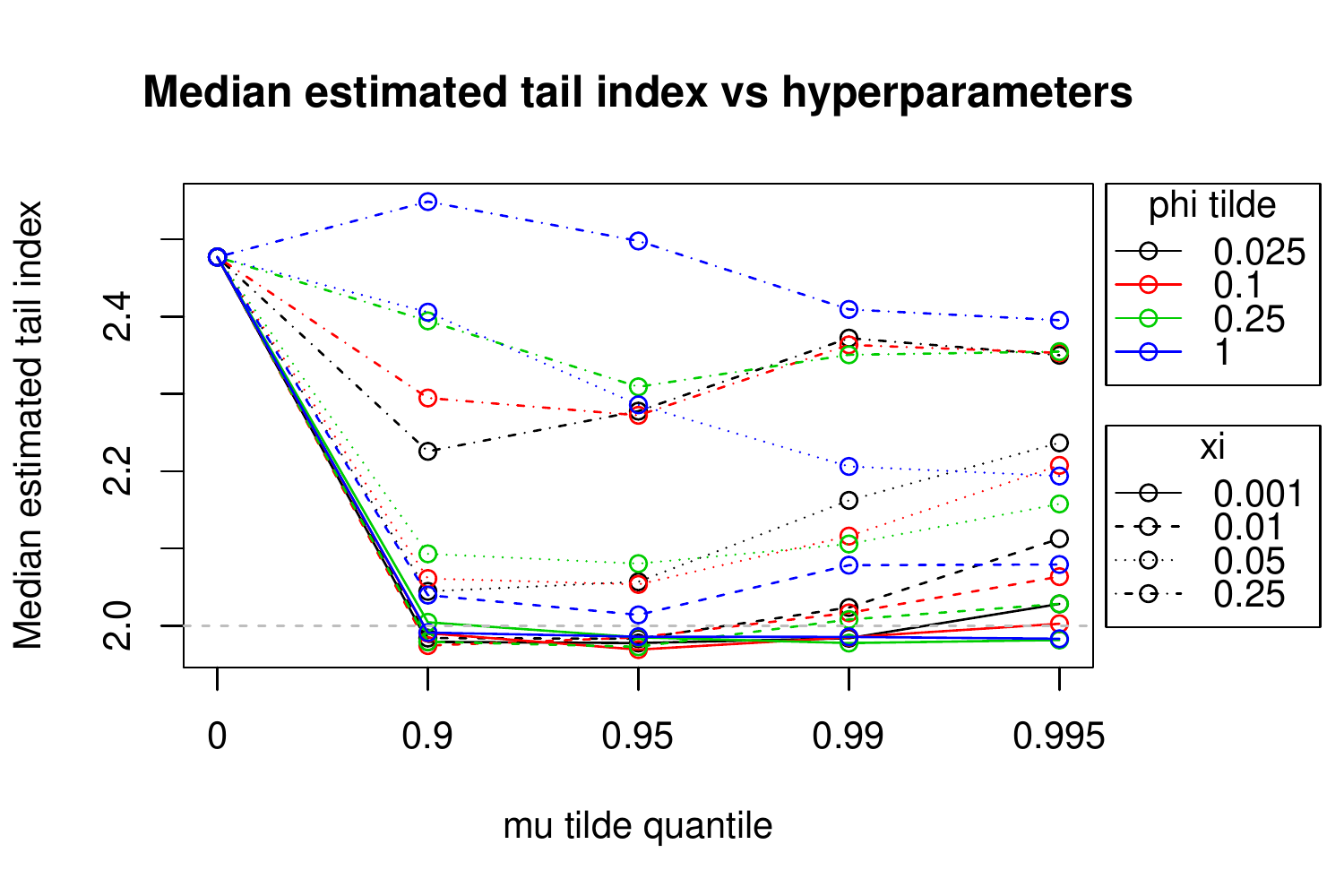}
\end{subfigure}
\hfill
\begin{subfigure}[h]{0.49\linewidth}
\includegraphics[width=\linewidth]{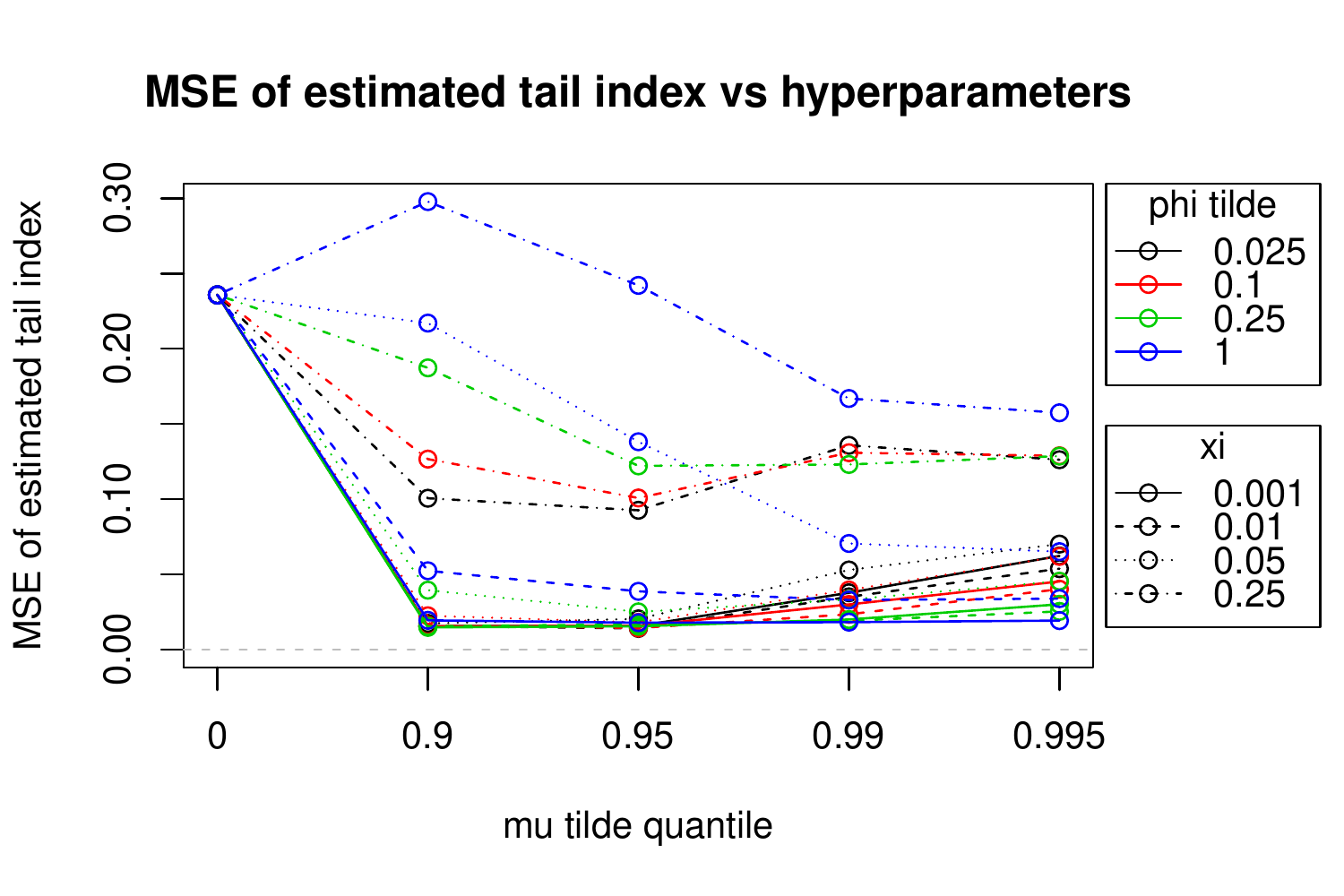}
\end{subfigure}
\end{center}
\caption{Median and MSE of the estimated tail index $\hat{\gamma}$ versus various weight function hyperparameters under the misspecified model.}
\label{fig:sim_mis_gamma}
\end{figure}

\begin{figure}[!h]
\begin{center}
\begin{subfigure}[h]{0.49\linewidth}
\includegraphics[width=\linewidth]{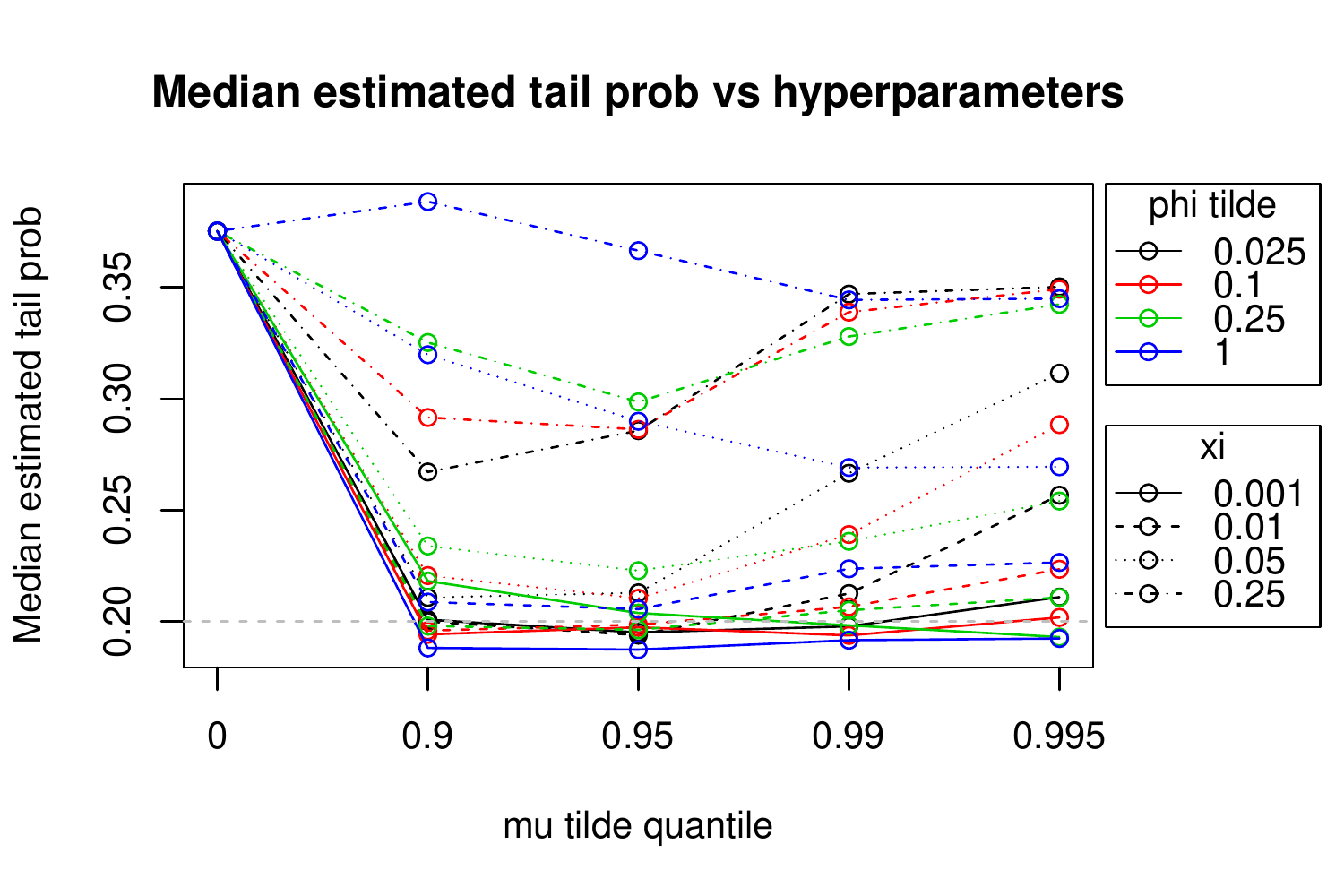}
\end{subfigure}
\hfill
\begin{subfigure}[h]{0.49\linewidth}
\includegraphics[width=\linewidth]{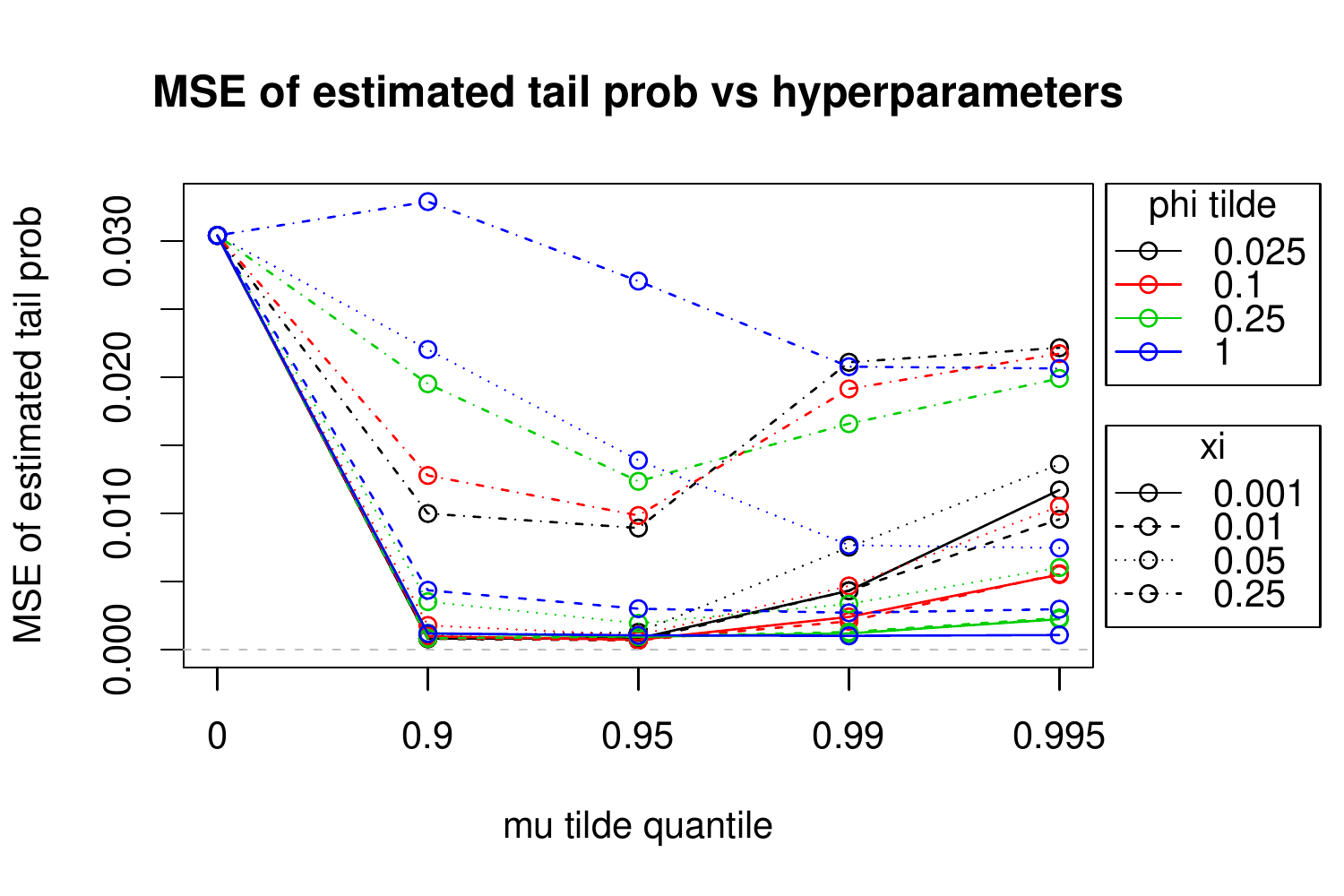}
\end{subfigure}
\end{center}
\caption{Median and MSE of the estimated tail probability $\hat{\pi}_{J+1}$ versus various weight function hyperparameters under the misspecified model.}
\label{fig:sim_mis_tailp}
\end{figure}

We now turn to the case where we fit a misspecified model (with $J=2$) to the simulated data generated from Model 2 (with $J=3$). The left panels of Figures \ref{fig:sim_mis_gamma} and \ref{fig:sim_mis_tailp} examine how the robustness of the estimated tail index $\hat{\gamma}$ and tail probability $\hat{\pi}_{J+1}$ differs among different choices of hyperparameters $(\xi,\tilde{\mu},\tilde{\phi})$. From the left panel, the MLE of the tail index is around $\hat{\gamma}=2.48$ which largely over-estimates the true tail index $\gamma=2$, indicating that the heavy-tailedness of the true distribution is under-extrapolated. On the other hand, with the incorporation of weight functions to under-weight the smaller claims, the biases of the MWLE of $\gamma$ are greatly reduced compared to that of the MLE under most choices of weight function hyperparameters. In particular, the bias reduction for tail index is more effective using smaller $\xi$ (i.e. $\xi\leq 0.05$). This is intuitive as smaller $\xi$ means smaller claims are under-weighted by a larger extent, reducing the impacts of smaller claims on the tail index estimations. Similarly from the right panel, the proposed MWLE approach effectively reduces the bias of the estimated tail probability $\hat{\pi}_{J+1}$. 

The analysis of bias-variance trade-off is also conducted through computing the mean-squared errors (MSE) of both estimated tail index $\hat{\gamma}$ and tail probability $\hat{\pi}_{J+1}$. From the right panels of Figures \ref{fig:sim_mis_gamma} and \ref{fig:sim_mis_tailp}, as evidenced by smaller MSEs under most choices of weight function hyperparameters, MWLE is much more preferable than MLE approach even after accounting for the increased parameter uncertainties through down-weighting the importance of smaller claims.

\subsubsection{Summary remark on the choice of weight function hyperparameters}
From the above two simulation studies, we find that under a wide range of choices of weight function hyperparameters, the proposed MWLE not only produces plausible model estimations under true model (Model 1), but is also effective in mitigating the bias of tail estimation inherited from model misspecifications (Model 2).

Among the three hyperparameters $(\xi,\tilde{\mu},\tilde{\phi})$, the choice of minimum weight hyperparameter $\xi$ plays a particularly vital role on the bias-variance trade-off of the estimated parameters. Under misspecified model (Model 2), smaller $\xi$ (i.e. $\xi\leq 0.05$) is more effective in reducing the biases of both estimated tail index $\hat{\gamma}$ and tail probability $\hat{\pi}_{J+1}$. However, as evidenced by the results produced under the true model (Model 1), the estimated parameters of the body distributions (i.e. $\hat{\bm{\mu}}$ and $\hat{\bm{\phi}}$) may become prohibitively unstable if $\xi$ is chosen to be extremely small (i.e. $\xi\leq 0.001$) such that smaller observations are effectively almost fully discarded. It is therefore important to compare parameter uncertainties of MWLE to that of MLE, and select/ consider only the weight function hyperparameters where the corresponding MWLE parameter uncertainties are within an acceptable range (i.e. not too off from the MLE parameter uncertainties). Overall, the choices of $\xi$ between 0.01 and 0.05 are deemed to be suitable.

\subsection{Real data analysis} \label{sec:ex:real}
\subsubsection{Data description and background} \label{sec:ex:real:dat}
In this section, we study an insurance claim severity dataset kindly provided by a major insurance company operating in Greece. It consists of 64,923 motor third-party liability (MTPL) insurance policies with non-zero property claims for underwriting years 2013 to 2017. This dataset is also analyzed by \cite{fung2021mixture} using a mixture composite model, with an emphasis on selecting various policyholder characteristics (explanatory variables) which significantly influence the claim severities. The empirical claim severity distribution exhibits several peculiar characteristics including multimodality and tail-heaviness. The excessive number of distributional nodes for small claims reflects the possibility of distributional contamination, which cannot be and should not be perfectly captured and over-fitted by parametric models like FMM. Preliminary analyses also suggest that the estimated tail index is around 1.3 to 1.4, but note that these are only rough and subjective estimates. The details of preliminary data analysis are provided in Section 9 of the supplementary materials. The key goals of this real data analysis are as follows:
\begin{enumerate}
\item Illustrate that MLE of FMM would produce highly unstable and unrobust estimates to the tail part of the claim severity distribution. This confirms that tail-robustness is an important research problem in real insurance claim severity modelling which needs to be properly addressed.
\item Demonstrate how the proposed MWLE approach leads to superior fittings to the tail and more reliable estimates of tail index as compared to MLE, without much sacrificing its ability to adequately capture the body.
\end{enumerate}

To avoid diverging the focus of this paper, in this analysis we solely examine the distributional fitting of the claim sizes without considering the explanatory variables. Note however that the proposed MWLE can be extended to a regression framework, with the discussions being leveraged to Section \ref{sec:discussion}.

\subsubsection{Fitting results}
The claim severity dataset is fitted to the mixture Gamma-Lomax distribution with density given by Equation \ref{eq:em:density_mixture} under the proposed MWLE approach. The fitting performances will be examined thoroughly across different number of Gamma (body) mixture components $J\in\{1,2,\ldots,10\}$ and various choices of weight function hyperparameters ($\tilde{\mu}\in\{q_{0},q_{0.9},q_{0.95},q_{0.99},q_{0.995}\}$, $\tilde{\phi}\in\{0.025,0.1,0.25,1\}$ and $\xi\in\{0.001,0.01,0.05,0.25\}$). The MWLE fitted parameters are also compared to the standard MLE across various $J$.

\begin{figure}[!h]
\begin{center}
\begin{subfigure}[h]{0.49\linewidth}
\includegraphics[width=\linewidth]{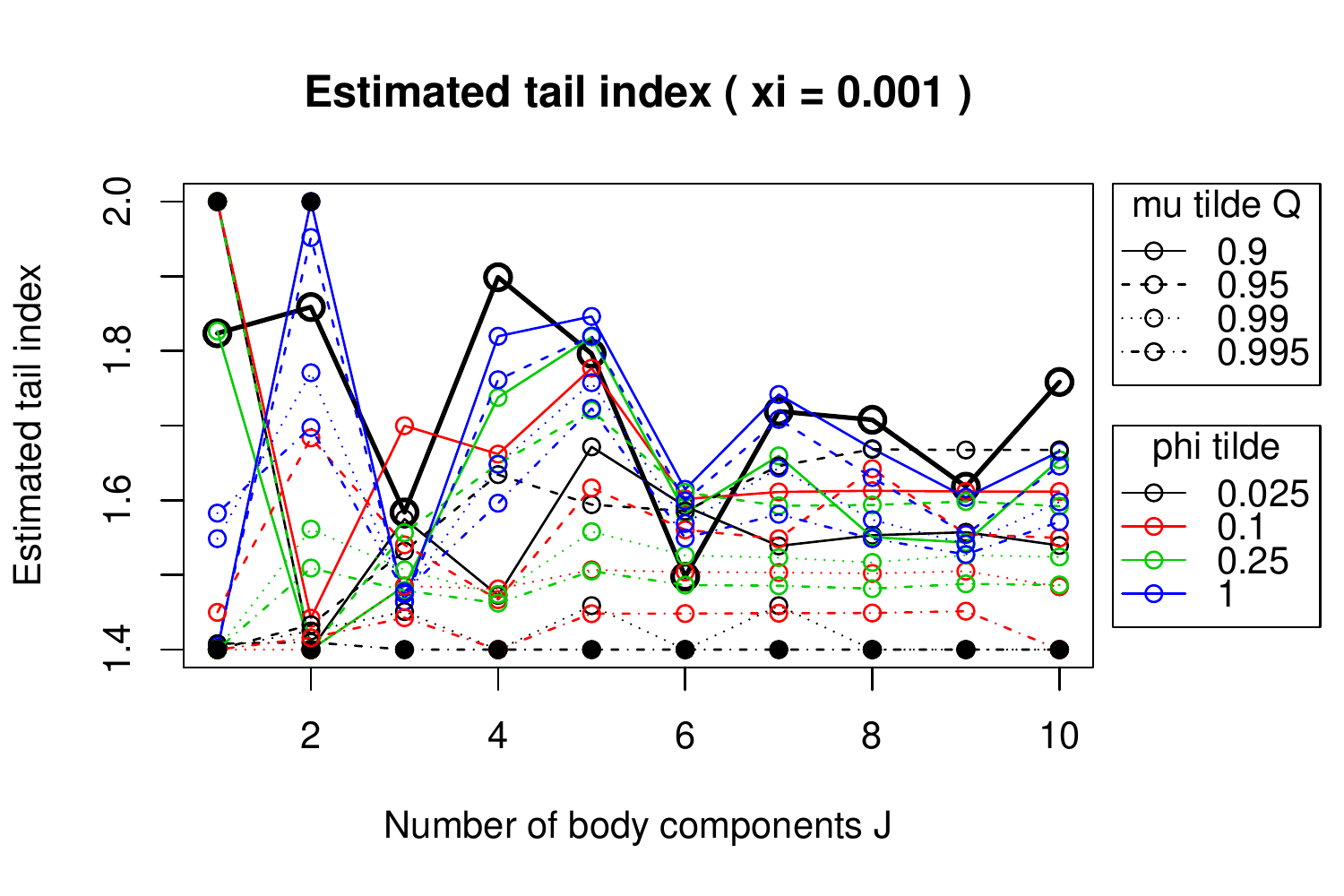}
\end{subfigure}
\begin{subfigure}[h]{0.49\linewidth}
\includegraphics[width=\linewidth]{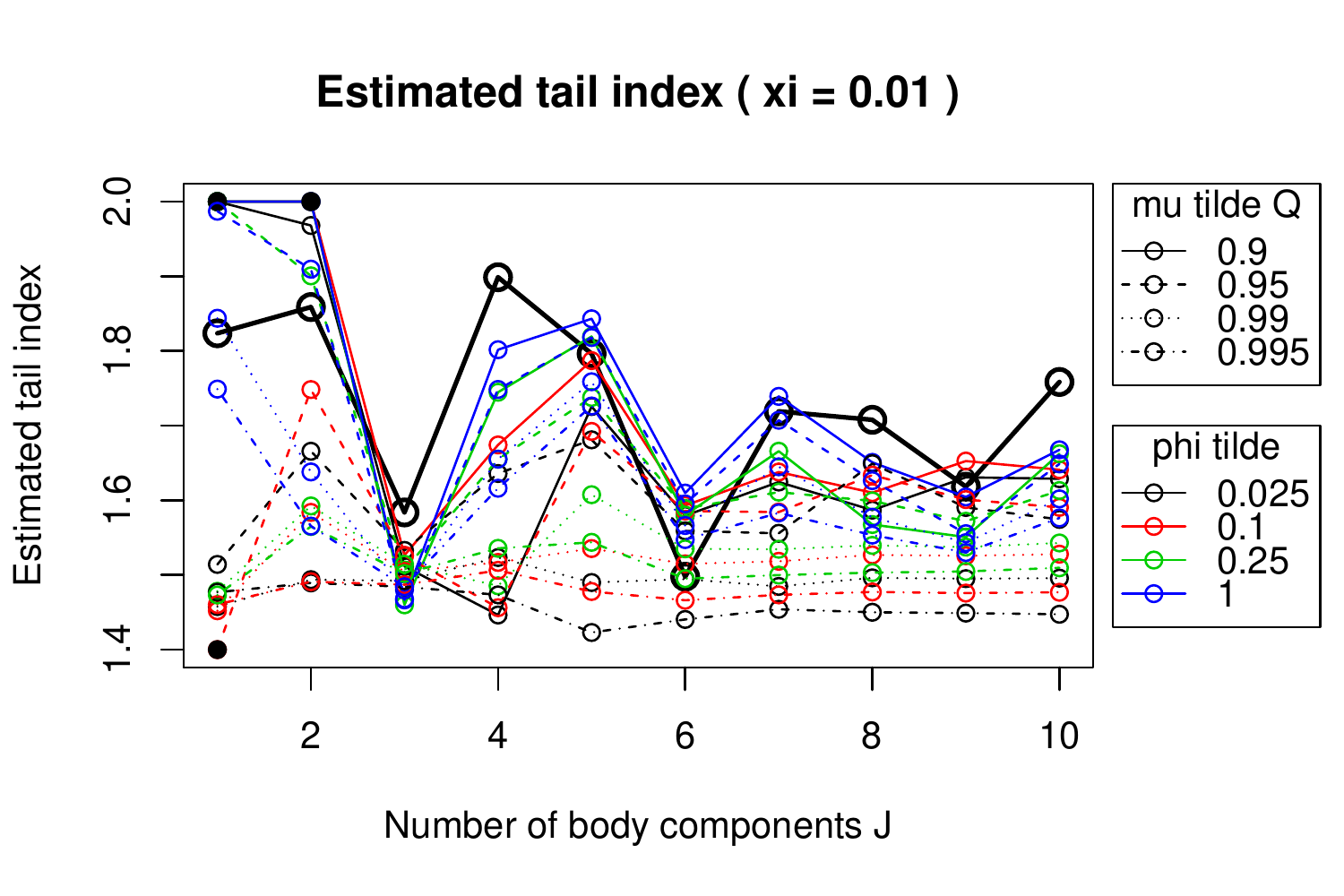}
\end{subfigure}
\begin{subfigure}[h]{0.49\linewidth}
\includegraphics[width=\linewidth]{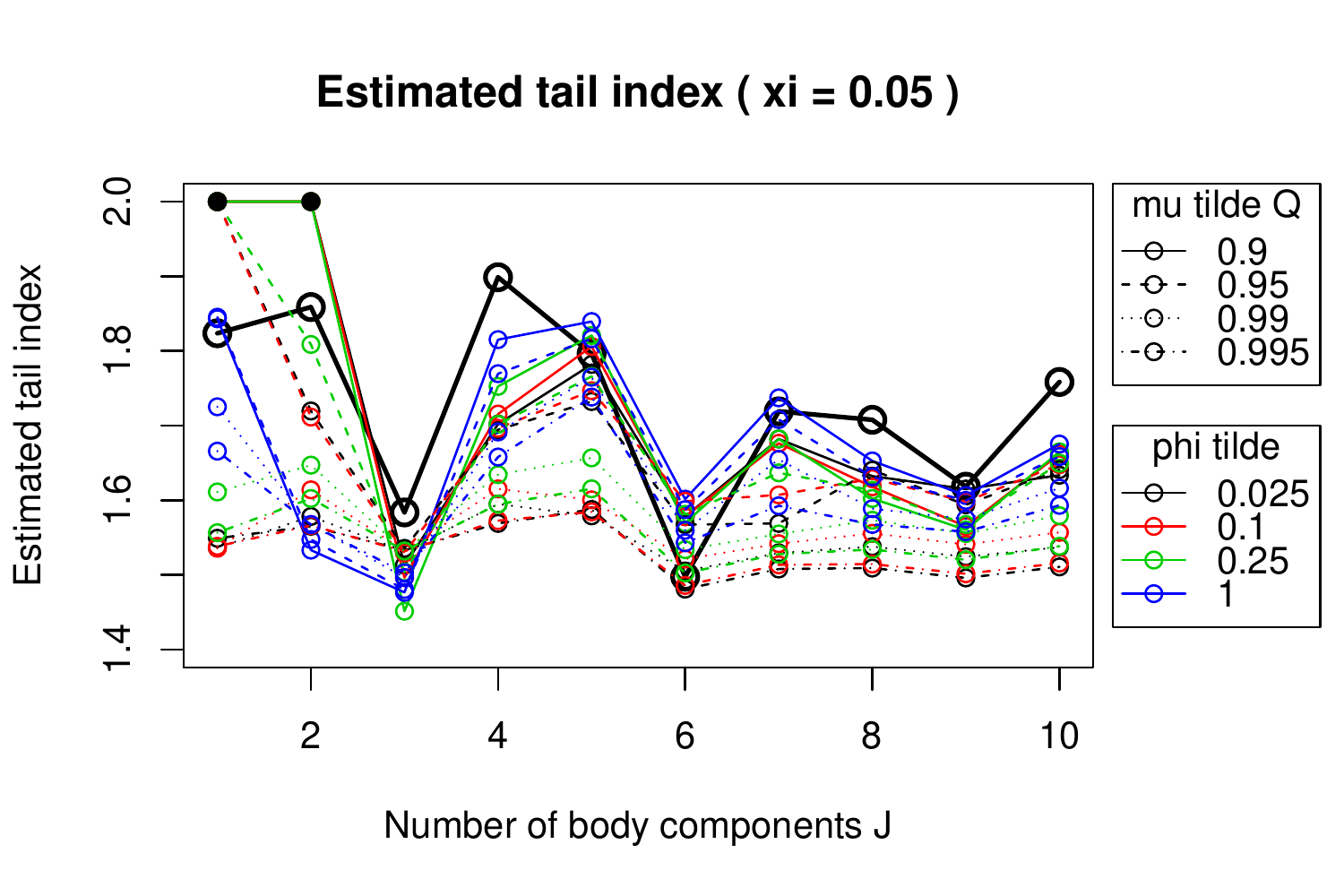}
\end{subfigure}
\begin{subfigure}[h]{0.49\linewidth}
\includegraphics[width=\linewidth]{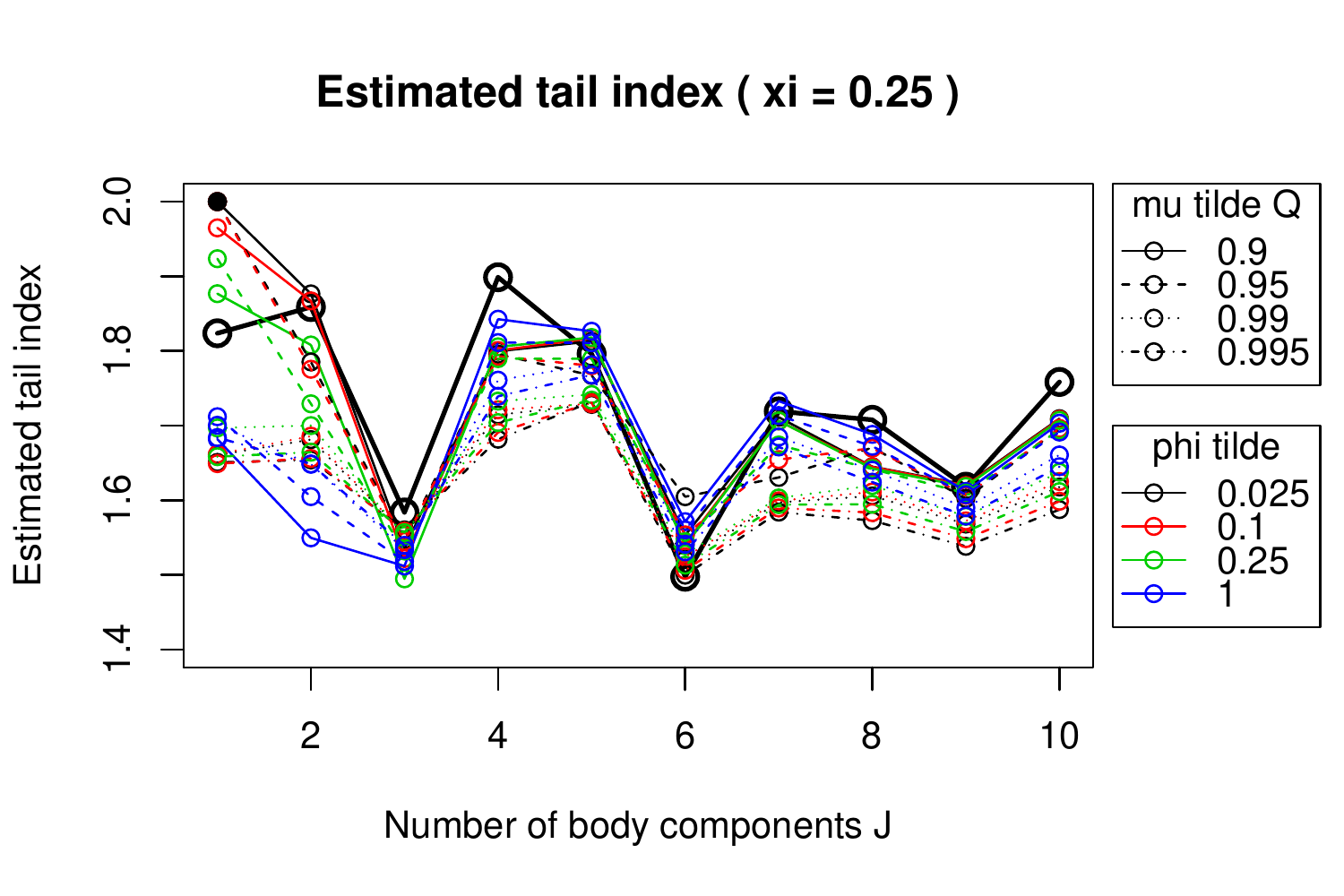}
\end{subfigure}
\end{center}
\caption{Estimated tail index versus the number of Gamma mixture components under MLE and MWLE with various choices of weight function hyperparameters.}
\label{fig:greek:gamma}
\end{figure}

We first present in Figure \ref{fig:greek:gamma} the fitted tail index $\hat{\gamma}$ versus the number of body components $J$ under all combinations of selected weight function hyperparameters. Each of the four sub-figures corresponds to a particular choice of $\xi\in\{0.001,0.01,0.05,0.25\}$. The black thick trends for each sub-figure are the MLE estimated tail indexes for comparison purpose. The MLE tail indexes are rather unstable as evidenced by great fluctuations across different number of body components $J$, showing that MLE may not be reliable in extrapolating the heavy-tailedness of complex claim distributions.
For instance, with a slight change of model specification from $J=5$ to $J=6$, the estimated tail index largely drops from about 1.8 to 1.5. This is rather unnatural because the change from $J=5$ to $J=6$ should only reflect a slight change in specifying the body.
The large drop of the estimated tail index reflects that the Lomax tail part of FMM is not specialized in extrapolating the tail-heaviness of the distribution, but instead is very sensitive to the small claims and the model specifications of the body part.
Therefore, we conclude that the mixture Gamma-Lomax FMM is not achieving its modelling purpose under the MLE.

On the other hand, looking individually at each path under MWLE, we find that the estimated $\hat{\gamma}$ is much more stable across different $J$ under most choices of weight function hyperparameters, especially when $J\geq 5$. Also, the estimated MWLE $\hat{\gamma}$ is in general smaller than the $\hat{\gamma}$ obtained by MLE, moving closer to the values roughly determined by the preliminary data analysis in Section \ref{sec:ex:real:dat}. Note in the figure that there are a few black solid dots, which appear when the estimated $\hat{\gamma}$ under MWLE is outside the range of the plots. These unstable estimates of $\hat{\gamma}$ are rare and only occur under one the following two situations: (i) $J$ is chosen be very small (i.e. $J\leq 2$) in the sense that the models would severely under-fit the distributional complexity of the dataset; (ii) extreme choices of weight function hyperparameters (very small $\xi$ and $\tilde{\phi}$) aligned to the results of the simulation studies.

The optimal choice of $J$ is tricky as evidenced by the excessive number of small distributional nodes for very small claim sizes described in Section \ref{sec:ex:real:dat}, which should not be over-emphasized or excessively modelled as these very small claims are almost irrelevant for pricing and risk management. However, both AIC and BIC decrease slowly and steadily for MLE models as $J$ increases. The optimal $J$ in this case goes way beyond $J=10$. Under the proposed MWLE approach with various choices of weight function hyperparameters, the same model selection problem exists using RAIC and RBIC, with the reasons already explained in Section \ref{sec:em:complex}. On the other hand, using TAIC and TBIC (especially for TBIC), a majority selections of weight function hyperparameters lead to an optimal $J=5$, aligning with the heuristic arguments by \cite{fung2021mixture} that $J=5$ is enough for capturing all the distributional nodes except for the very small claims which are smoothly approximated by a single mixture component. 

To better understand how the use of proposed MWLE affects the estimations of all parameters (not just the tail index but also parameters affecting the body distributions such as $\bm{\mu}$), we showcase in Table \ref{tab:greek:est_prm} all the estimated parameters and their standard errors (based on Equation (\ref{eq:asym_var})) using MWLE under two distinguishable example choices of hyperparameters (MWLE 1: $(\xi,\tilde{\mu},\tilde{\phi})=(0.01,q_{0.99},0.1)$; MWLE 2: $(\xi,\tilde{\mu},\tilde{\phi})=(0.05,q_{0.995},0.25)$) as compared to MLE parameters with $J=5$, the optimal number of body components under TBIC for both MWLE 1 and MWLE 2. Note that the two selected examples are for demonstration purpose -- generally the following findings and conclusions are also valid for other choices of weight function hyperparameters under the proposed MWLE.

We first observe that the estimated parameters influencing the body (i.e. $\bm{\pi}$, $\bm{\mu}$ and $\bm{\phi}$) under MWLE are very close to those under MLE, even if the smaller claims are greatly down-weighted. MWLE generally results to larger parameter uncertainties as compared with MLE -- reflecting a bias-variance trade-off, but these standard errors are of the same order of magnitude and are still relatively immaterial compared to the estimates as the sample size $n=64,923$ is large. 

Comparing between the above two MWLE examples, we further notice that the parameter uncertainties under MWLE 1 are greater than those under MWLE 2. This is expected because the influences of smaller claims are down-weighted more under MWLE 1 than those under MWLE 2 (as reflected by smaller minimum weight hyperparameter $\xi$ chosen under MWLE 1). On the other hand, the estimated tail index $\hat{\gamma}$ under MWLE 1 is slightly closer to the heuristic values (i.e. 1.3 to 1.4) than MWLE 2. These may also reflect the bias-variance trade-off among various choices of weight function hyperparameters.

\begin{table}[!h]
\centering
\begin{tabular}{lrrrrrr}
\toprule
 & \multicolumn{2}{c}{MWLE 1} & \multicolumn{2}{c}{MWLE 2} & \multicolumn{2}{c}{MLE} \\
\cmidrule(l{3pt}r{3pt}){2-3} \cmidrule(l{3pt}r{3pt}){4-5} \cmidrule(l{3pt}r{3pt}){6-7}
 & \multicolumn{1}{c}{Estimates} & \multicolumn{1}{c}{Std. Error} & \multicolumn{1}{c}{Estimates} & \multicolumn{1}{c}{Std. Error} & \multicolumn{1}{c}{Estimates} & \multicolumn{1}{c}{Std. Error} \\
\midrule
$\pi_1$ & 0.3787 & 0.0053 & 0.3829 & 0.0031 & 0.3878 & 0.0022 \\
$\pi_2$ & 0.0380 & 0.0036 & 0.0404 & 0.0021 & 0.0444 & 0.0014 \\
$\pi_3$ & 0.1117 & 0.0024 & 0.1134 & 0.0020 & 0.1161 & 0.0017 \\
$\pi_4$ & 0.0221 & 0.0059 & 0.0192 & 0.0021 & 0.0153 & 0.0008 \\
$\pi_5$ & 0.2173 & 0.0036 & 0.2163 & 0.0022 & 0.2130 & 0.0019 \\
\hline
$\mu_1$ & 1,303.21 & 50.30 & 1,322.10 & 16.70 & 1,348.68 & 11.11 \\
$\mu_2$ & 9,171.42 & 145.83 & 9,165.92 & 64.64 & 9,165.36 & 49.38 \\
$\mu_3$ & 27,590.46 & 125.68 & 27,571.75 & 64.36 & 27,538.88 & 52.41 \\
$\mu_4$ & 317,274.90 & 2,410.93 & 323,827.70 & 2,159.37 & 322,872.40 & 2,372.68 \\
$\mu_5$ & 89,007.07 & 170.41 & 88,979.12 & 112.01 & 88,895.92 & 99.20 \\
\hline
$\phi_1$ & 0.9945 & 0.0175 & 0.9996 & 0.0121 & 1.0062 & 0.0113 \\
$\phi_2$ & 0.0264 & 0.0089 & 0.0284 & 0.0030 & 0.0324 & 0.0020 \\
$\phi_3$ & 0.0154 & 0.0015 & 0.0158 & 0.0007 & 0.0164 & 0.0005 \\
$\phi_4$ & 0.0472 & 0.0033 & 0.0333 & 0.0025 & 0.0186 & 0.0020 \\
$\phi_5$ & 0.0127 & 0.0007 & 0.0126 & 0.0003 & 0.0122 & 0.0002 \\
\hline
$\gamma$ & 1.5353 & 0.0707 & 1.6153 & 0.0586 & 1.7963 & 0.0471 \\
$\theta$ & 62,637.42 & 7,829.79 & 73,604.51 & 6,630.65 & 101,107.20 & 5,088.29\\
\hhline{=======}
\end{tabular}
\caption{\label{tab:greek:est_prm}Estimated parameters and standard errors under MLE and MWLE approaches with $J=5$.}
\end{table}

\begin{figure}[!h]
\begin{center}
\begin{subfigure}[h]{0.49\linewidth}
\includegraphics[width=\linewidth]{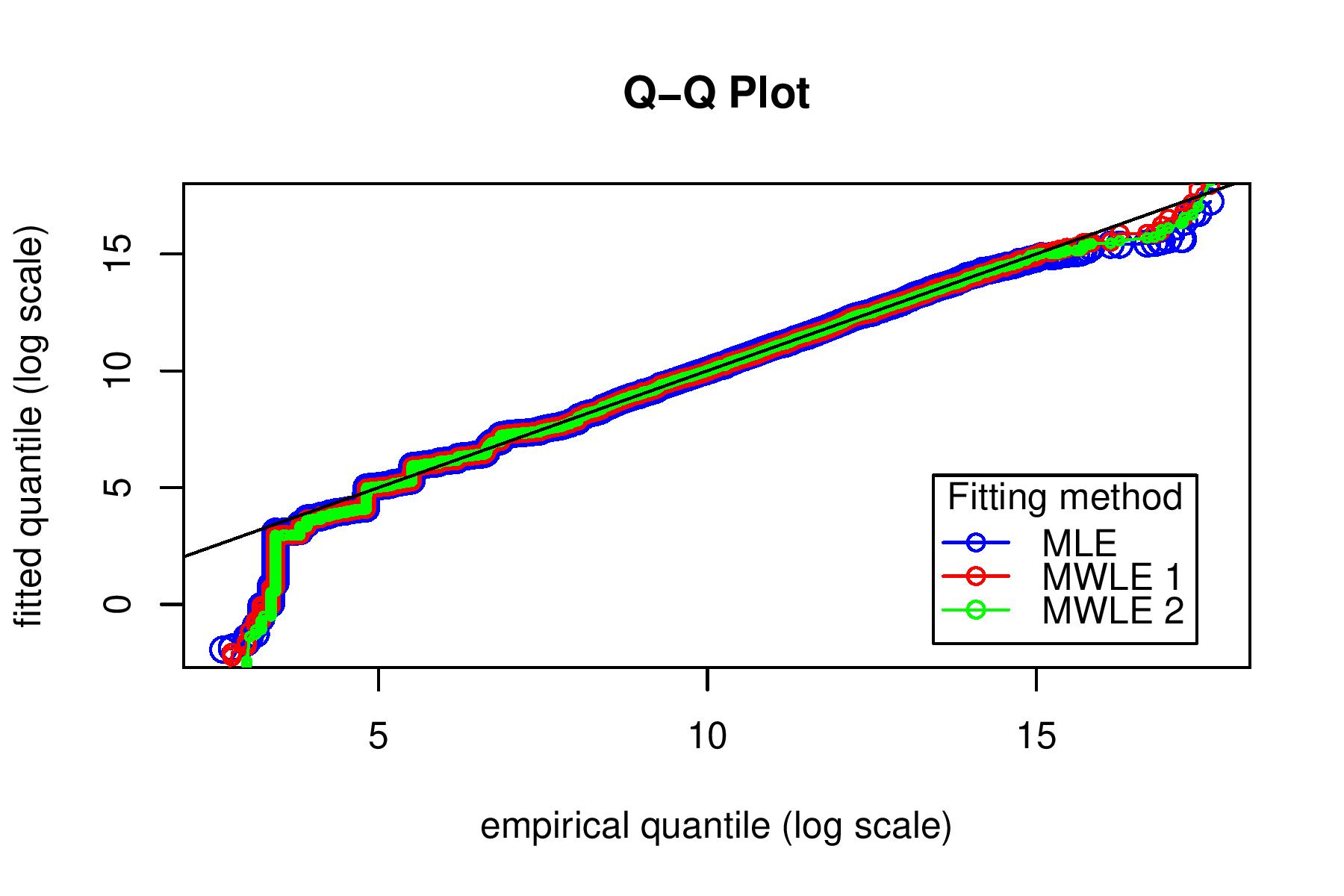}
\end{subfigure}
\end{center}
\caption{Q-Q plot under MLE and MWLE with two selected combinations of weight function hyperparameters.}
\label{fig:greek:qq}
\end{figure}

\begin{figure}[!h]
\begin{center}
\begin{subfigure}[h]{0.49\linewidth}
\includegraphics[width=\linewidth]{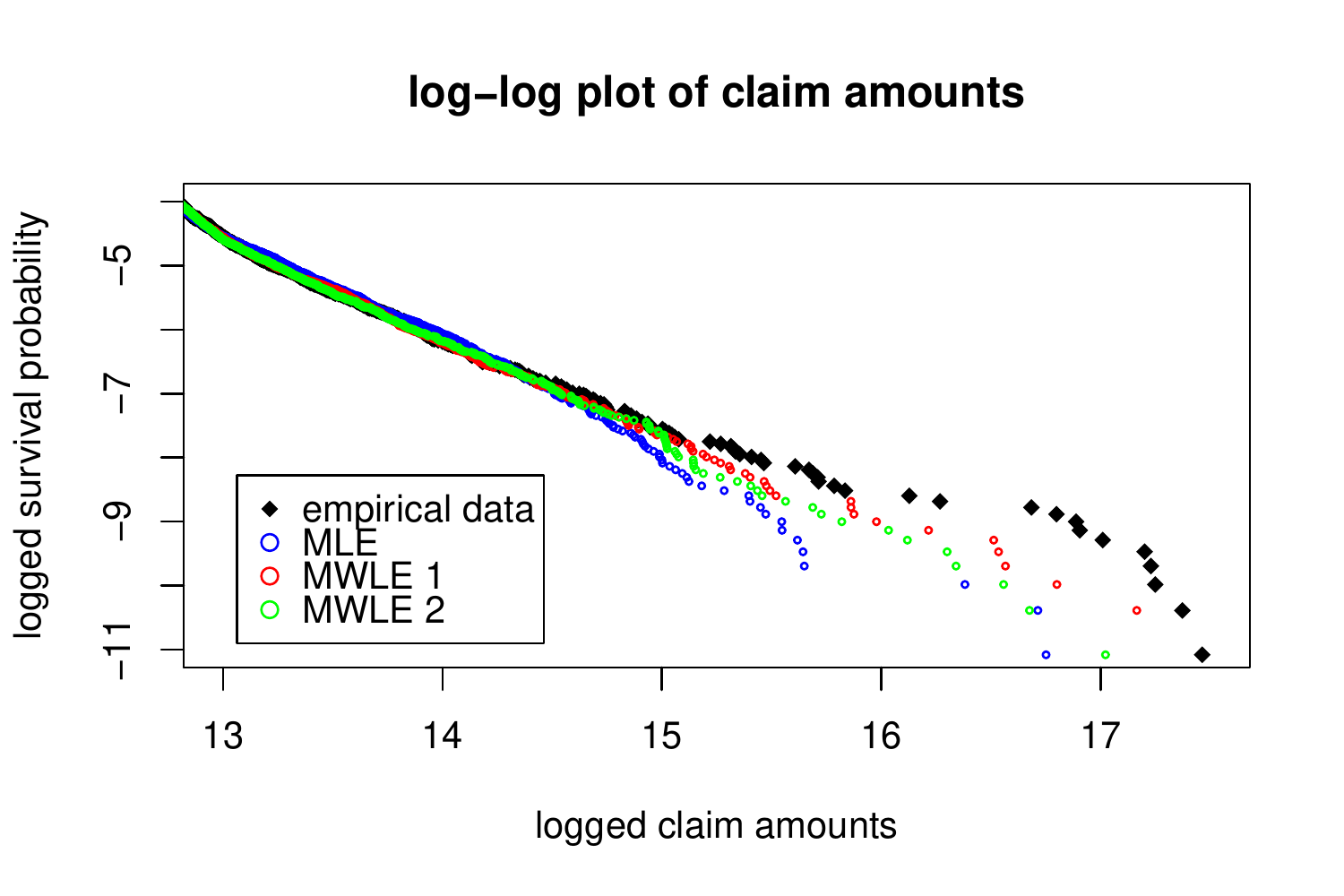}
\end{subfigure}
\begin{subfigure}[h]{0.49\linewidth}
\includegraphics[width=\linewidth]{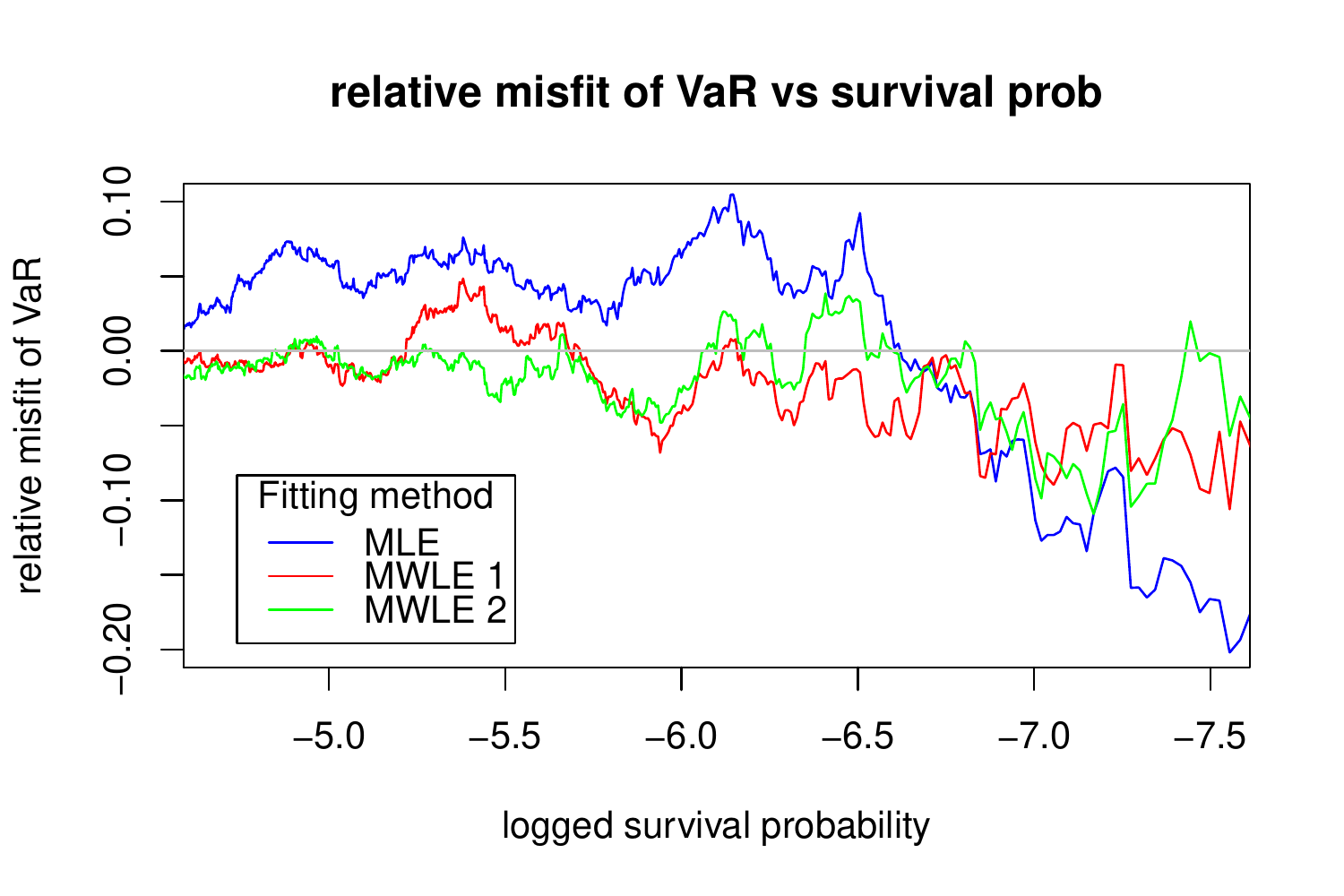}
\end{subfigure}
\end{center}
\caption{Left panel: log-log plot of fitted models compared to empirical data; Right panel: misfit of logged claim amounts versus logged survival probabilities under three fitted models.}
\label{fig:greek:loglog}
\end{figure}

The Q-Q plot in Figure \ref{fig:greek:qq} suggests that the fitting results are satisfactory under both MWLE and MLE except for the very immaterial claims (i.e. $y<100$). Note however that due to the log-scale nature of the Q-Q plot, it is hard to examine from the plot how well the fitted models extrapolate the tail-heaviness of the claim severity data. To examine the tail behavior of the fitted models, we present the log-log plot in the left panel of Figure \ref{fig:greek:loglog}, with the axis shifted to include large claims only. We observe that for extreme claims (i.e. claim amounts greater than about 0.5 millions, or $\log y>13$), the logged survival probability produced by MLE fitted model diverges quite significantly from that of empirical observations. Such a divergence can effectively be mitigated by using MWLE with either of the hyperparameter settings. 

We further compute the value-at-risk (VaR) and conditional tail expectation (CTE) at $100q^{\text{th}}$ security level (denoted as $\text{VaR}_q(Y;\hat{\bm{\Phi}})$ and $\text{CTE}_q(Y;\hat{\bm{\Phi}})$ respectively) from the fitted models, and compare them to the empirical values from the severity data (denoted as $\widehat{\text{VaR}}_q(Y)$ and $\widehat{\text{CTE}}_q(Y)$ respectively). The results are summarized in Table \ref{tab:greek:risk}. Both MLE and MWLE produce plausible estimates of VaR and CTE up to security levels of 95\% and 75\% respectively, reflecting the ability of both approaches in capturing the body part of severity distribution. Nonetheless, the MLE fitted model shows significant divergences of VaR and CTE from the empirical data at higher security levels. In particular, the 99\%-CTE and 99.5\%-CTE are largely underestimated by the MLE approach. Such a divergence is effectively reduced by the proposed MWLE approach where superior fittings to the tail are obtained. Further, MWLE 1 seems to perform slightly better than MWLE 2 in terms of tail fitting, as reflected by smaller underestimations of CTEs at high security levels. This provides a plausible trade-off to the increased parameter uncertainties under MWLE 1 as previously mentioned.

To visualize the results, we further plot the relative misfit of VaR, given by $\log(\text{VaR}_q(Y;\hat{\bm{\Phi}})/\widehat{\text{VaR}}_q(Y))$, versus the log survival probability $\log (1-q)\in(-4.7,-7.5)$, equivalent to the range of security level from 99\% to 99.95\%, in the right panel of Figure \ref{fig:greek:loglog}. We observe that the MLE fitted model over-estimates the VaR of large claims (security level between 99\% to 99.8\%) but then largely under-extrapolates the extreme claims (security level beyond 99.8\%). This issue is well mitigated by the MWLE where the misfits of VaR are smaller in both regions. Therefore, we conclude that the proposed MWLE effectively improves the goodness-of-fit on the tail part of distribution (as compared to the MLE) without much sacrificing its flexibly to adequately capture the body part. 

\begin{table}[!h]
\centering
\begin{tabular}{rrrrrlrrrrr}
\toprule
 \multicolumn{5}{c}{VaR ('000)} &  & \multicolumn{5}{c}{CTE ('000)} \\
\cmidrule(l{3pt}r{3pt}){1-5} \cmidrule(l{3pt}r{3pt}){7-11}
 \multicolumn{1}{c}{Level} & \multicolumn{1}{c}{MLE} & \multicolumn{1}{c}{MWLE 1} & \multicolumn{1}{c}{MWLE 2} & \multicolumn{1}{c}{Empirical} & \multicolumn{1}{c}{} & \multicolumn{1}{c}{Level} & \multicolumn{1}{c}{MLE} & \multicolumn{1}{c}{MWLE 1} & \multicolumn{1}{c}{MWLE 2} & \multicolumn{1}{c}{Empirical} \\
\midrule
50\% & 21 & 21 & 21 & 21 &  & 0\% & 109 & 112 & 111 & 116 \\
75\% & 83 & 82 & 82 & 82 &  & 50\% & 174 & 180 & 177 & 187 \\
95\% & 190 & 191 & 187 & 182 &  & 75\% & 468 & 505 & 489 & 536 \\
99\% & 461 & 450 & 445 & 452 &  & 90\% & 1,140 & 1,326 & 1,242 & 1,498 \\
99.5\% & 719 & 693 & 674 & 676 &  & 95\% & 1,711 & 2,115 & 1,948 & 2,455 \\
99.75\% & 1,149 & 1,031 & 1,046 & 1,075 &  & 99\% & 2,533 & 3,379 & 3,063 & 4,057 \\
99.95\% & 2,787 & 3,163 & 3,220 & 3,348 &  & 99.5\% & 5,956 & 10,243 & 8,572 & 13,329\\
\hhline{===========}
\end{tabular}
\caption{VaR and CTE (in thousands) estimated by MLE, MWLE and empirical approaches.}
\label{tab:greek:risk}
\end{table}

\section{Discussions} \label{sec:discussion}
In this paper, we introduce a maximum weighted log-likelihood estimation (MWLE) approach to robustly estimate the tail part of finite mixture models (FMM) while preserving the capability of FMM to flexibly capture the complex distributional phenomena from the body part. Asymptotic theories justify the unbiasedness and robustness of the proposed estimator. In computational aspect, the applicability of EM-based algorithm for efficient estimation of parameters makes the proposed MWLE distinctive compared to the existing literature on weighted likelihood approach. Through several simulation studies and real data analyses, we empirically confirm that the proposed MWLE approach is more appropriate in specifying the tail part of the distribution compared to MLE, and at the same time it still preserves the flexibility of FMM in fitting the smaller observations.

Another advantage of the MWLE not yet mentioned throughout this paper is its extensibility. First, it is obvious that the proposed MWLE is not restricted to FMM but it is also applicable to any continuous or discrete distributions. Second, MWLE can be easily extended to regression settings, which is crucial for insurance pricing perspective as insurance companies often determine different premiums across policyholders based on individual attributes (e.g. age, geographical location and past claim history). In regression settings, we define $\bm{x}=(\bm{x}_1,\ldots,\bm{x}_n)$ as the covariates vectors for each of the $n$ observations. Then, the weighted log-likelihood function in Equation (\ref{eq:loglik_weight}) is then re-expressed as
\begin{equation} \label{eq:loglik_weight_reg}
\mathcal{L}^*_n(\bm{\Phi};\bm{y},\bm{x})=\sum_{i=1}^{n}W(y_i)\log \frac{h(y_i;\bm{\Phi},\bm{x}_i)W(y_i)}{\int_{0}^{\infty}h(u;\bm{\Phi},\bm{x}_i)W(u)du}
\end{equation}
for some regression models with density function $h(y_i;\bm{\Phi},\bm{x}_i)$. Obviously, the asymptotic properties still hold subject further to some regularity conditions on covariates $\bm{x}_i$. For parameter estimations using the GEM algorithm, only the hypothetical data approach (Method 1, which converge slower than Method 2 in Section \ref{sec:ex}) works, because the transformed mixing probabilities in Equation (\ref{eq:em:pi_trans}) under Method 2 are assumed to be homogeneous across all observations. We leave all theoretical details with more empirical studies and applications to the future research direction.


\bibliographystyle{abbrvnat}
\bibliography{reference}

\end{document}


\title{Supplementary materials for ``Maximum weighted likelihood estimator for robust heavy-tail modelling of finite mixture models"}

\author{Tsz Chai Fung}

\maketitle

\section{Regularity conditions for asymptotic theory} \label{apx:asym_reg}
Let $h(y;\bm{\Phi})$ be the density function of $Y$ with parameter space of $\bm{\Phi}\in\bm{\Omega}$. For a more concise presentation on the regularity conditions, we here write $\bm{\Phi}=(\psi_1,\ldots,\psi_P)$ where $P$ is the total number of parameters in the model. The regularity conditions are:

\begin{enumerate}[font={\bfseries},label={R\arabic*.}]
\item $h(y;\bm{\Phi})$ has common support in $y$ for all $\bm{\Phi}\in\bm{\Omega}$, $h(y;\bm{\Phi})$ is identifiable in $\bm{\Phi}$ up to a permutation of mixture components.
\item $h(y;\bm{\Phi})$ admits third partial derivatives with respect to $\bm{\Phi}$ for each $\bm{\Phi}\in\bm{\Omega}$ and for almost all $y$.
\item For all $j_1,j_2=1,\ldots,P$, the first two derivatives of $h(y;\bm{\Phi})$ satisfy
\begin{equation}
E\left[\frac{\partial}{\partial\psi_{j_1}}\log h(y;\bm{\Phi})\right]=0;
\end{equation}
\begin{equation}
E\left[\frac{\partial}{\partial\psi_{j_1}}\log h(y;\bm{\Phi})\frac{\partial}{\partial\psi_{j_2}}\log h(y;\bm{\Phi})\right]=E\left[-\frac{\partial^2}{\partial\psi_{j_1}\partial\psi_{j_2}}\log h(y;\bm{\Phi})\right].
\end{equation}
\item The Fisher information matrix is finite and positive definite at $\bm{\Phi}=\bm{\Phi}_0$:
\begin{equation}
\mathcal{I}(\bm{\Phi})=E\left[\left(\frac{\partial}{\partial\bm{\Phi}}\log h(y;\bm{\Phi})\right)\left(\frac{\partial}{\partial\bm{\Phi}}\log h(y;\bm{\Phi})\right)^T\right].
\end{equation}
\item There exists an integrable function $\mathcal{M}(y)$ such that
\begin{equation}
\hspace{-1cm}
\left|\frac{\partial}{\partial\psi_{j_1}}\log h(y;\bm{\Phi})\right|\leq \mathcal{M}(y),\quad
\left|\frac{\partial^2}{\partial\psi_{j_1}\partial\psi_{j_2}}\log h(y;\bm{\Phi})\right|\leq \mathcal{M}(y),\quad
\left|\frac{\partial^3}{\partial\psi_{j_1}\partial\psi_{j_2}\partial\psi_{j_3}}\log h(y;\bm{\Phi})\right|\leq \mathcal{M}(y),
\end{equation}
\begin{equation}
\hspace{-1cm}
\left|\frac{\partial}{\partial\psi_{j_1}}\log h(y;\bm{\Phi})\frac{\partial}{\partial\psi_{j_2}}\log h(y;\bm{\Phi})\right|\leq \mathcal{M}(y),\quad
\left|\frac{\partial^2}{\partial\psi_{j_1}\partial\psi_{j_2}}\log h(y;\bm{\Phi})\frac{\partial}{\partial\psi_{j_3}}\log h(y;\bm{\Phi})\right|\leq \mathcal{M}(y),
\end{equation}
\begin{equation}
\hspace{-1cm}
\left|\frac{\partial}{\partial\psi_{j_1}}\log h(y;\bm{\Phi})\frac{\partial}{\partial\psi_{j_2}}\log h(y;\bm{\Phi})\frac{\partial}{\partial\psi_{j_3}}\log h(y;\bm{\Phi})\right|\leq \mathcal{M}(y).
\end{equation}
\end{enumerate}

\section{Proof of Theorems 1 and 2} \label{apx:asym_proof1}
We first focus on Theorem 1. Denote the weighted log-likelihood of a single observation
\begin{equation}
\mathcal{L}^{*}(\bm{\Phi};y)=W(y)\log \frac{h(y;\bm{\Phi})W(y)}{\int_{0}^{\infty}h(u;\bm{\Phi})W(u)du}.
\end{equation}

The consistency and asymptotic normality can be proved by applying Theorems 5.41 and 5.42 of \cite{van2000asymptotic}. The theorems require the regularity conditions that $E\left[\|\partial/\partial\bm{\Phi}\mathcal{L}^{*}(\bm{\Phi};Y)\|^2\right]<\infty$, the matrix $E\left[\partial^2/\partial\bm{\Phi}\partial\bm{\Phi}^T\mathcal{L}^{*}(\bm{\Phi};Y)\right]$ exists and that $|\partial^3/\partial\psi_{j_1}\partial\psi_{j_2}\partial\psi_{j_3}\mathcal{L}^{*}(\bm{\Phi};y)|$ is dominated by a fixed integrable function of $y$, $j_1,j_2,j_3=1,\ldots,P$ and $\psi_j$ is the $j^{\text{th}}$ element of $\bm{\Phi}$. Through a direct computation of differentiations, the aforementioned equations can all be expressed as functions of $\kappa(u;\bm{\Phi})$ and $\int_{0}^{\infty}\kappa(u;\bm{\Phi})h(u;\bm{\Phi})W(u)du$ only, where $\kappa(\bm{\Phi})$ can be the six terms presented in regularity condition \textbf{R5} (the left hand side of the six equations underneath \textbf{R5} without the absolute sign). Given \textbf{R5} that $\kappa(u;\bm{\Phi})$ is bounded by an integrable function and since $|\int_{0}^{\infty}\kappa(u;\bm{\Phi})h(u;\bm{\Phi})W(u)du|\leq \int_{0}^{\infty}|\kappa(u;\bm{\Phi})|h(u;\bm{\Phi})du$, the aforementioned regularity conditions required by \cite{van2000asymptotic} hold.

\medskip

For consistency, it suffices from Theorem 5.42 of \cite{van2000asymptotic} to show that $\bm{\Phi}_0$ is the maximizer of
\begin{align}
E_{\bm{\Phi}_0}\left[\mathcal{L}^{*}(\bm{\Phi};Y)\right]
&=\int_{0}^{\infty}W(y)\log \frac{h(y;\bm{\Phi})W(y)}{\int_{0}^{\infty}h(u;\bm{\Phi})W(u)du}h(y;\bm{\Phi}_0)dy\nonumber\\
&=c_1\int_{0}^{\infty}\tilde{h}(y;\bm{\Phi}_0)\log\frac{\tilde{h}(y;\bm{\Phi})}{\tilde{h}(y;\bm{\Phi}_0)}dy+c_2\nonumber\\
&=-c_1D_{\text{KL}}\left(\tilde{h}(y;\bm{\Phi})\|\tilde{h}(y;\bm{\Phi}_0)\right)+c_2,
\end{align}
where $c_1= \int_0^\infty h(y; \bm{\Phi}_0)W(y) dy >0$ and $c_2=c_1 \int_0^\infty \tilde{h}(y;\bm{\Phi}_0) \log \tilde{h}(y;\bm{\Phi}_0) dy$ are constants and $D_{\text{KL}}(Q_1\|Q_2)\geq 0$ is the KL divergence between $Q_1$ and $Q_2$. Since $D_{\text{KL}}\left(\tilde{h}(y;\bm{\Phi})\|\tilde{h}(y;\bm{\Phi}_0)\right)=0$ as $\bm{\Phi}=\bm{\Phi}_0$, the result follows.

\medskip

For asymptotic normality, from Theorem 5.41 of \cite{van2000asymptotic}, we have $\sqrt{n}(\hat{\bm{\Phi}}_n-\bm{\Phi}_0)\overset{d}{\rightarrow}\mathcal{N}(\bm{0},\bm{\Sigma})$ with 
$\bm{\Sigma}=\bm{\Gamma}^{-1}\bm{\Lambda}\bm{\Gamma}^{-1}$, where 
\begin{equation} \label{eq:asym:lambda_proof}
\bm{\Lambda}=E_{\bm{\Phi}_0}\left[\left[\frac{\partial}{\partial\bm{\Phi}}\mathcal{L}^{*}(\bm{\Phi};Y)\right]\left[\frac{\partial}{\partial\bm{\Phi}}\mathcal{L}^{*}(\bm{\Phi};Y)\right]^T\Bigg|_{\bm{\Phi}=\bm{\Phi}_0}\right]
\end{equation}
and
\begin{equation} \label{eq:asym:gamma_proof}
\bm{\Gamma}=-E_{\bm{\Phi}_0}\left[\frac{\partial^2}{\partial\bm{\Phi}\partial\bm{\Phi}^T}\mathcal{L}^{*}(\bm{\Phi};Y)\Bigg|_{\bm{\Phi}=\bm{\Phi}_0}\right].
\end{equation}

Performing the derivatives and algebra manipulations from Equations (2.3) and (2.4) would result to Equations (4.2) and (4.3) respectively, which prove the asymptotic normality result.

\medskip

Proof idea of Theorem 2 is completely identical as the above, except that the expectations in Equations (2.3) and (2.4) are taken as $\tilde{E}[\cdot]$ instead of $E_{\bm{\Phi}_0}[\cdot]$.

\section{Proof of Theorem 3} \label{apx:asym_proof2}
We begin with the following lemmas:
\begin{lemma} \label{apx:lem:asym1}
To prove Theorem 3, it suffices to show that
\begin{equation}
T_n(\bm{\Phi}):=\frac{\partial}{\partial\gamma}\tilde{E}_n[\log \tilde{h}_{n}(Y;\bm{\Phi})]
\end{equation}
is asymptotically a strictly decreasing function of $\gamma$ as $n\rightarrow\infty$, with $T_n(\bm{\Phi})|_{\gamma=\gamma_0}\rightarrow 0$ as $n\rightarrow\infty$.
\end{lemma}

\begin{proof}
If we keep the weight function $W(\cdot)$ fixed (independent of $n$), applying Theorem 5.7 of \cite{van2000asymptotic} we have that maximizing the weighted log-likelihood function $\mathcal{L}_n^{*}(\bm{\Phi};\bm{y})$ is asymptotically equivalent to maximizing $\tilde{E}_n[\log \tilde{h}_{n}(Y;\bm{\Phi})]$ (which is indeed independent of $n$).

Now that the weight function $W_n(\cdot)$ depends on $n$, and as $n$ increases, the increasing distortion (more down-weighting) of the relative importance of observations would cause reduction of the effective number of observations. Heuristically, we need the number of observations $n$ to increase faster than the distortion impacts of $W_n(\cdot)$, so that effective number of observations grows to infinity and large sample theory still applies. Quantitatively, we require that the variance of (scaled) empirical weighted log-likelihood
\begin{equation}
V_n(\bm{\Phi}):=\text{Var}\left(\frac{1}{n\int_{0}^{\infty}W_n(u)g(u)du}\sum_{i=1}^{n}W_n(Y)\log\tilde{h}_{n}(Y;\bm{\Phi})\right)\rightarrow 0
\end{equation}
as $n\rightarrow\infty$, such that the (scaled) empirical weighted log-likelihood function converges to its expectation which is $\tilde{E}_n[\log \tilde{h}_{n}(Y;\bm{\Phi})]$. Now, $V_n(\bm{\Phi})$ is evaluated as follows:
\begin{align}
V_n(\bm{\Phi})
&=\frac{1}{n(\int_{0}^{\infty}W_n(u)g(u)du)^2}\text{Var}\left(W_n(Y)\log\tilde{h}_{n}(Y;\bm{\Phi})\right)\nonumber\\
&\leq\frac{1}{n(\int_{0}^{\infty}W_n(u)g(u)du)^2}\tilde{E}\left[W_n(Y)(\log\tilde{h}_{n}(Y;\bm{\Phi}))^2\right]\nonumber\\
&=\frac{1}{n\int_{0}^{\infty}W_n(u)g(u)du}\int_{0}^{\infty}\frac{W_n(y)g(y)}{\int_{0}^{\infty}W_n(u)g(u)du}(\log\tilde{h}_{n}(y;\bm{\Phi}))^2dy\nonumber\\
&=\frac{\tilde{E}_n[(\log \tilde{h}_{n}(Y;\bm{\Phi}))^2]}{n\tilde{E}[W_n(Y)]}\rightarrow 0,
\end{align}
where the convergence is based on Assumption \textbf{A4}.
\end{proof}

\begin{lemma} \label{apx:lem:asym2}
(Monotone density theorem -- Theorem 1.7.2 of \cite{bingham1989regular}) Denote $H$ as a probability distribution function with $h$ being the corresponding probability density function. Assume $h$ is ultimately monotone (i.e. $h$ is monotone on $(z,\infty)$ for some $z>0$). If
\begin{equation}
\bar{H}(y)\sim y^{-\gamma}L(y)
\end{equation}
as $y\rightarrow\infty$ for some $\gamma>0$ and slowly varying functions $L$, then
\begin{equation}
h(y)\sim \gamma y^{-\gamma-1}L(y)
\end{equation}
as $y\rightarrow\infty$.
\end{lemma}

We proceed to the proof of Theorem 3 as follows. Using the result from Lemma \ref{apx:lem:asym1}, it suffices to evaluate
\begin{align} \label{apx:eq:proof2:Tn}
T_n(\bm{\Phi})
&=\frac{\partial}{\partial\gamma}\int_{0}^{\infty}\tilde{g}_n(y)\log \tilde{h}_{n}(y;\bm{\Phi})dy\nonumber\\
&=\frac{\partial}{\partial\gamma}\int_{\tau_n}^{\infty}\tilde{g}^{*}_n(y)\log\tilde{h}^{*}_{n}(y;\bm{\Phi})dy + o(1)\nonumber\\
&=\frac{\partial}{\partial\gamma}\log\tilde{h}^{*}_{n}(\tau_n;\bm{\Phi})
+\frac{\partial}{\partial\gamma}\int_{\tau_n}^{\infty}\left[\frac{\partial}{\partial y}\log\tilde{h}^{*}_{n}(y;\bm{\Phi})\right]\times\bar{\tilde{G}}^{*}_{n}(y)dy + o(1)\nonumber\\
&:=M_1(\tau_n;\bm{\Phi})+M_2(\tau_n;\bm{\Phi}) + o(1),
\end{align}
where
\begin{equation}
\tilde{g}^{*}_{n}(y)=\frac{g(y)W_n(y)}{\int_{\tau_n}^{\infty}g(u)W_n(u)du}1\{y\geq\tau_n\},\qquad \tilde{h}^{*}_{n}(y;\bm{\Phi})=\frac{h(y;\bm{\Phi})W_n(y)}{\int_{\tau_n}^{\infty}h(u;\bm{\Phi})W_n(u)du}1\{y\geq\tau_n\},
\end{equation}
are the proper transformed density functions, and $\tilde{G}^{*}_{n}$ and $\tilde{H}_{n}^{*}$ are the corresponding distribution functions. The second equality of Equation (\ref{apx:eq:proof2:Tn}) is resulted from Assumption \textbf{A3}, while the third equality is followed by integration by parts. Now, we evaluate $M_1(\tau_n;\bm{\Phi})$ and $M_2(\tau_n;\bm{\Phi})$ as follows:

\begin{align}
M_1(\tau_n;\bm{\Phi})
&=\frac{\partial}{\partial\gamma}\log\tilde{h}^{*}_{n}(\tau_n;\bm{\Phi})\nonumber\\
&=\frac{\partial}{\partial\gamma}\left[\log\gamma-(\gamma+1)\log\tau_n+\log L(\tau_n;\bm{\Phi})\right]\nonumber\\
&\hspace{3em}-\int_{\tau_n}^{\infty}\frac{\partial}{\partial\gamma}\left[\log\gamma-(\gamma+1)\log y+\log L(y;\bm{\Phi})\right]\tilde{h}^{*}_{n}(y;\bm{\Phi})dy+o(1)\nonumber\\
&=\frac{1}{\gamma}-\log\tau_n-\frac{1}{\gamma}+\int_{\tau_n}^{\infty}(\log y)\times\tilde{h}^{*}_{n}(y;\bm{\Phi})dy
-\frac{\partial}{\partial\gamma}\int_{\tau_n}^{\infty}\log\frac{L(y;\bm{\Phi})}{L(\tau_n;\bm{\Phi})}\times\tilde{h}^{*}_{n}(y;\bm{\Phi})dy+o(1)\nonumber\\
&=-\log\tau_n+\log\tau_n
+\int_{\tau_n}^{\infty}\frac{1}{y}\bar{\tilde{H}}^{*}_{n}(y;\bm{\Phi})dy
-\frac{\partial}{\partial\gamma}\int_{1}^{\infty}\log\frac{L(\tau_nt;\bm{\Phi})}{L(\tau_n;\bm{\Phi})}\times\tau_n\tilde{h}^{*}_{n}(\tau_nt;\bm{\Phi})dt+o(1)\nonumber\\
&=\int_{\tau_n}^{\infty}\frac{1}{y}\bar{\tilde{H}}^{*}_{n}(y;\bm{\Phi})dy + o(1),
\end{align}
where dominated convergence theorem and integration by parts are repeatedly used. The second equality involves monotone density theorem (Lemma \ref{apx:lem:asym2}) with Assumption \textbf{A5} being satisfied. The last term of the second last equality converges to zero uniformly on $\bm{\Phi}$ due to dominated convergence theorem and the uniform convergence conditions in Assumption \textbf{A2}. Using similar techniques as the above, $M_2(\tau_n;\bm{\Phi})$ can be evaluated as
\begin{align}
M_2(\tau_n;\bm{\Phi})
&=-\int_{\tau_n}^{\infty}\frac{1}{y}\bar{\tilde{G}}^{*}_{n}(y)dy
+\frac{\partial}{\partial\gamma}\int_{\tau_n}^{\infty}\frac{\partial}{\partial y}(\log L(y;\bm{\Phi}))\times\bar{\tilde{G}}^{*}_{n}(y)dy\nonumber\\
&=-\int_{\tau_n}^{\infty}\frac{1}{y}\bar{\tilde{G}}^{*}_{n}(y)dy-\frac{\partial}{\partial\gamma}\int_{\tau_n}^{\infty}\log\frac{L(y;\bm{\Phi})}{L(\tau_n;\bm{\Phi})}\times\tilde{g}^{*}_{n}(y)dy\nonumber\\
&=-\int_{\tau_n}^{\infty}\frac{1}{y}\bar{\tilde{G}}^{*}_{n}(y)dy + o(1).
\end{align}

To sum up, we have
\begin{equation}
T_n(\bm{\Phi})
=\int_{\tau_n}^{\infty}\frac{1}{y}\left[\bar{\tilde{H}}^{*}_{n}(y;\bm{\Phi})-\bar{\tilde{G}}^{*}_{n}(y)\right]dy
=\int_{1}^{\infty}\frac{1}{t}\left[\bar{\tilde{H}}^{*}_{n}(\tau_nt;\bm{\Phi})-\bar{\tilde{G}}^{*}_{n}(\tau_nt)\right]dt.
\end{equation}

Investigating each term inside the integrand, we have
\begin{align}
\bar{\tilde{H}}^{*}_{n}(\tau_nt;\bm{\Phi})
&=\frac{\int_t^{\infty}h(\tau_nv;\bm{\Phi})W_n(\tau_nv)dv}{\int_1^{\infty}h(\tau_nv;\bm{\Phi})W_n(\tau_nv)dv}\nonumber\\
&=\frac{\int_t^{\infty}v^{-\gamma-1}\tilde{W}_n(v)[L(\tau_nv;\bm{\Phi})/L(\tau_n;\bm{\Phi})]dv}{\int_1^{\infty}v^{-\gamma-1}\tilde{W}_n(v)[L(\tau_nv;\bm{\Phi})/L(\tau_n;\bm{\Phi})]dv} + o(1)\nonumber\\
&=\frac{\int_t^{\infty}v^{-\gamma-1}\tilde{W}_n(v)dv}{\int_1^{\infty}v^{-\gamma-1}\tilde{W}_n(v)dv} + o(1),
\end{align}
and
\begin{align}
\bar{\tilde{G}}^{*}_{n}(\tau_nt)
&=\frac{\int_t^{\infty}g(\tau_nv)W_n(\tau_nv)dv}{\int_1^{\infty}g(\tau_nv)W_n(\tau_nv)dv}\nonumber\\
&=\frac{\int_t^{\infty}v^{-\gamma_0-1}\tilde{W}_n(v)[L_0(\tau_nv)/L_0(\tau_n)]dv}{\int_1^{\infty}v^{-\gamma_0-1}\tilde{W}_n(v)[L_0(\tau_nv)/L_0(\tau_n)]dv} + o(1)\nonumber\\
&=\frac{\int_t^{\infty}v^{-\gamma_0-1}\tilde{W}_n(v)dv}{\int_1^{\infty}v^{-\gamma_0-1}\tilde{W}_n(v)dv} + o(1),
\end{align}
where $\tilde{W}_n(v)=W_n(\tau_nv)$. Therefore, it is clear that
\begin{equation}
T_n(\bm{\Phi})=\int_{1}^{\infty}\frac{1}{t}\left[\frac{\int_t^{\infty}v^{-\gamma-1}\tilde{W}_n(v)dv}{\int_1^{\infty}v^{-\gamma-1}\tilde{W}_n(v)dv}-\frac{\int_t^{\infty}v^{-\gamma_0-1}\tilde{W}_n(v)dv}{\int_1^{\infty}v^{-\gamma_0-1}\tilde{W}_n(v)dv}\right]dt+o(1)
\end{equation}
converges to zero for $\gamma=\gamma_0$ as $n\rightarrow\infty$. To show that $T_n(\bm{\Phi})$ is a strictly decreasing function of $\gamma$ as $n\rightarrow\infty$, it suffices to evaluate
\begin{align}
\frac{\partial}{\partial\gamma}\log\frac{\int_t^{\infty}v^{-\gamma-1}\tilde{W}_n(v)dv}{\int_1^{\infty}v^{-\gamma-1}\tilde{W}_n(v)dv}
&=-\frac{\int_t^{\infty}(\log v)v^{-\gamma-1}\tilde{W}_n(v)dv}{\int_t^{\infty}v^{-\gamma-1}\tilde{W}_n(v)dv}+\frac{\int_1^{\infty}(\log v)v^{-\gamma-1}\tilde{W}_n(v)dv}{\int_1^{\infty}v^{-\gamma-1}\tilde{W}_n(v)dv},
\end{align}
which is negative if and only if
\begin{equation}
\int_{1}^{t}(\log v)k_{n,1,t}(v;\gamma)dv<\int_{t}^{\infty}(\log v)k_{n,t,\infty}(v;\gamma)dv,
\end{equation}
where
\begin{equation}
k_{n,t_1,t_2}(v;\gamma)=\frac{v^{-\gamma-1}\tilde{W}_n(v)}{\int_{t_1}^{t_2}v^{-\gamma-1}\tilde{W}_n(v)dv}1\{t_1<v\leq t_2\}
\end{equation}
is a proper probability density function with $1\leq t_1<t_2\leq\infty$. Since $k_{n,1,t}(v;\gamma)$ and $k_{n,t,\infty}(v;\gamma)$ are both proper densities, it is clear that $\int_{1}^{t}(\log v)k_{n,1,t}(v;\gamma)dv=\log v_1$ for some $v_1\in (1,t)$ and $\int_{t}^{\infty}(\log v)k_{n,t,\infty}(v;\gamma)dv=\log v_2$ for some $v_2\in (t,\infty]$. The result then follows.

\section{GEM algorithm for MWLE under J-Gamma Lomax mixture model: Hypothetical data approach} 

\subsection{Construction of complete data}
The complete data is given by
\begin{equation}
\mathcal{D}^{\text{com}}=\{(y_i,\bm{z}_i,k_i,\{\bm{z}'_{is},y'_{is}\}_{s=1,\ldots,k_i})\}_{i=1,\ldots,n},
\end{equation}
where $\bm{z}_i=(z_{i1},\ldots,z_{i(J+1)})$ with $z_{ij}=1$ if observation $i$ belongs to the $j^{\text{th}}$ latent class and $z_{ij}=0$ otherwise. Similarly, $\bm{z}'_i=(z'_{is1},\ldots,z'_{is(J+1)})$ and $z'_{isj}=1$ if the $s^{\text{th}}$ missing sample generated by observation $i$ belongs to the $j^{\text{th}}$ latent class, and $z'_{isj}=0$ otherwise. The complete data weighted log-likelihood function is given by

\begin{align}
\tilde{\mathcal{L}}^{*}(\bm{\Phi};\mathcal{D}^{\text{com}})
&=\sum_{i=1}^nW(y_i)\left\{\left[\sum_{j=1}^{J}z_{ij}\log \pi_jf_b(y_i;\mu_j,\phi_j)\right]+z_{i(J+1)}\log\pi_{J+1} f_t(y_i;\theta,\gamma)\right\}\nonumber\\
&\quad +\sum_{i=1}^{n}\sum_{s=1}^{k_i}W(y_i)\left\{\left[\sum_{j=1}^J z'_{ijs}\log\pi_j f_b(y'_{is};\mu_j,\phi_j)\right]+z'_{i(J+1)s}\log\pi_{J+1} f_t(y'_{is};\theta,\gamma)\right\}.
\end{align}

\subsection{E-step} \label{supp:sec:em_e}
The expectation of the complete data weighted log-likelihood is given by the following for the $l^{\text{th}}$ iteration:
\begin{align}
&Q^{*}(\bm{\Phi}|\bm{\Phi}^{(l-1)})\nonumber\\
&=\sum_{i=1}^nW(y_i)\Bigg\{\sum_{j=1}^{J}z^{(l)}_{ij}\left\{\log\pi_j-\frac{1}{\phi_j}\log\phi_j-\frac{1}{\phi_j}\log\mu_j-\log\Gamma(\frac{1}{\phi_j})+(\frac{1}{\phi_j}-1)\log y_i-\frac{y_i}{\phi_j\mu_j}\right\}\nonumber\\
&\hspace{8em}+ z^{(l)}_{i(J+1)}\left\{\log\pi_{J+1}+\log\gamma+\gamma\log\theta-(\gamma+1)\log(y_i+\theta)\right\}\Bigg\}\nonumber\\
&\quad +k^{(l)}\left(\sum_{i=1}^{n}W(y_i)\right)\Bigg\{\sum_{j=1}^J z^{'(l)}_{j}\left\{\log\pi_j-\frac{1}{\phi_j}\log\phi_j-\frac{1}{\phi_j}\log\mu_j-\log\Gamma(\frac{1}{\phi_j})+(\frac{1}{\phi_j}-1)\widehat{\log y'}^{(l)}_j-\frac{\widehat{y'}^{(l)}_j}{\phi_j\mu_j}\right\}\nonumber\\
&\hspace{12em}+z^{'(l)}_{(J+1)}\left\{\log\pi_{J+1}+\log\gamma+\gamma\log\theta-(\gamma+1)\widehat{\log(y'+\theta)}^{(l)}_{J+1}\right\}\Bigg\},
\end{align}
where
\begin{equation}
z^{(l)}_{ij}=P(z_{ij}=1|\bm{y},\bm{\Phi}^{(l-1)})=
\begin{cases}
\dfrac{\pi_j^{(l-1)}f_b(y_i;\mu_j^{(l-1)},\phi_j^{(l-1)})}{h(u;\bm{\Phi}^{(l-1)})},\quad j=1,\ldots,J\\
\dfrac{\pi^{(l-1)}_{J+1}f_t(y_i;\theta,\gamma^{(l-1)})}{h(u;\bm{\Phi}^{(l-1)})},\quad j=J+1,
\end{cases}
\end{equation}
\begin{equation}
k^{(l)}=E(k_i|\bm{y},\bm{\Phi}^{(l-1)})=\frac{\int_0^{\infty}f_t(u;\bm{\Phi}^{(l-1)})(1-W(u))du}{\int_0^{\infty}h(u;\bm{\Phi}^{(l-1)})W(u)du},
\end{equation}
\begin{equation}
z_{j}^{'(l)}=P(z'_{ijs}=1|\bm{y},\bm{\Phi}^{(l-1)})
\begin{cases}
\dfrac{\pi_j^{(l-1)}\int_0^{\infty}f_b(u;\mu_j^{(l-1)},\phi_j^{(l-1)})(1-W(u))du}{\int_0^{\infty}h(u;\bm{\Phi}^{(l-1)})(1-W(u))du},\quad j=1,\ldots,J\\
\dfrac{\pi_j^{(l-1)}\int_0^{\infty}f_t(u;\theta,\gamma^{(l-1)})(1-W(u))du}{\int_0^{\infty}h(u;\bm{\Phi}^{(l-1)})(1-W(u))du}, \quad j=J+1,
\end{cases}
\end{equation}
\begin{equation}
\widehat{y'}^{(l)}_j=E(y_{is}'|\bm{y},\bm{\Phi}^{(l-1)},z'_{ijs}=1)
=\frac{\int_0^{\infty}uf_b(u;\mu_j^{(l-1)},\phi_j^{(l-1)})(1-W(u))du}{\int_0^{\infty}f_b(u;\mu_j^{(l-1)},\phi_j^{(l-1)})(1-W(u))du}, \quad j=1,\ldots,J,
\end{equation}
\begin{equation}
\widehat{\log y'}^{(l)}_j=E(\log y_{is}'|\bm{y},\bm{\Phi}^{(l-1)},z'_{ijs}=1)
=\frac{\int_0^{\infty}\log uf_b(u;\mu_j^{(l-1)},\phi_j^{(l-1)})(1-W(u))du}{\int_0^{\infty}f_b(u;\mu_j^{(l-1)},\phi_j^{(l-1)})(1-W(u))du}, \quad j=1,\ldots,J,
\end{equation}
\begin{equation}
\widehat{\log(y'+\theta)}^{(l)}_{J+1}=E(\log (y_{is}'+\theta)|\bm{y},\bm{\Phi}^{(l-1)},z'_{i(J+1)s}=1)
=\frac{\int_0^{\infty}\log (u+\theta) f_t(u;\theta,\gamma^{(l-1)})(1-W(u))du}{\int_0^{\infty}f_t(u;\theta,\gamma^{(l-1)})(1-W(u))du}.
\end{equation}

Note that the above integrals all have analytical form of solutions under the example choice of the following weight functions (for the generalized weight function form as presented in the main paper, we do a numerical integration instead):
\begin{itemize}
    \item Case 1: Exponential distribution with $W(y;\tilde{\mu})=1-\exp\{-y/\tilde{\mu}\}$;
    \item Case 2: Two-point discrete distribution with $W(y;\tilde{\mu},\tilde{\phi})=(1-\tilde{\phi})1\{y>\tilde{\mu}\}+\tilde{\phi}$.
\end{itemize}

Re-parameterize the gamma distribution with $\alpha=1/\phi_j$ and $\beta=1/\phi_j\mu_j$, we are to compute
\begin{equation}
\int_{0}^{\infty}q(u)\frac{\beta^{\alpha}}{\Gamma(\alpha)}u^{\alpha-1}\exp\{-\beta u\}(1-W(u))du
\end{equation}
for $q(u)=1$, $q(u)=u$ and $q(u)=\log u$; and
\begin{equation}
\int_{0}^{\infty}r(u)\frac{\gamma\theta^{\gamma}}{(u+\theta)^{\gamma+1}}(1-W(u))du
\end{equation}
for $r(u)=1$ and $r(u)=\log(u+\theta)$.

\textbf{Case 1}. We have the following analytical results:
\begin{equation}
\int_{0}^{\infty}\frac{\beta^{\alpha}}{\Gamma(\alpha)}u^{\alpha-1}\exp\{-\beta u\}(1-W(u))du=\left(\frac{\beta}{\beta+1/\tilde{\mu}}\right)^{\alpha},
\end{equation}
\begin{equation}
\int_{0}^{\infty}u\frac{\beta^{\alpha}}{\Gamma(\alpha)}u^{\alpha-1}\exp\{-\beta u\}(1-W(u))du=\frac{\alpha\beta^{\alpha}}{(\beta+1/\tilde{\mu})^{\alpha+1}},
\end{equation}
\begin{equation}
\int_{0}^{\infty}\log u\frac{\beta^{\alpha}}{\Gamma(\alpha)}u^{\alpha-1}\exp\{-\beta u\}(1-W(u))du=\frac{\beta^{\alpha}}{\Gamma(\alpha)}\frac{\partial}{\partial\alpha}\frac{\Gamma(\alpha)}{(\beta+1/\tilde{\mu})^{\alpha}},
\end{equation}
\begin{equation}
\int_{0}^{\infty}\frac{\gamma\theta^{\gamma}}{(u+\theta)^{\gamma+1}}(1-W(u))du=\gamma\left(\frac{\theta}{\tilde{\mu}}\right)^{\gamma}\exp\left\{\frac{\theta}{\tilde{\mu}}\right\}\Gamma(-\gamma;\frac{\theta}{\tilde{\mu}},\infty),
\end{equation}
\begin{equation}
\int_{0}^{\infty}\log(u+\theta)\frac{\gamma\theta^{\gamma}}{(u+\theta)^{\gamma+1}}(1-W(u))du=-\gamma\theta^{\gamma}\exp\left\{\frac{\theta}{\tilde{\mu}}\right\}\frac{\partial}{\partial\gamma}\Gamma(-\gamma;\frac{\theta}{\tilde{\mu}},\infty),
\end{equation}
where $\Gamma(m;c_1,c_2)=\int_{c_1}^{c_2}u^{m-1}\exp\{-u\}du$ is an incomplete gamma function.

\textbf{Case 2}. We have the following analytical results:
\begin{equation}
\int_{0}^{\infty}\frac{\beta^{\alpha}}{\Gamma(\alpha)}u^{\alpha-1}\exp\{-\beta u\}(1-W(u))du=\tilde{\phi}\frac{\Gamma(\alpha;\beta\tilde{\mu},\infty)}{\Gamma(\alpha)}+(1-\tilde{\phi}),
\end{equation}
\begin{equation}
\int_{0}^{\infty}u\frac{\beta^{\alpha}}{\Gamma(\alpha)}u^{\alpha-1}\exp\{-\beta u\}(1-W(u))du=\frac{\alpha}{\beta}\left[\tilde{\phi}\Gamma(\alpha+1;\beta\tilde{\mu},\infty)+(1-\tilde{\phi})\right],
\end{equation}
\begin{equation}
\int_{0}^{\infty}\log u\frac{\beta^{\alpha}}{\Gamma(\alpha)}u^{\alpha-1}\exp\{-\beta u\}(1-W(u))du=\frac{\beta^{\alpha}}{\Gamma(\alpha)}\left[\tilde{\phi}\frac{\partial}{\partial\alpha}\frac{\Gamma(\alpha;\beta\tilde{\mu},\infty)}{\beta^{\alpha}}+(1-\tilde{\phi})\frac{\partial}{\partial\alpha}\frac{\Gamma(\alpha)}{\beta^{\alpha}}\right],
\end{equation}
\begin{equation}
\int_{0}^{\infty}\frac{\gamma\theta^{\gamma}}{(u+\theta)^{\gamma+1}}(1-W(u))du=\tilde{\phi}\left(\frac{\theta}{\tilde{\mu}+\theta}\right)^{\gamma}+(1-\tilde{\phi}),
\end{equation}
\begin{equation}
\int_{0}^{\infty}\log(u+\theta)\frac{\gamma\theta^{\gamma}}{(u+\theta)^{\gamma+1}}(1-W(u))du=-\gamma\theta^{\gamma}\frac{\partial}{\partial\gamma}\left[\tilde{\phi}\frac{1}{\gamma(\tilde{\mu}+\theta)^{\gamma}}+(1-\tilde{\phi})\frac{1}{\gamma\theta^{\gamma}}\right].
\end{equation}

\subsection{M-step} \label{supp:sec:em_m}
Maximizing $Q^{*}(\bm{\Phi}|\bm{\Phi}^{(l-1)})$ with respect to $\bm{\Phi}$ yields the following parameter updates:
\begin{equation}
\pi_j^{(l)}=\frac{\sum_{i=1}^{n}W(y_i)z_{ij}^{(l)}+\left(\sum_{i=1}^{n}W(y_i)\right)k^{(l)}z_j^{'(l)}}{\sum_{j'=1}^{J+1}\left\{\sum_{i=1}^{n}W(y_i)z_{ij'}^{(l)}+\left(\sum_{i=1}^{n}W(y_i)\right)k^{(l)}z_{j'}^{'(l)}\right\}},\quad j=1,\ldots,J+1,
\end{equation}
\begin{equation}
\mu_j^{(l)}=\frac{\sum_{i=1}^{n}W(y_i)z_{ij}^{(l)}y_i+\left(\sum_{i=1}^{n}W(y_i)\right)k^{(l)}z_j^{'(l)}\widehat{y'}^{(l)}_j}{\sum_{i=1}^{n}W(y_i)z_{ij}^{(l)}+\left(\sum_{i=1}^{n}W(y_i)\right)k^{(l)}z_j^{'(l)}},\quad j=1,\ldots,J,
\end{equation}
\begin{align}
\phi_j^{(l)}
&=\underset{\phi_j>0}{\text{argmax}}\Bigg\{\sum_{i=1}^nW(y_i)z^{(l)}_{ij}\left\{-\frac{1}{\phi_j}\log\phi_j-\frac{1}{\phi_j}\log\mu_j^{(l)}-\log\Gamma(\frac{1}{\phi_j})+(\frac{1}{\phi_j}-1)\log y_i-\frac{y_i}{\phi_j\mu_j}\right\}\nonumber\\
&\hspace{5em} +k^{(l)}\left(\sum_{i=1}^{n}W(y_i)\right)z^{'(l)}_{j}\left\{-\frac{1}{\phi_j}\log\phi_j-\frac{1}{\phi_j}\log\mu_j^{(l)}-\log\Gamma(\frac{1}{\phi_j})+(\frac{1}{\phi_j}-1)\widehat{\log y'}^{(l)}_j-\frac{\widehat{y'}^{(l)}_j}{\phi_j\mu_j^{(l)}}\right\}\Bigg\},\nonumber\\
\end{align}
\begin{equation}
\gamma^{(l)}=\frac{\sum_{i=1}^{n}W(y_i)z_{i(J+1)}^{(l)}+\left(\sum_{i=1}^{n}W(y_i)\right)k^{(l)}z_{J+1}^{'(l)}}{\sum_{i=1}^{n}W(y_i)z_{i(J+1)}^{(l)}\left[\log(y_i+\theta)-\log\theta\right]+\left(\sum_{i=1}^{n}W(y_i)\right)k^{(l)}z_{J+1}^{'(l)}\left[\widehat{\log(y'+\theta)}^{(l)}_{J+1}-\log\theta\right]}.
\end{equation}

Note here that $\theta$ is treated as a fixed hyperparameter not involved in estimation procedure. To estimate $\theta$ as a parameter, we may need to take a further step to numerically maximize the observed data weighted log-likelihood $\mathcal{L}^{*}_n(\bm{\Phi};\bm{y})$ w.r.t. $\theta$.

\section{GEM algorithm for MWLE under J-Gamma Lomax mixture model: Parameter transformation approach}
\subsection{Construction of complete data}
The complete data is given by
\begin{equation}
\mathcal{D}^{\text{com}}=\{(y_i,\bm{z}_i^{*})\}_{i=1,\ldots,n},
\end{equation}
where $\bm{z}_i^{*}=(z_{i1}^{*},\ldots,z_{i(J+1)}^{*})$ are the labels where $z_{ij}^{*}=1$ if observation $i$ belongs to the $j^{\text{th}}$ (transformed) latent mixture component and $z_{ij}^{*}=0$ otherwise. The complete data weighted log-likelihood function is given by
\begin{align}
\tilde{\mathcal{L}}^{*}_n(\bm{\Phi};\mathcal{D}^{\text{com}})
&=\sum_{i=1}^{n}W(y_i)\Bigg\{\left[\sum_{j=1}^{J}z_{ij}^{*}\left(\log\pi_j^{*}+\log f_b(y_i;\mu_j,\phi_j) -\log\int_0^{\infty}f_b(u;\mu_j,\phi_j)W(u)du\right)\right]\nonumber\\
&\hspace{8em}+z_{i(J+1)}^{*}\left(\log\pi_{J+1}^{*}+\log f_t(y_i;\theta,\gamma)W(y_i)-\log\int_0^{\infty}f_t(u;\theta,\gamma)W(u)du\right)\Bigg\}.
\end{align}

\subsection{E-step} \label{supp:sec:em_e2}
The expectation of the complete data weighted log-likelihood is given by the following for the $l^{\text{th}}$ iteration:
\begin{align}
Q^{*}(\bm{\Phi}|\bm{\Phi}^{(l-1)})
&=\sum_{i=1}^{n}W(y_i)\Bigg\{\Bigg[\sum_{j=1}^{J}z_{ij}^{*(l)}\Bigg(\log\pi_j^{*}-\frac{1}{\phi_j}\log\phi_j-\frac{1}{\phi_j}\log\mu_j-\log\Gamma(\frac{1}{\phi_j})+(\frac{1}{\phi_j}-1)\log y_i-\frac{y_i}{\phi_j\mu_j}\nonumber\\
&\hspace{12em}-\log\int_0^{\infty}f_b(u;\mu_j,\phi_j)W(u)du\Bigg)\Bigg]\nonumber\\
&\hspace{8em}+z_{i(J+1)}^{*(l)}\Bigg(\log\pi_{J+1}^{*}+\log\gamma+\gamma\log\theta-(\gamma+1)\log(y_i+\theta)\nonumber\\
&\hspace{13em}-\log\int_0^{\infty}f_t(u;\theta,\gamma)W(u)du\Bigg)\Bigg\},
\end{align}
where
\begin{equation}
z^{*(l)}_{ij}=P(z^{*}_{ij}=1|\bm{y},\bm{\Phi}^{(l-1)})=
\begin{cases}
\dfrac{\pi_j^{*(l-1)}f_b(y_i;\mu_j^{(l-1)},\phi_j^{(l-1)})W(y_i)}{\int_0^{\infty}f_b(u;\mu_j^{(l-1)},\phi_j^{(l-1)})W(u)du\times h(y_i;\bm{\Phi}^{(l-1)})},\quad j=1,\ldots,J\\
\dfrac{\pi^{*(l-1)}_{J+1}f_t(y_i;\theta,\gamma^{(l-1)})W(y_i)}{\int_0^{\infty}f_t(u;\theta,\gamma^{(l-1)})W(u)du\times h(y_i;\bm{\Phi}^{(l-1)})},\quad j=J+1.
\end{cases}
\end{equation}

\subsection{M-step} \label{supp:sec:em_m2}
Maximizing $Q^{*}(\bm{\Phi}|\bm{\Phi}^{(l-1)})$ with respect to $\bm{\Phi}$ yields the following parameter updates:
\begin{equation}
\pi_j^{*(l)}=\frac{\sum_{i=1}^{n}W(y_i)z_{ij}^{*(l)}}{\sum_{j'=1}^{J+1}\sum_{i=1}^{n}W(y_i)z_{ij'}^{*(l)}},\quad j=1,\ldots,J+1,
\end{equation}
and the other parameters $(\bm{\mu},\bm{\phi},\theta,\gamma)$ are sequentially updated by numerically maximizing $Q^{*}(\bm{\Phi}|\bm{\Phi}^{(l-1)})$ w.r.t. each of the parameters.

\section{Proof of Proposition 3} \label{supp:sec:ascend}
Write $\mathcal{L}^{*}_n(\bm{\Phi};\bm{y})=\sum_{i=1}^{n}W(y_i)\log p(y_i|\bm{\Phi})$ and $\tilde{\mathcal{L}}^{*}_n(\bm{\Phi}^{(l)};\mathcal{D}^{\text{com}})=\sum_{i=1}^{n}W(y_i)\left[\log p(y_i|\bm{\Phi}) +\log p(\mathcal{D}^{\text{mis}}_i|\bm{\Phi},y_i)\right]$ for some probability density $p$ and missing data from sample $i$ given by $\mathcal{D}^{\text{mis}}_i$. Then, we have
\begin{align}
\mathcal{L}^{*}_n(\bm{\Phi};\bm{y})
&=\tilde{\mathcal{L}}^{*}_n(\bm{\Phi}^{(l)};\mathcal{D}^{\text{com}})-\sum_{i=1}^{n}W(y_i)\log p(\mathcal{D}^{\text{mis}}_i|\bm{\Phi},y_i)\nonumber\\
&=Q^{*}(\bm{\Phi}|\bm{\Phi}^{(l-1)})-\sum_{i=1}^{n}W(y_i)\int p(\bm{v}_i|\bm{\Phi}^{(l-1)},y_i)\log p(\bm{v}_i|\bm{\Phi},y_i)d\bm{v}_i,
\end{align}
where the second equality results from expectation of both sides on the missing data under parameters $\bm{\Phi}^{(l-1)}$. Then, we have
\begin{align}
\mathcal{L}^{*}_n(\bm{\Phi}^{(l)};\bm{y})-\mathcal{L}^{*}_n(\bm{\Phi}^{(l-1)};\bm{y})
&=Q^{*}(\bm{\Phi}^{(l)}|\bm{\Phi}^{(l-1)})-Q^{*}(\bm{\Phi}^{(l-1)}|\bm{\Phi}^{(l-1)})\nonumber\\
&\quad+\sum_{i=1}^{n}W(y_i)\int p(\bm{v}_i|\bm{\Phi}^{(l-1)},y_i)\log\frac{p(\bm{v}_i|\bm{\Phi}^{(l-1)},y_i)}{p(\bm{v}_i|\bm{\Phi}^{(l)},y_i)}d\bm{v}_i\geq 0.
\end{align}

\section{Initialization of parameters} \label{apx:em:init}
As briefly described in Section 5.3 of the paper, parameter initialization $\bm{\Phi}^{(0)}$ is done using the CMM approach by \cite{gui2018fit}. This comes with the following steps:
\begin{enumerate}
\item Determine a threshold $\tau$ which classifies observations $y_i$ into either body (when $y_i\leq\tau$) or tail (when $y_i>\tau$) part of distribution. This can be done by plotting the log of empirical data survival function against $\log y_i$, which is called the log-log plot. For regular varying distributions, the log-log plot is asymptotically linear. $\tau$ is approximated by the point where the curve turns linear onwards.
\item Perform K-means clustering on $\{y_i\}_{i:y_i\leq\tau}$ with $J$ clusters, and obtain the clustering mean $\{\mu^{\text{cluster}}_j\}_{j=1,\ldots,J}$, variance $\{(\sigma^{\text{cluster}}_j)^2\}_{j=1,\ldots,J}$ and weights $\{\tilde{\pi}_j^{\text{cluster}}\}_{j=1,\ldots,J}$.
\item Set $\mu_j^{(0)}=\mu^{\text{cluster}}_j$, $\phi_j^{(0)}=({\sigma^{\text{cluster}}_j})^2/{\mu^{\text{cluster}}_j}^2$.
\item Obtain $\theta^{(0)}$ and $\gamma^{(0)}$ by matching the first two moments of observations belonging to the tail component (i.e. $\{y_i\}_{i:y_i>\tau}$).
\item Set $\pi_{J+1}^{(0)}$ as the proportion of observations satisfying $y_i>\tau$.
\item Set the remaining weight parameters as $\pi_{j}^{(0)}=\tilde{\pi}_j^{\text{cluster}}(1-\pi_{J+1}^{(0)})$.
\end{enumerate}

\section{Truncated log-likelihood function} \label{sec:supp:tll}
This section includes more details for Remark 6 in the paper. Denote $g(y)$ as the true distribution generating the observations and $\tilde{h}(y;\bm{\Phi})=\frac{h(y;\bm{\Phi})W(y)}{\int_0^{\infty}h(u;\bm{\Phi})W(u)du}$ as the truncated distribution. The expected weighted log-likelihood can be alternatively written as
\begin{align}
n\times\tilde{E}[\mathcal{L}^{*}(\bm{\Phi};\bm{Y})]
&=n\int_{0}^{\infty}W(u)\log \tilde{h}(u;\bm{\Phi})\times g(u)du\nonumber\\
&=n\int_{0}^{\infty}g(u)W(u)du\times\int_{0}^{\infty}\log \tilde{h}(u;\bm{\Phi})\times\frac{g(u)W(u)}{\int_{0}^{\infty}g(t)W(t)dt}du\nonumber\\
&=n\int_{0}^{\infty}g(u)W(u)du\times\tilde{E}^*[\log \tilde{h}(u;\bm{\Phi})],
\end{align}
where the expectation $\tilde{E}^*$ is taken on $Y$ under the random truncated distribution $\frac{g(u)W(u)}{\int_{0}^{\infty}g(t)W(t)dt}$. Next, denote a random set $S_n=\{i:V_i(y_i)=1\}$, such that $\mathcal{L}^{**}_n(\bm{\Phi};\bm{y})$ can be written as
\begin{equation}
\mathcal{L}^{**}_n(\bm{\Phi};\bm{y})=\sum_{i\in S_n}\log \tilde{h}(u;\bm{\Phi}),
\end{equation}
with effective number of terms $\|S_n\|\approx n\int_{0}^{\infty}g(u)W(u)du\approx \sum_{i=1}^{n}W(y_i)$ in probability as $n\rightarrow\infty$. Comparing the above two equations, they simply correspond to standard MLE with bias term of $P$.

\section{Preliminary analysis of the motivating Greek dataset} \label{apx:prelim_data}
\begin{figure}[!h]
\begin{center}
\begin{subfigure}[h]{0.49\linewidth}
\includegraphics[width=\linewidth]{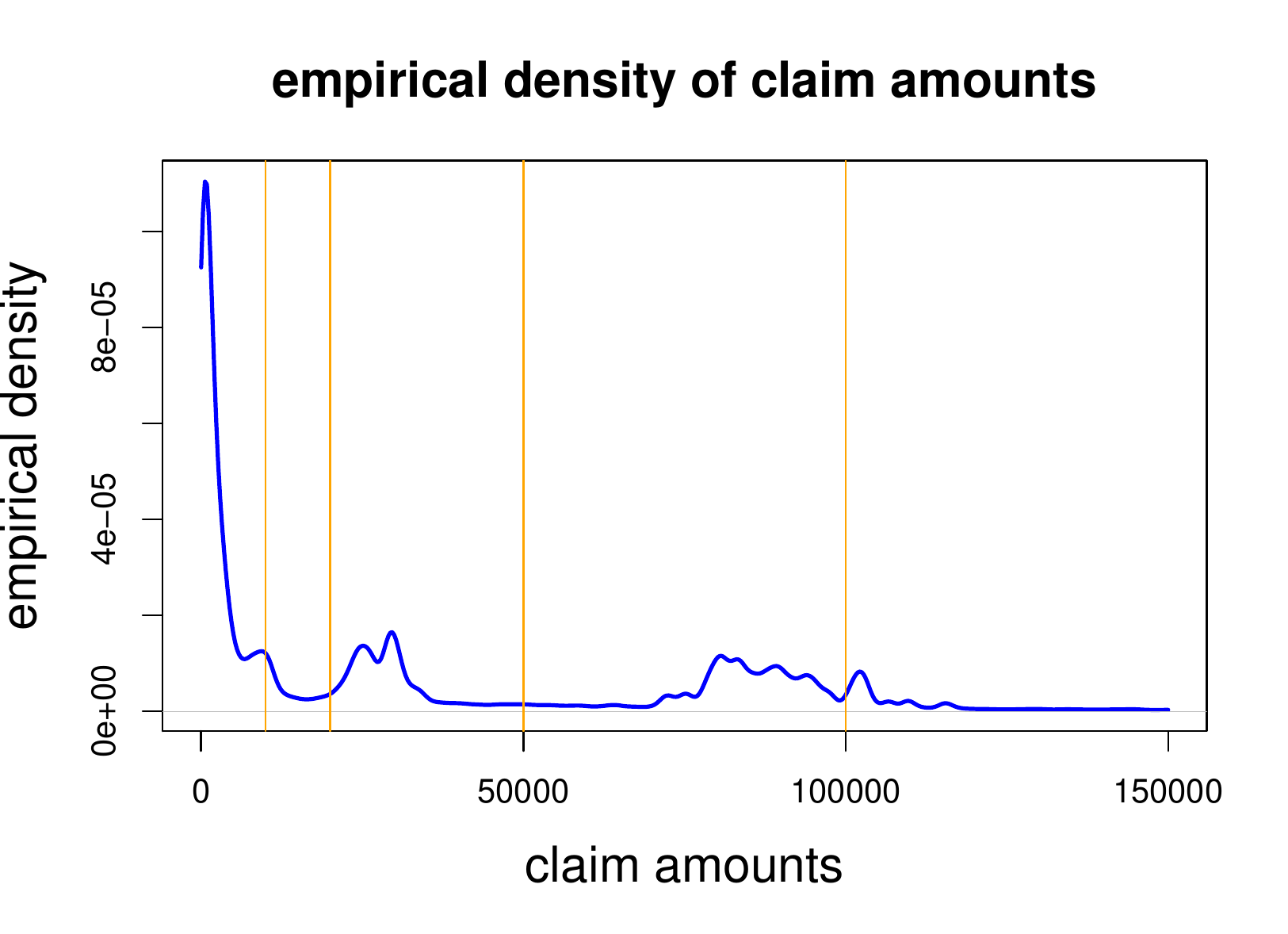}
\end{subfigure}
\hfill
\begin{subfigure}[h]{0.49\linewidth}
\includegraphics[width=\linewidth]{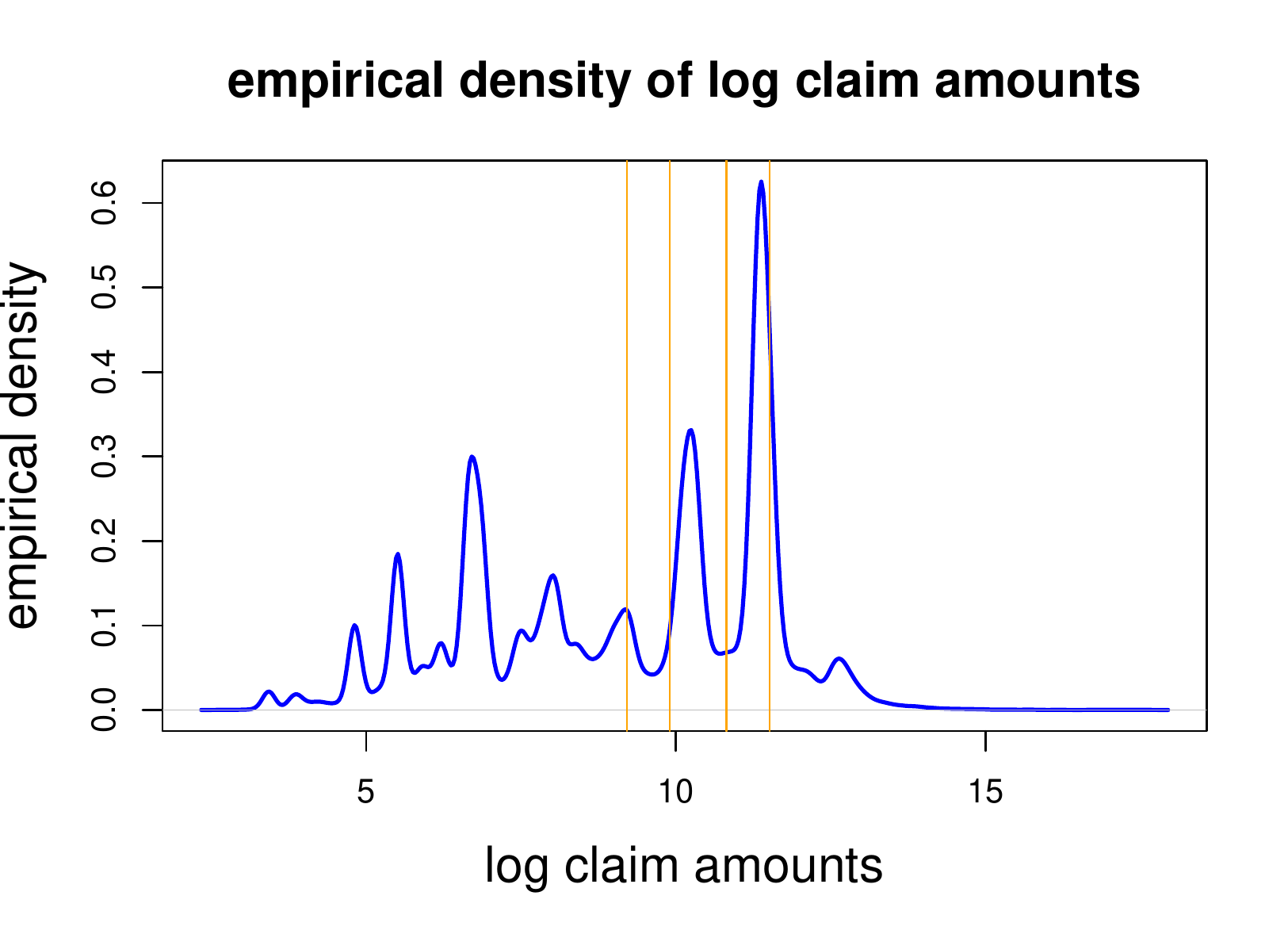}
\end{subfigure}
\end{center}
\caption{Empirical density of claim amounts (left panel) and log claim amounts (right panel); the orange vertical lines represent amounts of 10,000, 20,000, 50,000 and 100,000 respectively.}
\label{fig:density}
\end{figure}

\begin{figure}[!h]
\begin{center}
\begin{subfigure}[h]{0.49\linewidth}
\includegraphics[width=\linewidth]{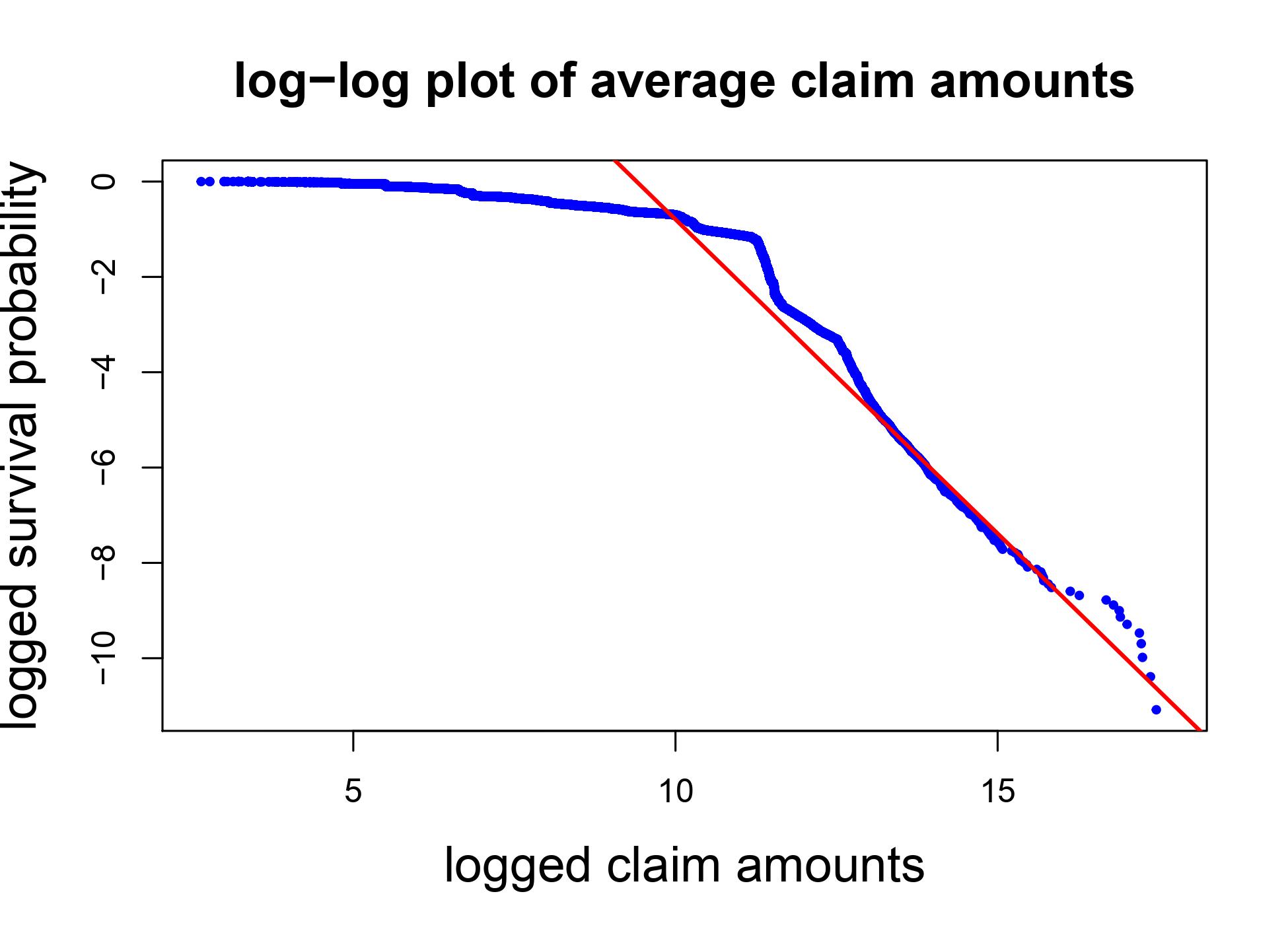}
\end{subfigure}
\begin{subfigure}[h]{0.49\linewidth}
\includegraphics[width=\linewidth]{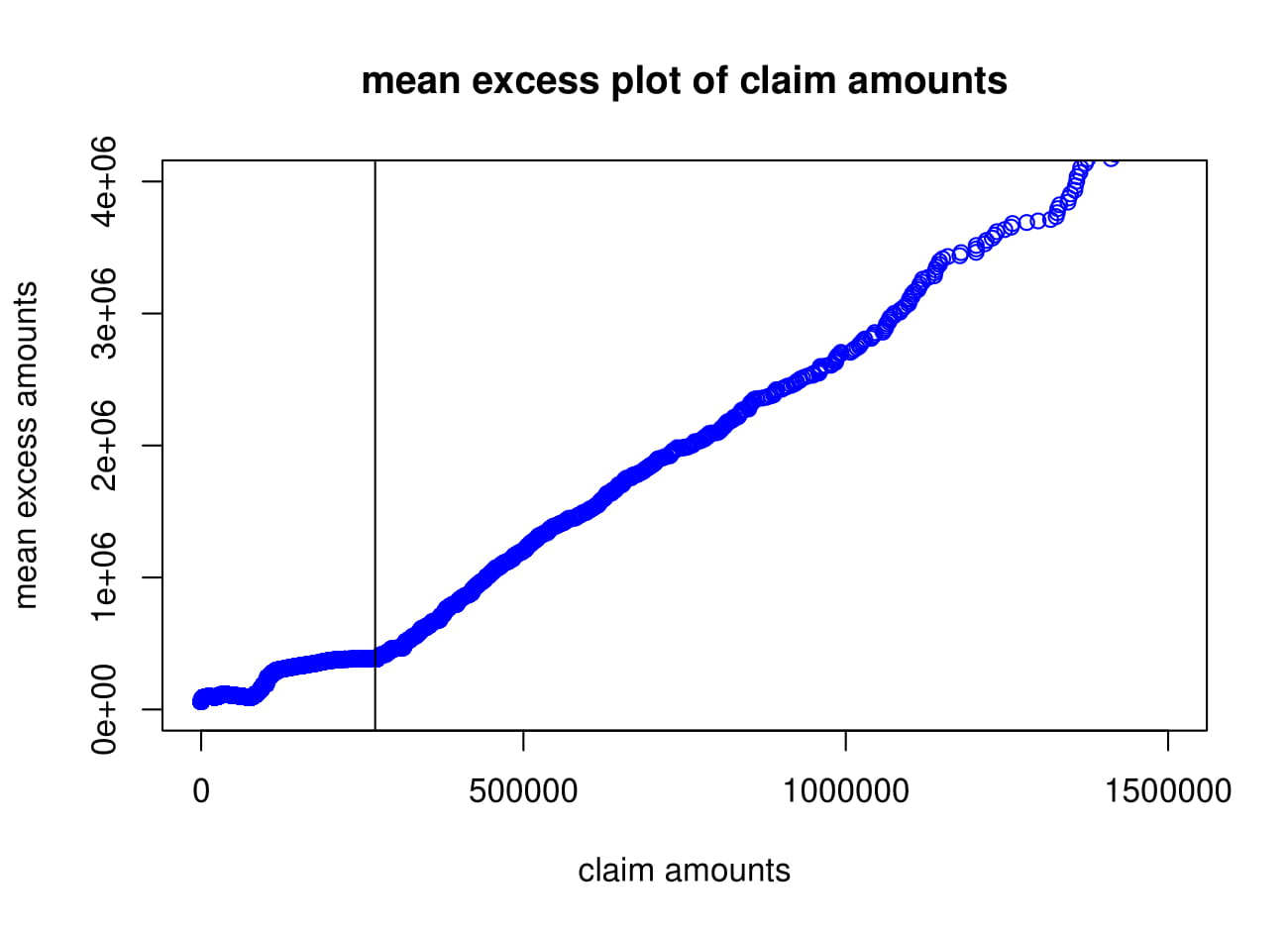}
\end{subfigure}
\end{center}
\caption{Left panel: log-log plot of the claim amounts; right panel: mean excess plot.}
\label{fig:loglogplot}
\end{figure}

Modelling the property damage claim size distribution is very challenging. Observing from Figures \ref{fig:density} and \ref{fig:loglogplot} which are also presented by \cite{fung2021mixture}, the claim size distribution is not only heavy-tailed but also multi-modal. The key complexity of the empirical distribution is that there are many small distributional nodes for smaller claims, as evidenced by the right panel of Figure \ref{fig:density}. On the other hand, it is undesirable to model all these nodes using excessive number of mixture components as (i) precise predictions of small claims are of less relevance of insurance pricing and risk management; (ii) this impedes the model interpretability. Further, the heavy-tailedness of empirical distribution is evidenced by asymptotic linearity of both log-log plot and mean excess plot in Figure \ref{fig:loglogplot}. The asymptotic slope of log-log plot suggests that the estimated tail index is $\gamma\approx 1.3$ while the Lomax tail index obtained by \cite{fung2021mixture} is about $\gamma=1.38$, under a subjective choice of splicing threshold. Note however that these only provide a very rough guidance on the true tail index.

Note that distributional multimodality and contamination are indeed prevalent not only to the aforementioned Greek dataset, but also to many publicly available insurance data sets. Notable examples include the French business interruption losses (\textbf{frebiloss}), French motor third party liability claims (\textbf{fremotor2sev9907} and \textbf{freMPL8}) and Norwegian car claims (\textbf{norauto}) which can all be retrived from the \textbf{R} package \textbf{CASdatasets}. This suggests that the modelling challenges emphasized in this paper is not only valid for the Greek data set we are analyzing, but is also applicable in many insurance claim severity data sets.

\bibliographystyle{abbrvnat}
\bibliography{reference}